\documentclass[traditabstract]{aa}  
\usepackage{graphicx}
\usepackage{txfonts}
\usepackage{natbib}
\usepackage{longtable}

\begin{document}

\title{Evidence for the concurrent growth of thick discs and central mass concentrations from S$^4$G imaging} 

   \author{S.~Comer\'on\inst{1,2}, 
           B.~G.~Elmegreen\inst{3},
           H.~Salo\inst{1},
           E.~Laurikainen\inst{1,2},
           B.~W.~Holwerda\inst{4},
           and J.~H.~Knapen\inst{5,6}
         }

   \institute{University of Oulu, Astronomy Division, Department of Physics, P.O.~Box 3000, FIN-90014, Finland\\
              \email{seb.comeron@gmail.com}
         \and
             Finnish Centre of Astronomy with ESO (FINCA), University of Turku, V\"ais\"al\"antie 20, FI-21500, Piikki\"o, Finland
             \and
             IBM Research Division, T.J.~Watson Research Center, Yorktown Hts., NY 10598, USA
             \and
             European Space Agency Research Fellow (ESTEC), Keplerlaan, 1, 2200 AG Noordwijk, The Netherlands
             \and
             Instituto de Astrof\'isica de Canarias, E-38200 La Laguna, Tenerife, Spain
             \and
             Departamento de Astrof\'isica, Universidad de La Laguna, E-38205 La Laguna, Tenerife, Spain
             }

\authorrunning{Comer\'on, S., et al.}

\abstract{We have produced $3.6\mu{\rm m}+4.5\mu{\rm m}$ vertically integrated radial luminosity profiles of 69 edge-on galaxies from the Spitzer Survey of Stellar Structure in Galaxies (S$^4$G). We decomposed the luminosity profiles into a disc and a central mass concentration (CMC). These fits, combined with thin/thick disc decompositions from our previous studies, allow us to estimate the masses of the CMCs, the thick discs, and the thin discs ($\mathcal{M}_{\rm CMC}$, $\mathcal{M}_{\rm T}$, and $\mathcal{M}_{\rm T}$). We obtained atomic disc masses ($\mathcal{M}_{\rm g}$) from the literature. We then consider the CMC and the thick disc to be dynamically hot components and the thin disc and the gas disc to be dynamically cold components. We find that the ratio between the mass of the hot components and that of the cold components, $(\mathcal{M}_{\rm CMC}+\mathcal{M}_{\rm T})/(\mathcal{M}_{\rm t}+\mathcal{M}_{\rm g})$, does not depend on the total galaxy mass as described by circular velocities ($v_{\rm c}$). We also find that the higher the $v_{\rm c}$, the more concentrated the hot component of a galaxy. We suggest that our results are compatible with having CMCs and thick discs built in a short and early high star forming intensity phase. These components were born thick because of the large scale height of the turbulent gas disc in which they originated. Our results indicate that the ratio between the star forming rate in the former phase and that of the formation of the thin disc is of the order of 10, but the value depends on the duration of the high star forming intensity phase.}


\maketitle

\section{Introduction}

Historically, the surface brightness profiles of edge-on galaxies perpendicular to their mid-planes have been fitted with a ${\rm sech}^2\,(z/z_0)$ function, where $z_0$ is the scale height \citep{SPITZ42, KRUIT81}. The roughly exponential excesses of light discovered by \citet{BURS79} and \citet{TSI79} at large heights and low surface brightnesses are now known as thick discs. They have been found in all edge-on galaxies where they have been sought \citep{DAL02, YOA06, CO11A}. Two thick discs have been found in at least one galaxy \citep{CO11C}.

Three main models have been proposed to explain the formation of thick discs. In the first, they have a secular origin caused by the vertical heating and/or radial migration of stars \citep[e.g.,][]{VILL85, SCHO09}. However, some studies show that radial migration does not heat the disc \citep{MIN12, VE14}. In the second thick disc formation mechanism, they were thick from the beginning as a result of a high velocity dispersion of the interstellar medium at high redshift; the thin disc accreted later with a lower dispersion \citep[e.g.,][]{BROOK04, EL06, BOUR09}. In the third model, they arise from interactions with satellite galaxies, either because of dynamical heating \citep[e.g.,][]{QUINN93, QU11} or because of accretion of stars \citep[e.g.,][]{AB03}. Whether this last  family of mechanisms is secular or not depends on the merging history of the galaxies.

In studies of the Milky Way thick disc, some authors favour a secular formation \citep[e.g.,][]{BO12}, while others favour an early origin \citep[e.g.,][]{GIL02, MIC13, BENS14}. Extragalactic studies based on photometric decompositions have been interpreted as evidence of an early origin \citep{YOA06, CO11B}, although it has also been suggested that the data for high-mass galaxies \citep[circular velocity $v_{\rm c}\gtrsim120\,{\rm km\,s^{-1}}$, which corresponds to a baryonic mass $\mathcal{M}\sim10^{10}\,\mathcal{M}_{\bigodot}$ using the Tully-Fisher relation in][]{ZAR14}, are compatible with an internal secular origin \citep[][hereafter CO12]{CO12}. The $v_{\rm c}=120\,{\rm km\,s^{-1}}$ limit is found to distinguish two groups of disc galaxies with different properties. Low-mass galaxies have thick discs with masses similar to those of their thin discs \citep{YOA06, CO11B}. Higher mass galaxies ($v_{\rm c}>120\,{\rm km\,s^{-1}}$) usually have thick discs that are significantly less massive than thin discs.

Central mass concentrations (CMCs) are divided into classical bulges and pseudobulges. The distinction between the two kinds of CMCs is made using a variety of criteria (S\'ersic index, stellar populations, the presence of substructures like rings or spiral arms, stellar kinematics, etc.), which often, but not always, provide an unambiguous classification. The picture is further complicated by CMCs that are the superposition of a classical bulge and a pseudobulge \citep{ER08B}. Classical bulges are believed to have formed violently early in a galaxy's history and pseudobulges are thought to have formed secularly \citep[e.g.,][]{KOR04}. However, recent numerical models have shown that some pseudobulges might also have formed quickly at high redshift \citep{SCAN11, IN12, IN14}. In these studies, the distinction between a classical bulge and a pseudobulge is made using the S\'ersic index.

Here, we consider classical bulges and pseudobulges to be part of the CMC, and we do not consider boxy/peanut inner structures to be part of the CMC; indeed, they are considered by some authors to be edge-on bars \citep{KUIJ95, LUT00}, so we consider them as a disc substructure.

In this paper, we explore the links between the masses of the gas, thin, and the thick discs and the CMCs in a sample of edge-on galaxies from the Spitzer Survey of Stellar Structure in Galaxies \citep[S$^4$G;][]{SHETH10}. The S$^4$G is a deep $3.6\mu$m and $4.5\mu$m survey made with the Infrared Array Camera \citep[IRAC;][]{FAZ04} of 2352 nearby ($v_{\rm radio}<3000\,{\rm km\,s^{-1}}$), bright ($m_{B,{\rm corr}}<15.5$), and large ($D>1^{\prime}$) galaxies away from the Galactic plane ($|b|>30^{\rm o}$). The S$^4$G data is public and can be downloaded from the NASA/IPAC Infrared Science Archive (IRSA) website\footnote{http://irsa.ipac.caltech.edu/}.

The paper is structured as follows. In Sect.~2 we describe our sample, in Sect.~3 we explain the data processing, and in Sect.~4 we give our results. In Sect.~5 we discuss and interpret our results and present our conclusions in Sect.~6. Appendix~\ref{appendix} presents the fits made with the procedure that is described in Sect.~3.

\section{The sample}

\begin{table*}
\caption{Properties of the galaxies in the sample}
\label{table}
\centering
\begin{tabular}{l c c c c c| l c c c c c}
\hline
\hline
ID & $v_{\rm c}$ & $T$ & $d$ & $M_{3.6\mu{\rm m}}({\rm AB})$& CMC &ID & $v_{\rm c}$ & $T$ & $d$ & $M_{3.6\mu{\rm m}}({\rm AB})$ & CMC\\
& (km/s) & & (Mpc)&&&& (km/s) & & (Mpc)&\\
(1)&(2)&(3)&(4)&(5)&(6)&(1)&(2)&(3)&(4)&(5)&(6)\\
\hline
ESO~157-49&115&3.0&23.8&-19.23&U&NGC~4081&103&1.0&30.2&-19.93&N\\
ESO~240-11&276&4.8&40.8&-21.94&C&NGC~4111&180&-1.4&15.6&-20.70&C?\\
ESO~292-14&131&6.5&28.7&-19.21&N&NGC~4330&126&6.3&20.4&-19.48&N\\
ESO~346-1&127&5.1&28.2&-18.87&P&NGC~4359&110&5.0&16.6&-18.29&N\\
ESO~440-27&132&6.6&20.9&-19.70&P&NGC~4437&149&6.0&11.1&-20.32&P\\
ESO~443-21&164&5.8&47.5&-20.75&U&NGC~4565&250&3.2&11.8&-21.83&CX?\\
ESO~466-14&124&3.8&47.7&-19.48&N&NGC~4607&86&3.4&22.2&-19.78&U\\
ESO~469-15&104&3.3&27.4&-18.63&N&NGC~4747&102&7.2&12.3&-18.35&N\\
ESO~533-4&159&5.2&40.2&-20.52&U&NGC~5470&121&3.1&24.7&-19.55&N\\
ESO~544-27&125&3.4&38.6&-19.27&U&NGC~5529&301&5.3&44.1&-22.20&UX\\
IC~217&104&5.9&26.0&-19.02&N&NGC~5981&225&4.3&29.2&-20.48&U\\
IC~610&152&3.9&38.1&-20.65&U&NGC~6010&158&0.4&21.6&-20.11&CU?\\
IC~1197&100&6.0&25.3&-18.73&N&NGC~7347&127&4.6&38.7&-19.96&U\\
IC~1553&128&5.2&33.4&-19.65&U&PGC~13646&179&5.1&34.4&-20.80&U\\
IC~1711&193&3.0&43.3&-21.19&CX?&PGC~28308&180&6.8&75.1&-21.72&U\\
IC~1913&79&3.4&21.0&-17.43&N&PGC~30591&122&6.8&60.3&-20.27&N\\
IC~2058&109&6.5&21.3&-18.48&N&PGC~32548&103&-0.2&49.6&-18.59&CU\\
IC~2135&113&6.0&24.6&-20.08&N&PGC~52809&112&5.9&24.2&-19.89&P\\
IC~5176&180&4.5&26.4&-20.86&U&UGC~903&182&3.9&38.6&-21.23&U\\
NGC~489&191&2.7&43.3&-20.93&C?&UGC~1970&109&5.9&36.0&-19.08&N\\
NGC~522&190&4.1&40.0&-21.03&UX&UGC~5347&104&6.5&50.8&-19.03&U\\
NGC~678&196&3.0&27.9&-21.06&CX?&UGC~5689&127&6.4&58.3&-20.01&U\\
NGC~1032&261&0.4&35.6&-21.76&C?&UGC~5958&83&4.0&29.4&-18.20&N\\
NGC~1163&147&4.1&33.6&-19.84&U&UGC~6526&90&7.0&32.3&-18.97&N\\
NGC~1422&72&2.3&21.5&-18.48&N&UGC~7086&136&3.1&37.0&-20.22&U\\
NGC~1495&114&5.0&16.7&-18.79&N&UGC~8737&169&4.0&39.1&-20.83&U\\
NGC~1596&194&-2.0&15.6&-20.22&C?&UGC~9448&123&3.2&46.7&-19.45&U\\
NGC~2732&162&-2.0&27.5&-20.52&P?&UGC~9665&146&4.0&39.6&-20.32&U\\
NGC~3098&173&-1.5&20.3&-19.72&CX?&UGC~10043&160&4.1&48.9&-20.75&C?\\
NGC~3279&174&6.5&36.7&-20.97&N&UGC~10288&180&5.3&30.8&-20.51&U\\
NGC~3454&103&5.5&26.6&-19.26&U&UGC~10297&112&5.1&38.8&-18.89&N\\
NGC~3501&146&5.9&23.8&-20.11&U&UGC~12518&176&3.0&33.9&-19.65&U\\
NGC~3592&94&5.3&26.6&-18.80&U&UGC~12692&113&3.9&49.6&-19.73&N\\
NGC~3600&111&1.0&17.0&-18.29&C?&UGC~12857&117&4.0&35.4&-19.66&U\\
NGC~3628&226&3.1&11.3&-21.73&PX\\
\hline
\end{tabular}
\tablefoot{ID (Col.~1) refers to the galaxy name and $v_{\rm c}$ (Col.~2) refers to the galaxy circular velocity from the EDD, except for ESO~466-14 \citep{MAT96}, NGC~1596 \citep{WILL10}, NGC~2732 \citep{SI97}, NGC~3098 \citep{BAL81}, and NGC~4111 \citep{SI98}. $T$ (Col.~3) indicates the galaxy stage from HyperLeda, $d$ (Col.~4) its distance from NED, and $M_{\rm 3.6\mu{\rm m}}(\rm{AB})$ (Col.~5) its absolute magnitude in $3.6\mu{\rm m}$ from the S$^4$G Pipeline~3 fluxes and Col.~4. CMC (Col.~6) indicates the kind of CMC as classified by the authors of this paper (see Section~\ref{bulge}). ``C'' stands for classical bulge, ``P'' for pseudo-bulge, ``U'' for unresolved CMC, ``X'' for boxy or X-shaped inner isophotes, and ``N'' for none. Doubtful cases are indicated by a question mark ``?''. Light from boxy or X-shaped structures has been treated as belonging to the disc in our analysis.}
\end{table*}

The galaxy sample is based on the sample used in CO12. CO12 studied 70 nearly edge-on galaxies with Hubble types $-2\leq T\leq7$, distances $11\,{\rm Mpc}\leq d\leq75\,{\rm Mpc}$ (redshift-independent distances from NED, or if not available, redshift distances corrected for the Virgo and Shapley Clusters and the Great Attractor effects and with a Hubble-Lema\^itre constant $H_0=73\,{\rm km\,s^{-1}\,Mpc^{-1}}$), and brightnesses $-17.4\leq M_{3.6\mu{\rm m}}(AB)\le-22.2$ (from the S$^4$G Pipeline~3; Mu\~noz-Mateos et al., in prep.).

In CO12 we performed thin/thick disc photometric decompositions. To ensure good fits, the galaxies were selected to have a high edge-on surface brightness, to be undistorted, and to have thin discs that could be vertically resolved. Additionally, we selected galaxies where the disc dominated in at least the outer $\sim50\%$ of the optical radius, so clean thin/thick disc decompositions could be obtained. Here, compared to the CO12 sample, we removed NGC~5084 for which we now think the bulge has a large influence on the thin/thick disc fits and is causing the antitruncation in the radial luminosity profile (therefore, the antitruncation in NGC~5084 seen in CO12 is not a true disc feature). This reduces the sample size to 69 galaxies. Our selection criteria \citep[and also those from the S$^4$G sample;][]{SHETH10} bias the sample against early-type galaxies. Only five galaxies out of 69 have $T<0$ according to HyperLeda \citep{PAT03}. The properties of the galaxies in the sample can be found in Table~\ref{table}.

The thin/thick disc photometric decompositions in CO12 give the relative masses of the thin and the thick discs. This is later used to obtain the thin and thick disc masses.

\section{Fits of the radial luminosity profiles}

\begin{figure}
  \includegraphics[width=0.45\textwidth]{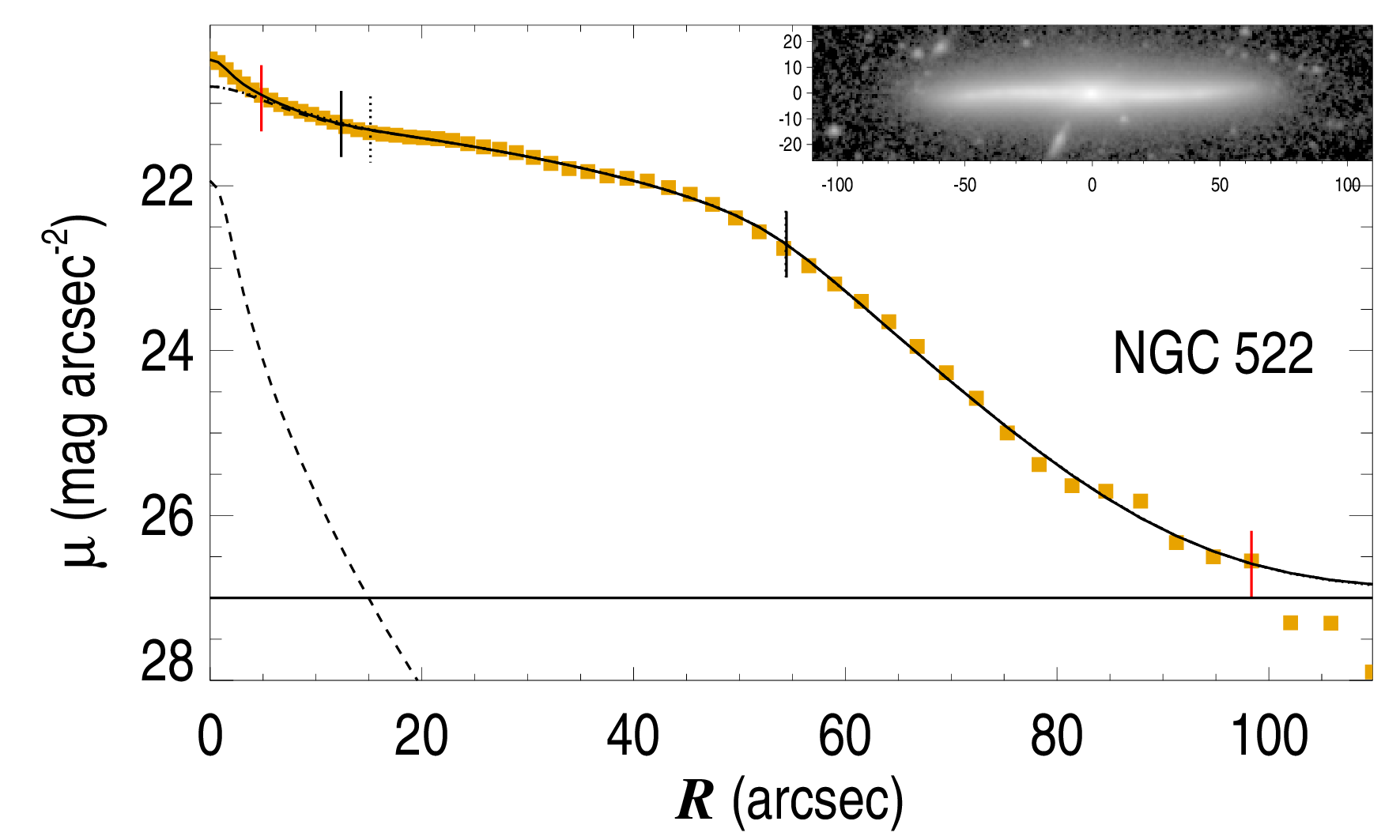}\\
  \caption{\label{plotexample} Radial luminosity profile of NGC~522 (yellow symbols) and fits to it. The vertical red bars indicate the limits of the fit in CO12. The dotted lines indicate the fitted profile and the break radii in CO12. The continuous lines indicate the fitted profile and the break radii in this paper. The dashed profiles indicate the CMC and the disc contributions. The units in the inset axes are arcseconds. The inset image is unmasked and covers the [$-z_{\rm u}$,$z_{\rm u}$] vertical range.}
\end{figure}

In CO12, we obtained radial luminosity profiles parallel to the mid-plane of the galaxies for the thin and thick discs and for the whole disc using the averages of the 3.6$\mu$m and $4.5\mu$m images. The profiles for the whole disc were averaged over ($-z_{\rm u}$,$z_{\rm u}$), where at $z_{\rm u}$ the surface brightness averaged over some radial extent is $\mu({\rm AB})=26\,{\rm mag\,arcsec^{-2}}$. This radial extent was usually between $0.2r_{25}$ and $0.8r_{25}$, but see CO12 for details. Here, we only use the luminosity profiles that correspond to the whole disc.

To produce the luminosity profiles we masked the foreground stars and the background galaxies. We expanded the masks in CO12 so the masking now also covers the region with $R<0.2r_{25}$.

The luminosity profiles in CO12 were fitted with a function that represents an edge-on disc with multiple radial truncations. Inner parts of the galaxies were excluded from the fits to avoid the CMC. Here, we are interested in both CMCs and discs, so we extend our fits all the way to the centre. The fitted function is the superposition of a function describing the disc and one describing the CMC:
\begin{equation}
 \label{total}
 J(R)=J_{\rm disc}(R)+J_{\rm CMC}(R),
\end{equation}
where $R$ is the projected distance from the centre of the galaxy in the direction of the disc plane.

We assume the disc face-on luminosity profile $I_{\rm disc}$ to be either exponential or that of a truncated disc \citep[][CO12]{ER08A} depending on the galaxy. The face-on luminosity profile of a disc with multiple truncations is
\begin{equation}
I(r^{\prime})=S\,I_0\,e^{-r^{\prime}/h_1}\prod_{i=2}^{i=n}\left\{\left[1+e^{\alpha_{i-1,i}\left(r^{\prime}-r^{\prime}_{i-1,i}\right)}\right]^{\frac{1}{\alpha_{i-1,i}}\left(\frac{1}{h_{i-1}}-\frac{1}{h_i}\right)}\right\},
\end{equation}
where $r^{\prime}$ is the 2D radius in the plane of the galaxy, $I_0$ is the central surface brightness, $S$ is a normalization factor, $n$ is the number of exponential sections, $h_i$ is the scale length of the exponential section $i$,  $r^{\prime}_{i-1,i}$ is the break radius between the sections $i-1$ and $i$, and $\alpha_{i-1,i}$ controls the sharpness of the break. As done in CO12 and \citet{LAI14}, we fix $\alpha_{i-1,i}=0\farcs5$ for all the breaks. This value is typical of the sharpness of the breaks in \citet{ER08A}. The maximum number of fitted exponential sections is $n_{\rm max}=4$. We integrate the face-on luminosity profiles along the line of sight to obtain edge-on luminosity profiles as follows:
\begin{equation}
 J_{\rm disc}(R)=F_{\rm disc}\frac{\int_{s=-s_{\rm f}}^{s=s_{\rm f}}I_{\rm disc}\left(\sqrt{R^2+s^2}\right)ds}{\int_{R=-5R_{\rm f}}^{R=5R_{\rm f}}\int_{s=-s_{\rm f}}^{s=s_{\rm f}}I_{\rm disc}\left(\sqrt{R^2+s^2}\right)ds\,dR},
\end{equation}
where the parameter $F_{\rm disc}$ describes the total flux from the disc. To ease the integration, we consider the discs to be truncated at a large 2D radius. We select this radius to be $r^{\prime}_{\rm f}=5R_{\rm f}$ where $R_{\rm f}$ is the projected radius over which the disc fits were produced in CO12, $s$ is the coordinate along the line of sight, and $s_{\rm f}=\sqrt{\left(5R_{\rm f}\right)^2-R^2}$. The free parameters of the fit are $F_{\rm disc}$, the scale lengths of the exponential disc sections, and the break radii.

We assume that the 2D surface brightness of the CMC follows a S\'ersic profile \citep{SER63}
\begin{equation}
 I_{\rm CMC}(r)=I_{\rm CMC}(r_{\rm e}){\rm e}^{\left[-b_{n}\left(r/r_{\rm e}\right)^{1/n}-1\right]}
\end{equation}
where $r$ is the 2D radius in the plane of the sky, $r_{\rm e}$ is the half-light radius, and $n$ is the S\'ersic index that describes the shape of the profile. We use $b_{n}\approx1.999n-0.327$ \citep{CA89, GRA97, GRA05}. In all cases $r_{\rm e}\ll R_{\rm f}$ and $z_{\rm u}<R_{\rm f}$, so we consider the CMCs to be truncated at a radius $r=R_{\rm f}$ to ease integration. Hence,
\begin{equation}
 J_{\rm CMC}(R)=F_{\rm CMC}\frac{{\int_{z=-{\rm min}(z_{\rm u},z_{\rm f})}^{z={\rm min}(z_{\rm u},z_{\rm f})}I_{\rm CMC}\left(\sqrt{R^2+z^2}\right)dz}}{\int_{R=-R_{\rm f}}^{R=R_{\rm f}}{\int_{z=-z_{\rm f}}^{z=z_{\rm f}}I_{\rm CMC}\left(\sqrt{R^2+z^2}\right)dz\,dR}}
 \label{bulge}
\end{equation}
where $z$ is the projected distance from the centre of the galaxy in the direction perpendicular to the galaxy mid-plane, $z_{\rm f}=\sqrt{R_{\rm f}^2-R^2}$, and $F_{\rm CMC}$ is the total flux from the CMC. The free parameters of this function are $F_{\rm CMC}$, $r_{\rm e}$, and $n$.

Because of the large number of free parameters involved, we do not fit them all at the same time. Instead, we divide the fitting process into two steps. In step one, we take the radial luminosity profile and subtract the light corresponding to the disc as fitted in CO12. In some cases, these disc fits have to be slightly changed to ease the convergence (e.g., by adding inner disc sections in regions that were not fitted in the previous paper due to the presence of the CMC). The residual is fitted with Equation~\ref{bulge} using {\sc IDL}'s {\sc curvefit}. Then, in step two, we fit the full radial profile with the function in Equation~\ref{total} using the $J_{\rm CMC}(R)$ found in step one and keeping it fixed. Then we repeat step one, but this time we subtract the disc function obtained in step two and so on. As in CO12, we used a Gaussian kernel with a $2\farcs2$ full width at half maximum (FWHM) to take into account point spread function (PSF) effects. We iterate until none of the free parameters varies by more than 1\%. In five cases with unresolved or barely resolved CMCs (NGC~3454, NGC~5981, UGC~7086, UGC~8737, and UGC~9448), we fix $n=1$ in order to prevent the S\'ersic index from diverging.

An example of a radial luminosity profile and its fit are presented in Figure~\ref{plotexample}. The fits for all the galaxies with a detectable CMC can be found in the Appendix~\ref{appendix}. The fits of radial luminosity profiles of galaxies with no detectable CMC are the same as presented in the on-line material in CO12.

The typical formal errors in the fluxes of CMCs and discs are $\sim5\%$. An additional source of error is that our fitting function considers CMCs to be spherical. To check what would happen if CMCs were in fact disky or boxy, we first create bulges with non-circular isophotes using the generalized ellipses defined in \citet{ATH90}. We then build a grid of oblate bulges with axis ratios $b/a=[0.6,0.8,1.0]$, diskyness/boxyness parameter $c=[1.7,2,2.3]$ \citep[$c$ is defined in][]{ATH90}, S\'ersic index $n=[1,2,3,4]$, effective radii $r_{\rm e}=[0.9,1.8,3.6,7.3]$ (in units of the PSF FWHM), and brightnesses $L_{\rm CMC}=[0.05,0.1,0.2,0.4]$ (in units of $L_{\rm disc}$). Then we embed these bulges into single-exponential discs with scale lengths $h=[6.8,13.6,20.5]$ (in the same units as $r_{\rm e}$). The total number of models in the grid is 1728. Then we fit the radial luminosity profiles of our synthetic galaxies with Equation~\ref{total}. We find that the fitted CMC flux is not very sensitive to the oblateness and boxiness of the CMC (typical flux errors are smaller than 10\%). However, we find that when the CMC has an effective radius comparable to the scale length of the disc, the CMC component often fits part of the disc, which causes the CMC flux to be overestimated. When this happens, the CMC effective radius is also overestimated by a large amount (a factor typically much larger than five). These cases are rare, because even in the case of S0 galaxies, the median ratio between the bulge effective radius and the disc scale length is $r_{\rm e}/h\sim0.2$ \citep{LAU10}. Also, we are confident that we would be able to detect these largely overestimated effective radii in the real galaxy fits. Once these cases are not considered, the typical uncertainty in the fitted CMC fluxes (accounting for the formal errors) is of the order of 10\%. The typical error in the fitted disc fluxes is 5\%.

The fits provide estimates of the fractions of the flux emitted by the CMC and the disc. Also, from CO12, we know the fraction of disc light that corresponds to the thin and thick disc components. By making assumptions on the value of mass-to-light ratios of the discs and the CMC, we can estimate the masses of each of these components.

Here, as in CO12, we assume that the ratio of $3.6\mu{\rm m}$ mass-to-light ratios of the thick and the thin discs is $\Upsilon_{\rm T}/\Upsilon_{\rm t}=1.2$. We assume that $\Upsilon_{\rm t}=0.55$, so the global mass-to-light ratio of the disc is compatible with the $\Upsilon\approx0.6$ value from \citet{MEIDT14}. There is some evidence that the Milky Way CMC might have a similar chemical history to that of the thick disc \citep{MEL08}. Therefore, we adopt a CMC mass-to-light ratio $\Upsilon_{\rm CMC}=\Upsilon_{\rm T}$. This also agrees with the hypothesis (discussed in Section~\ref{discussion}) that CMCs and thick discs formed at the same time. We assume that $M_{\bigodot,3.6\mu{\rm m}}({\rm AB})=6.06$ \citep{OH08}.

As in CO12 we obtain atomic gas masses from the HyperLeda fluxes with the method proposed by \citet{ZWAAN97} and by multiplying the atomic gas mass by a 1.4 factor to account for He and metals. Fluxes are not available for two galaxies, namely ESO~466-14 and PGC~30591. In what follows, we considered $\mathcal{M}_{\rm g}=0$ for these two galaxies. Molecular gas is not considered.

The masses of the gas disc, thin disc, the thick disc, and the CMC are designated as $\mathcal{M}_{\rm g}$, $\mathcal{M}_{\rm t}$, $\mathcal{M}_{\rm T}$, and $\mathcal{M}_{\rm CMC}$, respectively.

\section{Results}

\subsection{Component masses as a function of $v_{\rm c}$}

\begin{table*}
\caption{Masses of the galaxy components}
\label{table2}
\centering
\begin{tabular}{l c c c c | l c c c c}
\hline
\hline
ID&$\mathcal{M}_{\rm g}$&$\mathcal{M}_{\rm t}$&$\mathcal{M}_{\rm T}$&$\mathcal{M}_{\rm CMC}$&ID&$\mathcal{M}_{\rm g}$&$\mathcal{M}_{\rm t}$&$\mathcal{M}_{\rm T}$&$\mathcal{M}_{\rm CMC}$\\
& ($10^7\,\mathcal{M}_{\bigodot}$)& ($10^7\,\mathcal{M}_{\bigodot}$)& ($10^7\,\mathcal{M}_{\bigodot}$)& ($10^7\,\mathcal{M}_{\bigodot}$)&& ($10^7\,\mathcal{M}_{\bigodot}$)& ($10^7\,\mathcal{M}_{\bigodot}$)& ($10^7\,\mathcal{M}_{\bigodot}$)& ($10^7\,\mathcal{M}_{\bigodot}$)\\
\hline
ESO~157-49&$190\pm10$&$350\pm70$&$380\pm80$&$54\pm5$&NGC~4081&$172\pm6$&$800\pm200$&$600\pm100$&$-$\\
ESO~240-11&$5300\pm300$&$7000\pm1000$&$1200\pm300$&$1400\pm100$&NGC~4111&$200\pm20$&$1500\pm300$&$500\pm100$&$1000\pm100$\\
ESO~292-14&$650\pm40$&$440\pm90$&$320\pm60$&$-$&NGC~4330&$110\pm10$&$480\pm100$&$500\pm100$&$-$\\
ESO~346-1&$160\pm20$&$250\pm50$&$190\pm40$&$130\pm10$&NGC~4359&$138\pm9$&$180\pm40$&$150\pm30$&$-$\\
ESO~440-27&$1170\pm90$&$390\pm80$&$380\pm80$&$480\pm50$&NGC~4437&$730\pm40$&$1500\pm300$&$430\pm90$&$110\pm10$\\
ESO~443-21&$810\pm50$&$2200\pm400$&$700\pm100$&$140\pm10$&NGC~4565&$2100\pm200$&$6000\pm1000$&$1700\pm300$&$470\pm50$\\
ESO~466-14&$-$&$600\pm100$&$390\pm80$&$-$&NGC~4607&$118\pm7$&$700\pm100$&$600\pm100$&$23\pm2$\\
ESO~469-15&$500\pm50$&$240\pm50$&$210\pm40$&$-$&NGC~4747&$106\pm5$&$160\pm30$&$200\pm40$&$-$\\
ESO~533-4&$680\pm20$&$1500\pm300$&$1000\pm200$&$19\pm2$&NGC~5470&$320\pm20$&$600\pm100$&$400\pm80$&$-$\\
ESO~544-27&$420\pm30$&$480\pm100$&$280\pm60$&$34\pm3$&NGC~5529&$3800\pm100$&$7000\pm1000$&$4100\pm800$&$1000\pm100$\\
IC~217&$300\pm10$&$380\pm80$&$250\pm50$&$-$&NGC~5981&$520\pm50$&$1700\pm300$&$500\pm100$&$140\pm10$\\
IC~610&$180\pm20$&$1800\pm400$&$900\pm200$&$93\pm9$&NGC~6010&$27\pm1$&$1100\pm200$&$350\pm70$&$250\pm20$\\
IC~1197&$320\pm10$&$160\pm30$&$350\pm70$&$-$&NGC~7347&$1200\pm200$&$900\pm200$&$500\pm100$&$45\pm4$\\
IC~1553&$250\pm20$&$800\pm200$&$220\pm50$&$58\pm6$&PGC~13646&$660\pm20$&$2300\pm500$&$700\pm100$&$220\pm20$\\
IC~1711&$520\pm20$&$1800\pm400$&$1200\pm200$&$1800\pm200$&PGC~28308&$1400\pm100$&$4800\pm1000$&$1900\pm400$&$910\pm90$\\
IC~1913&$165\pm5$&$90\pm20$&$60\pm10$&$-$&PGC~30591&$-$&$1400\pm300$&$600\pm100$&$-$\\
IC~2058&$430\pm20$&$160\pm30$&$250\pm50$&$-$&PGC~32548&$690\pm20$&$230\pm50$&$120\pm20$&$83\pm8$\\
IC~2135&$480\pm30$&$700\pm100$&$1000\pm200$&$-$&PGC~52809&$1260\pm60$&$600\pm100$&$700\pm100$&$250\pm30$\\
IC~5176&$2300\pm200$&$2200\pm400$&$1100\pm200$&$84\pm8$&UGC~903&$1340\pm50$&$3300\pm700$&$1300\pm300$&$160\pm20$\\
NGC~489&$246\pm9$&$2200\pm500$&$700\pm100$&$760\pm80$&UGC~1970&$450\pm10$&$440\pm90$&$220\pm40$&$-$\\
NGC~522&$110\pm7$&$2500\pm500$&$1400\pm300$&$140\pm10$&UGC~5347&$1000\pm100$&$420\pm80$&$180\pm40$&$31\pm3$\\
NGC~678&$500\pm30$&$1800\pm400$&$700\pm100$&$1800\pm200$&UGC~5689&$880\pm30$&$1000\pm200$&$500\pm100$&$38\pm4$\\
NGC~1032&$26\pm1$&$3700\pm800$&$1500\pm300$&$2900\pm300$&UGC~5958&$110\pm10$&$150\pm30$&$160\pm30$&$-$\\
NGC~1163&$560\pm30$&$700\pm100$&$480\pm100$&$170\pm20$&UGC~6526&$120\pm10$&$350\pm70$&$260\pm50$&$-$\\
NGC~1422&$79\pm3$&$180\pm40$&$220\pm40$&$-$&UGC~7086&$900\pm40$&$1100\pm200$&$700\pm100$&$67\pm7$\\
NGC~1495&$160\pm10$&$330\pm70$&$180\pm40$&$-$&UGC~8737&$174\pm6$&$1900\pm400$&$1300\pm300$&$94\pm9$\\
NGC~1596&$210\pm20$&$700\pm100$&$280\pm60$&$1100\pm100$&UGC~9448&$820\pm40$&$600\pm100$&$370\pm70$&$9\pm1$\\
NGC~2732&$35\pm5$&$1000\pm200$&$700\pm200$&$810\pm80$&UGC~9665&$1700\pm200$&$1500\pm300$&$460\pm90$&$120\pm10$\\
NGC~3098&$140\pm20$&$500\pm100$&$260\pm50$&$510\pm50$&UGC~10043&$2600\pm200$&$1500\pm300$&$1100\pm200$&$660\pm70$\\
NGC~3279&$360\pm30$&$2400\pm500$&$1400\pm300$&$-$&UGC~10288&$1800\pm60$&$1400\pm300$&$900\pm200$&$180\pm20$\\
NGC~3454&$150\pm10$&$500\pm100$&$260\pm50$&$21\pm2$&UGC~10297&$410\pm20$&$360\pm70$&$200\pm40$&$-$\\
NGC~3501&$520\pm20$&$1000\pm200$&$700\pm100$&$38\pm4$&UGC~12518&$260\pm30$&$500\pm100$&$360\pm70$&$300\pm30$\\
NGC~3592&$82\pm8$&$240\pm50$&$270\pm50$&$19\pm2$&UGC~12692&$290\pm20$&$800\pm200$&$400\pm80$&$-$\\
NGC~3600&$680\pm50$&$100\pm20$&$130\pm30$&$120\pm10$&UGC~12857&$980\pm80$&$800\pm200$&$290\pm60$&$37\pm4$\\
NGC~3628&$740\pm30$&$3000\pm600$&$4400\pm900$&$650\pm70$\\
\hline
\end{tabular}
\tablefoot{$\mathcal{M}_{\rm g}$ from HyperLeda. The remaining masses are from fits in this paper and CO12.}
\end{table*}

\begin{figure*}
\begin{center}
  \includegraphics[width=0.9\textwidth]{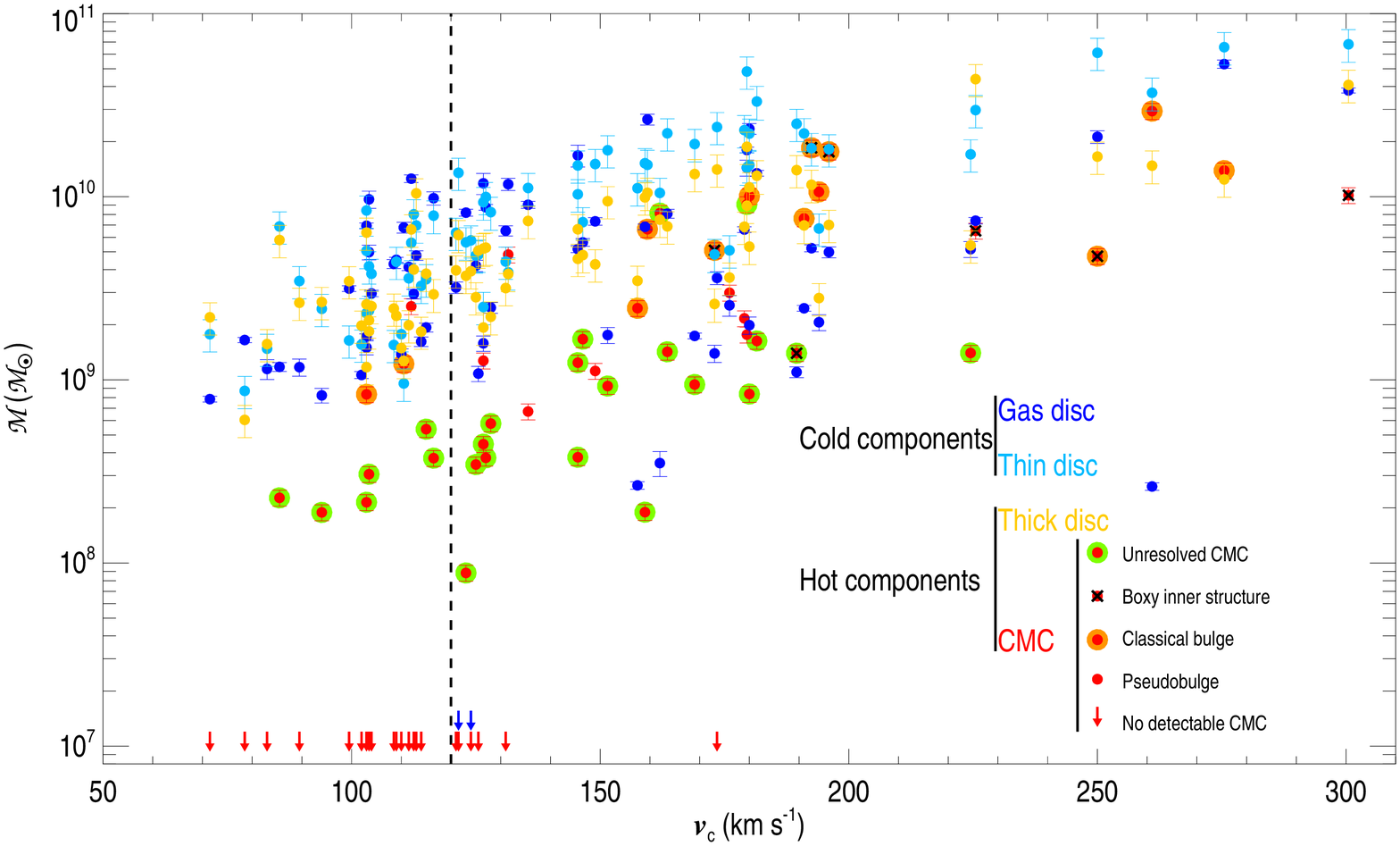}\\
\end{center}
  \caption{\label{thinthickmasses} Mass of the gas discs (deep blue circles), thin discs (blue circles), thick discs (orange circles), and CMCs (red circles) as a function of the galaxy circular velocity. Red arrows indicate galaxies for which no significant CMC is found. Red circles in large green circles indicate unresolved CMCs, red circles in large dark orange circles indicate classical bulges, and simple red circles correspond to pseudobulges. Deep blue arrows indicate galaxies for which the 21-cm flux is not available. Crosses indicate boxy/X-shaped inner structures that are superimposed onto a given CMC. The vertical dashed line corresponds to $v_{\rm c}=120\,{\rm km\,s^{-1}}$.}
\end{figure*}

In Figure~\ref{thinthickmasses} we plot the gas disc, thin disc, the thick disc, and the CMC masses as a function of the galaxy circular velocity. The galaxy circular velocity is a good proxy for the total galaxy mass. In most cases, we use circular velocities based on gas rotation from the Extragalactic Distance Database \citep[EDD;][]{COUR09}. In two cases for which EDD velocities are not available (ESO~466-14, NGC~3098), we use data from \citet{MAT96} and \citet{BAL81}. However, in the case of perturbed or truncated gas discs, their circular velocities might not be representative of the true mass of the system. We compared H{\sc I} circular velocities with those from stellar velocities based on absorption lines in \citet{SI97, SI98} (four galaxy overlap) and those from dynamical models in \citet{WILL10} (four galaxy overlap). We find them to be the same within 10\% except for NGC~1596, NGC~2732, and NGC~4111. These three galaxies are S0s which are likely not to have much gas. For these three galaxies, we adopt the circular velocities not based on gas.

In Figure~\ref{thinthickmasses}, the error bars for the CMC masses correspond to a 10\% fitting uncertainty. The thin and thick disc masses are calculated using the fraction of light corresponding to the disc and by using the ratio of thick to thin disc masses from CO12 ($\mathcal{M}_{\rm T}/\mathcal{M}_{\rm t}$). The errors in the thin and thick disc masses come from a quadratic sum of two factors. The first one is the 5\% fitting uncertainty in the fraction of light assigned to the combined disc from our luminosity profile fit. The second one is a fixed 20\% of the thin or thick disc mass. This factor is a reasonable estimate of the error caused by the biases in the thin/thick disc fits (Section 4.3.2 in CO12). The fitted $\mathcal{M}_{\rm t}$, $\mathcal{M}_{\rm T}$, and $\mathcal{M}_{\rm CMC}$ values are presented in Table~\ref{table2}. This table also includes the atomic gas mass estimates, $\mathcal{M}_{\rm g}$. The errors in the gas masses come from the flux error estimates from HyperLeda.

At least two kinds of CMCs may have their measured masses different from the real value for reasons other than those accounted by the error bars. First, some CMCs might actually be nuclear clusters with a small $\Upsilon_{\rm CMC}$. This is most likely to happen in unresolved CMCs which are marked in Figure~\ref{thinthickmasses} with green circles. Here we have marked as unresolved those CMCs with $r_{\rm e}<3^{\prime\prime}$. Second, in seven cases, the CMC is superposed to a boxy or X-shaped structure which could be caused by an edge-on bar. These structures are detected visually. Although the light of these structures seems to be mostly fitted by the inner disc sections, they could slightly perturb the fitting of the CMC. In Figure~\ref{thinthickmasses}, we indicate CMC fits that could be perturbed by X-shaped structures with crosses. Because these structures are fitted as part of the disc, their light is assigned into the thin and the thick disc components in the same proportions as for the rest of the disc light.

CMCs that are resolved are visually classified as classical bulges or pseudobulges according to their morphology. CMCs protruding the disc or with round isophotes (axis ratio $b/a\gtrsim0.6$) are considered to be classical bulges. This is because classical bulges are generally thought to be pressure supported, while pseudobulges are often thought to have a significant amount of rotation and therefore be more flattened than classical bulges. All protruding CMCs have round isophotes. This classification is uncertain in many cases due to its subjectivity, and also to the fact that some of the CMCs are poorly resolved.

The CMC classifications are presented in Table~\ref{table} and doubtful cases are indicated with a question mark.

The masses of the thin discs, the thick discs, and the CMCs increase with $v_{\rm c}$. Thick discs generally have masses in the range $10^9\mathcal{M}_{\bigodot}-10^{10}\mathcal{M}_{\bigodot}$  for low-mass galaxies ($v_{\rm c}\lesssim120\,{\rm km\,s^{-1}}$) but they can be as massive as several times $10^{10}\mathcal{M}_{\bigodot}$ for galaxies with a larger mass. As shown in \citet{CO11B} and CO12, thin discs have masses similar to those of thick discs for $v_{\rm c}<120\,{\rm km\,s^{-1}}$ and are on average more massive than thick discs for galaxies with larger rotation velocities. CMCs are almost always less massive than both the thin and the thick discs. All but five of the 23 galaxies without a CMC are found in galaxies with $v_{\rm c}\lesssim120\,{\rm km\,s^{-1}}$. Gas disc masses have little or no dependence with $v_{\rm c}$ and are most often in the $10^9\mathcal{M}_{\bigodot}-10^{10}\mathcal{M}_{\bigodot}$ range.

\subsection{Mass ratios of the different components}

\begin{figure}
  \includegraphics[width=0.45\textwidth]{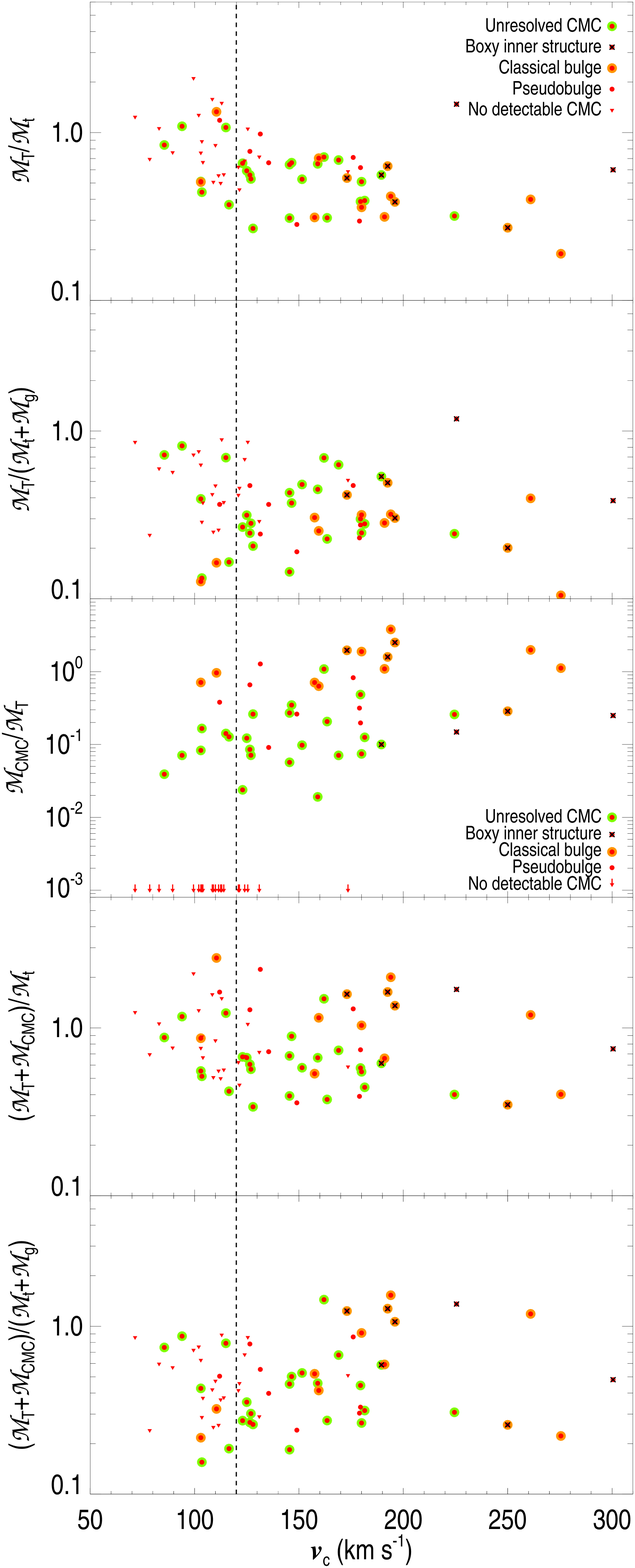}\\
  \caption{\label{ratios} Mass of the thick disc divided by that of the thin disc (first panel), mass of the thick disc divided by that of the cold component (second panel), mass of the CMC divided by that of the thick disc (third panel), and mass of the hot component divided by that of the thin disc (fourth panel), and mass of the hot component divided by that of the cold component (fifth panel), as a function of the circular velocity. The symbols indicate the properties of the CMC of each galaxy and follow the same conventions as in Figure~\ref{thinthickmasses} except for the triangles in the first, second, third, and fourth and the bottom panels that indicate galaxies with no detectable CMC. In the second and fifth panels, the two galaxies with no available 21-cm flux measurement are considered to have $\mathcal{M}_{\rm g}=0$.}
\end{figure}

In this paper, we consider that the dynamically ``hot component'' is the sum of the CMC and the thick disc. This is based on the observation that young galaxies are highly turbulent \citep{FORS09, KAS12}, which means that their gas discs are thicker than today's gas discs. Stars that formed in this turbulent gas should have shared the same large thickness as the gas, just like today's young stars have the same thickness as today's gas. This means that all stars in a young galaxy were in a relatively thick disc; there was no thin disc component for either the stars or the gas \citep{BOUR09}. Consequently, any angular momentum exchange inside the disc at that time, leading to disc contraction or clump migration to the centre, will have made a thick central concentration. This differs from disc migration in a later stage cool disc, which can make a thin inner concentration. It also differs from the later formation of a bar out of a cool inner stellar disc, and from the thickening produced by the bar through vertical instabilities and bar buckling.  Thus we consider both thick discs and CMCs formed in an early disc with a large scale height compared to that of the thin disc. The dynamically ``cold component'' is made of the thin disc and the gas disc (for which the atomic gas mass is a lower mass boundary).

Even though some of our CMCs might actually be unresolved nuclear clusters unrelated to classical bulges and pseudobulges, we are not making a large error at measuring the mass of the hot component. This is because the unresolved CMCs usually are the least massive ones. Also, we are ignoring the fact that at least part of the mass of pseudobulges is a product of secular evolution and are therefore a disc-like structure. We visually checked whether the six pseudobulges identified in the sample were thicker than the thin disc and could therefore be considered a ``hot component''. We find that three of them (NGC~3628, NGC~4437, and PGC~52809) have thicknesses comparable to that of the thin disc. The three have relatively small masses ($\mathcal{M}_{\rm CMC}/(\mathcal{M}_{\rm CMC}+\mathcal{M}_{\rm T}+\mathcal{M}_{\rm t}+\mathcal{M}_{\rm g})<0.1$). These three ``flat'' CMCs might be the product of secular evolution and actually be ``disky'' pseudobulges.

Previous studies showed that $\mathcal{M}_{\rm T}/\mathcal{M}_{\rm t}$ decreases with $v_{\rm c}$ \citep[][CO12]{YOA06, CO11B}. We show this trend in the top panel of Figure~\ref{ratios}. A similar but milder trend is found for the ratio between the thick disc mass and the cold component mass, $\mathcal{M}_{\rm T}/(\mathcal{M}_{\rm t}+\mathcal{M}_{\rm g})$ \citep[second panel in Figure~\ref{ratios}, also shown in][and in CO12]{YOA06}. These two trends might be related to that found  in a sample of moderately inclined galaxies at intermediate redshifts by \citet{SHETH12}: their Figure 4 relates kinematic properties of the galaxies (from optical spectroscopy) to their masses and shows that low-mass galaxies tend to be dynamically hotter than those with a larger mass. In other words, the smaller the mass, the larger the ratio $\sigma/v_{\rm c}$, where $\sigma$ is the velocity dispersion. The trend presented here and the one displayed in \citet{SHETH12} might be related because the component of the velocity dispersion perpendicular to the galaxy plane ($\sigma_{z}$) and the scale height ($z_0$) have a positive correlation. \citet{SHETH12}, however, measured gas kinematics and not stellar kinematics.

The third panel of Figure~\ref{ratios} shows the ratio of the masses of the CMC and the thick disc as a function of $v_{\rm c}$. It shows a trend between $\mathcal{M}_{\rm CMC}/\mathcal{M}_{\rm T}$ and $v_{\rm c}$. This indicates that the hot component is more centrally concentrated in more massive galaxies.

In the fourth and fifth panels of Figure~\ref{ratios}, we show that the ratio between the mass of the hot component (sum of the thick disc and the CMC) and that of the thin disc and the ratio between the mass of the hot and the cold components have no clear trend with the circular velocity, although the plots have a large scatter. The lack of a clear trend in these ratios might be caused by large intrinsic scatter and fitting uncertainties which might mask any correlation. However, a more interesting possibility is that these ratios are indeed mass-invariant. This is because, as seen in the first, the second and the third panels, $\mathcal{M}_{\rm T}/\mathcal{M}_{\rm t}$ and $\mathcal{M}_{\rm T}/(\mathcal{M}_{\rm t}+\mathcal{M}_{\rm g})$ decrease with $v_{\rm c}$ whereas $\mathcal{M}_{\rm CMC}/\mathcal{M}_{\rm t}$ increases with $v_{\rm c}$ in a way that the two effects roughly cancel each other.

\section{Discussion}

\label{discussion}

\subsection{Star formation rates}

The star formation intensity (SFI) in a galaxy disc (the star formation rate, SFR, per unit of area) can be expressed as $\dot{\Sigma}_{*}=\epsilon_{\rm orb}\Sigma/t_{\rm orb}$, where $\Sigma$ is the total surface density, $\epsilon_{\rm orb}$ is the fraction of mass converted into stars in a galactic rotation, and $t_{\rm orb}$ is the rotation period of the studied region \citep[here we use the formalism from][]{KRUM12}. In a single-component disc in equilibrium \citep{SPITZ42} the scale height is $z_0=\sigma_{z}^2/(\pi G\Sigma)$. In a gas-rich disc, the Toomre parameter is $Q=\sqrt{2(\beta+1)}\sigma\Omega/(\pi G\Sigma)$, where $\beta$ describes the slope of the rotation curve, and $\Omega$ is the angular rotation velocity \citep{KRUM12}. If we assume that $\sigma_z$ is porportional to $\sigma$, in a gas-rich galaxy with a single-component disc
\begin{equation}
 \dot{\Sigma}_{*}\propto\frac{\epsilon_{\rm orb}}{t_{\rm orb}}\frac{2(\beta+1)\Omega^2}{\pi GQ^2}z_0
\end{equation}
and the SFI is proportional to the scale height ($\dot{\Sigma}_{*}\propto z_0$). Dynamically hot components, if born already hot, must have formed at a time when the SFI was larger than it is today. Thick discs being a consequence of turbulence caused by a high SFI was already proposed by \citet{LEN14}. High turbulence could also be caused by accretion energy \citep{EL10} and by gravitational instabilities and stirring from the massive clumps that form in those instabilities, even without star formation feedback \citep{BOUR09}.

The conditions for a large scale height were fulfilled during the early disc phase of galaxies when the cosmic star formation rate was larger than it is now \citep[see][for a recent review]{MA14}. This epoch overlaps with that of clumpy discs described in \citet{EL06}. The size of a clump formed by gravitational instabilities is about the Jeans' length, and this is about the same as the disc thickness because both depend on the ratio of the square of the velocity dispersion, $\sigma$, to the disc mass column density, $\Sigma$, so the inverse Jeans' wavenumber is $\lambda_{\rm J}=\sigma^2/\left(\pi G \Sigma\right)$. Thus the observed large sizes of star-forming clumps in young galaxies serve as an indirect measure for large disc thicknesses: clumpy discs should also be thick discs \citep{EL09}. If the hot components formed at that time, then the ratio the hot component mass ($\mathcal{M}_{\rm T}+\mathcal{M}_{\rm CMC}$) to the thin disc mass ($\mathcal{M}_{\rm t}$) should be equal to the ratio of the SFR multiplied by the duration of star formation for the hot and the cool phases:
\begin{equation}
\label{eq}
 \frac{\mathcal{M}_{\rm T}+\mathcal{M}_{\rm CMC}}{\mathcal{M}_{\rm t}}\approx\frac{\dot{\Sigma}_{\rm *h}\times t_{\rm h}}{\dot{\Sigma}_{\rm *t}\times t_{\rm t}}.
\end{equation}
From Figure~\ref{ratios} we find that $(\mathcal{M}_{\rm T}+\mathcal{M}_{\rm CMC})/\mathcal{M}_{\rm t}\sim1$. Therefore, because $\dot{\Sigma}_{\rm *h}>\dot{\Sigma}_{\rm *t}$ (this is a consequence of $\dot{\Sigma}_{*}\propto z_0$), the time taken to build the hot component was shorter than the time to build the cool thin disc, $t_{\rm h}<t_{\rm t}$. Assuming that $t_{\rm h}\sim1\,{\rm Gyr}$ \citep[the approximate duration of the disc clumpy phase;][]{BOUR07} and $t_{\rm t}\sim10\,{\rm Gyr}$ (the time elapsed since then), we find $\dot{\Sigma}_{\rm *h}/\dot{\Sigma}_{\rm *t}$ to be $\sim10$ with a large spread. However, this ratio is highly dependent on the duration of the intense star formation phase. Longer or shorter hot component formation times can change it significantly. To exemplify this, we can calculate this ratio for the Milky Way by using star formation histories in the literature. \citet{SNAITH14} claim that for the Galaxy $\mathcal{M}_{\rm T}/\mathcal{M}_{\rm t}\sim1$. They find that the thick disc formed in an intense star formation phase that lasted for $\sim3.5\,{\rm Gyr}$ and that the thin disc has formed later during $\sim9\,{\rm Gyr}$. Assuming that the mass of the CMC of the Milky Way is small compared to that of the disc components, we obtain $\dot{\Sigma}_{\rm *h}/\dot{\Sigma}_{\rm *t}\sim2.5$.

\subsection{Thick disc and CMC formation}

The observations presented here suggest that about half of the stars in a galaxy are in a hot and thick component. This is a sensible fraction considering the variation of the star formation rate and mass build-up over a Hubble time.  According to the toy model in \citet{DE13}, half of the mass in a galaxy is accreted before $z\sim0.9$, when the accretion rate exceeded by $2.4$ times the current rate ($\mathcal{M}\propto{\rm exp}(-0.79z)$ and $\dot{\mathcal{M}}\propto{\rm exp}(-0.79z)(1+z)^{5/2}$; the universe is assumed to be matter-dominated for most of a galaxy's accretion history). In an equilibrium situation, and assuming that the surface of the galaxy does not change dramatically, the SFI is roughly proportional to the accretion rate \citep{DE13}. With the assumption of a constant Toomre $Q$ and pressure equilibrium in the disc \citep{KRUM12}, a SFI that scales with the gas column density also scales with the velocity dispersion and the scale height. Therefore, one would expect the scale height for instantaneous star formation to scale with the accretion rate. This means that by the time the first half of the stellar disc formed, at $z\sim0.9$, the thickness had always been larger than 2.4 times the present thickness, just from the higher accretion and SFI at these earlier times. The second half of the disc formed thinner because of the lower SFI. If we identify the first half of the assembly with the thick disc and CMC components of today's galaxies and the second half with the thin disc and gas components, then this minimum theoretical thickness for the first half is consistent with our observations in \citet{CO11B}, where the thick-to-thin scale height ratio is $z_{\rm T}/z_{\rm t}\sim4$, increasing slightly for earlier types. In this assembly model, the old and thick discs of spiral galaxies are direct results of the high accretion rates at earlier cosmic times. 

This picture is consistent with the formation of the thick disc and CMC in a relatively short time in a clumpy disc \citep{BOUR07, BOUR09} at high redshift and a subsequent slow formation of the thin disc. It is also consistent with the $\alpha$-enhancement found for the Galaxy CMC and thick disc \citep{MEL08}, and with the age differences between extragalactic thin and thick discs found by \citet{YOA08}. Different simulations have shown that the inward migration of clumps in clumpy discs can build classical bulges \citep{EL08, PER13} and/or pseudobulges \citep{IN12, IN14}. The mechanisms causing the loss of angular momentum of clumps (ultimately leading to their merger and the formation of the CMC) would then be more efficient in the most massive galaxies, where the CMC mass is the largest.

The exact relationship between the SFI and the SFR is hard to asses. \citet{LEN14}, in their Milky Way SFI study, assume that for the Milky Way the thick disc scale length is shorter than the thin disc scale length as advocated by many \citep{BENS11, BO12, CHENG12} using chemical properties to divide stellar populations. Thus, the reduction in the SFI would be both a consequence of the increase in the surface of the disc and the reduction of the gas accretion. Interestingly, studies based on structural decompositions show the opposite, namely that the scale length of the thick disc is larger than that of the thin disc \citep{JUR08, CHANG11, POL13, LOP14}. Regarding other galaxies, structural decompositions show that in general thick discs have large scale lengths compared to thin discs \citep[][CO12]{YOA06}, which is in uncomfortable tension with some of the studies regarding the Milky Way morphology. In this case, the SFI evolution would be driven by two processes working in oposite directions, namely a reduction in the accretion rate and a reduction of the radius within which the star formation is found. The fact that thin discs are indeed thin shows that the effect of the decreased gas accretion would be more important than that of the shrinking disc.

The mean CMC-to-total mass ratio in our study of edge-on galaxies is $B/T=0.09$. This is comparable to the observed $B/T$ in not very inclined Sab-Sc galaxies \citep{LAU10}, but is smaller than that found in the simulation papers where a CMC is formed in a clumpy disc. \citet{EL08} fit a double-disc to the outer parts of galaxies and define the CMC as the central excess of light. They find $B/T$ values in the range between 0.12 and 0.36. \citet{PER13} make S\'ersic+exponential disc radial fits to their models and obtain $B/T$ values in the range between 0.24 and 0.39. \citet{IN14} also use S\'ersic+exponential fits and obtain $B/T$ in the range between 0.23 and 0.35. A reason for the difference between our $B/T$ values and those in simulations is that the latter correspond to the ratio before the thin disc formed. Hence, it corresponds to $\mathcal{M}_{\rm CMC}/(\mathcal{M}_{\rm CMC}+\mathcal{M}_{\rm T})$ rather than to $\mathcal{M}_{\rm CMC}/(\mathcal{M}_{\rm CMC}+\mathcal{M}_{\rm T}+\mathcal{M}_{\rm t})$. The average $\mathcal{M}_{\rm CMC}/(\mathcal{M}_{\rm CMC}+\mathcal{M}_{\rm T})$ vtalue for our sample is 0.19, which is closer to what is found in simulations.

\subsection{Are all CMCs hot components?}

A caveat of our study is that we consider the whole mass of CMCs to be part of a hot component made at high redshift. This assumption may not always be true, especially for pseudobulges, often described as the consequence of a secular process.

A recent study of the colours of pseudobulges (defined to have $n<2.5$) in a sample of isolated galaxies has shown that $\sim2/3$ of them are found in the red sequence \citep{FER14}. They show that those red pseudobulges have colours compatible with a single star formation burst 8\,Gyr ago ($z\sim1$ in a standard cosmology). \citet{ART09} also find for a sample of CMCs with $n<2.5$ that the fraction of the stellar mass in the CMC with ages $>10\,{\rm Gyr}$ is at least 50\%. However, pseudobulges have a fraction of mass built secularly either in a continuous or an episodic way: \citet{FISH09} define the CMC growth time as its stellar divided by the present-day SFR and find that $\sim50\%$ of their pseudobulges have a growth time of less than 10\,Gyr (the current SFR is larger than the past average SFR which is the CMC mass divided by a Hubble-Lema\^itre time) and thus still have a significant SFR. In their study, pseudobulges are defined to have $n<2$ and/or disc morphology features such as spiral arms.

Another piece of evidence for the building of a large fraction of the CMC mass during the large SFI phase is found in \citet{DOK13}. They find that after $z\sim1$, the SFR of Milky Way-like galaxies drops, and that this drop is stronger in the central 2\,kpc. Therefore, while the disc continues to be slowly built after $z\sim1$, the growth of the CMC has almost stopped.

Whatever is the fraction of mass in CMCs built at high redshift, the total mass of the CMC is only an upper limit to the mass of the CMC built as a hot component at high $z$.

\section{Conclusion}

In this paper, we study vertically integrated radial luminosity profiles from mid-infrared images of 69 edge-on galaxies from the S$^4$G. We produce photometric decompositions of the profiles to distinguish the light that comes from the central mass concentration (CMC) from that coming from the disc. Here we assume CMCs to be classical bulges and pseudobulges, and we exclude boxy and X-shaped features because we consider them to be related to disc features (bars). Hence, the light from boxy and X-shaped features is assigned to discs in our analysis. We combine this information with that in CO12 and assumptions on components' mass-to-light ratios to obtain the masses of the CMCs, thin discs, and thick discs ($\mathcal{M}_{\rm CMC}$, $\mathcal{M}_{\rm t}$, and $\mathcal{M}_{\rm T}$, respectively). We obtain atomic gas disc masses, $\mathcal{M}_{\rm g}$, from 21-cm fluxes in HyperLeda. We consider the sum of the CMC and the thick disc to be the dynamically hot component of a galaxy and the sum of the gas disc and the thin disc to be the dynamically cold component.

We confirm a previous result that the ratio of the thick to thin disc masses, $\mathcal{M}_{\rm T}/\mathcal{M}_{\rm t}$, decreases with increasing $v_{\rm c}$. A similar but milder trend is found between $\mathcal{M}_{\rm T}/(\mathcal{M}_{\rm t}+\mathcal{M}_{\rm g})$ and $v_{\rm c}$. However, the ratio of hot component to thin disc masses, $(\mathcal{M}_{\rm CMC}+\mathcal{M}_{\rm T})/\mathcal{M}_{\rm t}$, and the ratio of the hot component to cold component masses, $(\mathcal{M}_{\rm CMC}+\mathcal{M}_{\rm T})/(\mathcal{M}_{\rm t}+\mathcal{M}_{\rm g})$, are roughly independent of the galaxy mass (although with a large scatter). We also find that the larger $v_{\rm c}$, the more centrally concentrated the hot component is (the ratio $\mathcal{M}_{\rm CMC}/\mathcal{M}_{\rm T}$ becomes larger).

We suggest that our results are compatible with the two hot components (the thick disc and the CMC) being built in a short time during the rapid accretion phase of disc galaxies when the discs were clumpy and the velocity dispersions, scale heights, and star formation rates were all high. In this picture, the thin disc slowly forms afterwards from gas accreted in cold flows. Moreover, we suggest that the CMC formed by concentration of the thick disc during this hot phase, and that is why both the thick disc and the CMC together give a fixed fraction of the total disc mass, and not just one or the other alone. We compare  the ratio of hot component to thin disc masses, $(\mathcal{M}_{\rm CMC}+\mathcal{M}_{\rm T})/\mathcal{M}_{\rm t}$, in our sample with the predictions from toy models. We estimate that the SFI during the formation of the hot components was roughly ten times larger than it was during the formation of the thin disc. The value of the f SFI during the former phase relative to that of the thin disc formation is a function of the duration of the high star formation intensity epoch.

When combined with observations of galaxies at intermediate to high redshift, our results suggest that the hot disc phase occurred during the high accretion phase of galaxy formation, when the relative gas mass was high following the high accretion rate, the star formation rate was high because of the large relative gas mass, and the gas turbulent speed was high as a result of several processes, including accretion energy input, violent gravitational instabilities, stirring by massive clumps that form in these instabilities, and feedback by rapid star formation.  In this model, the high turbulent speed for the gas is the single most important reason for the thick disc and for any thick central mass concentration that forms in it. Highly turbulent discs are automatically thick because of their large pressures. Any stars that form in a thick gas disc will share the high random speeds and thick distributions of the gas. Bulge formation is a secondary step involving angular momentum transfer in a disc that is already thick from high turbulence.

\acknowledgements

We thank our referee, Matthew Lehnert, who helped to substantially improve the discussion section of this paper. We thank Sim\'on D\'iaz-Garc\'ia, Mirian Fern\'andez Lorenzo, Mart\'in Herrera-Endoqui, Joannah L.~Hinz, Luis C.~Ho, and Kartik Sheth for commenting on the manuscript. This work is based on observations and archival data made with the Spitzer Space Telescope, which is operated by the Jet Propulsion Laboratory, California Institute of Technology under a contract with NASA. We are grateful to the dedicated staff at the Spitzer Science Center for their help and support in planning and execution of this Exploration Science program. We also gratefully acknowledge support from NASA JPL/Spitzer grant RSA 1374189 provided for the S$^4$G project. We acknowledge financial support to the DAGAL network from the People Programme (Marie Curie Actions) of the European Union's Seventh Framework Programme FP7/2007-2013/ under REA grant agreement number PITN-GA-2011-289313. This research has made use of the NASA/IPAC Extragalactic Database (NED) which is operated by the Jet Propulsion Laboratory, California Institute of Technology, under contract with the National Aeronautics and Space Administration.

\bibliographystyle{aa}
\bibliography{thickbulge}

\appendix

\section{Fits of the radial luminosity profiles}
\label{appendix}

This Appendix presents the radial fits to the luminosity profiles of the galaxies with a detectable CMC. The fits are displayed as in Fig.~\ref{plotexample}.

\clearpage

\begin{figure}
  \includegraphics[width=0.45\textwidth]{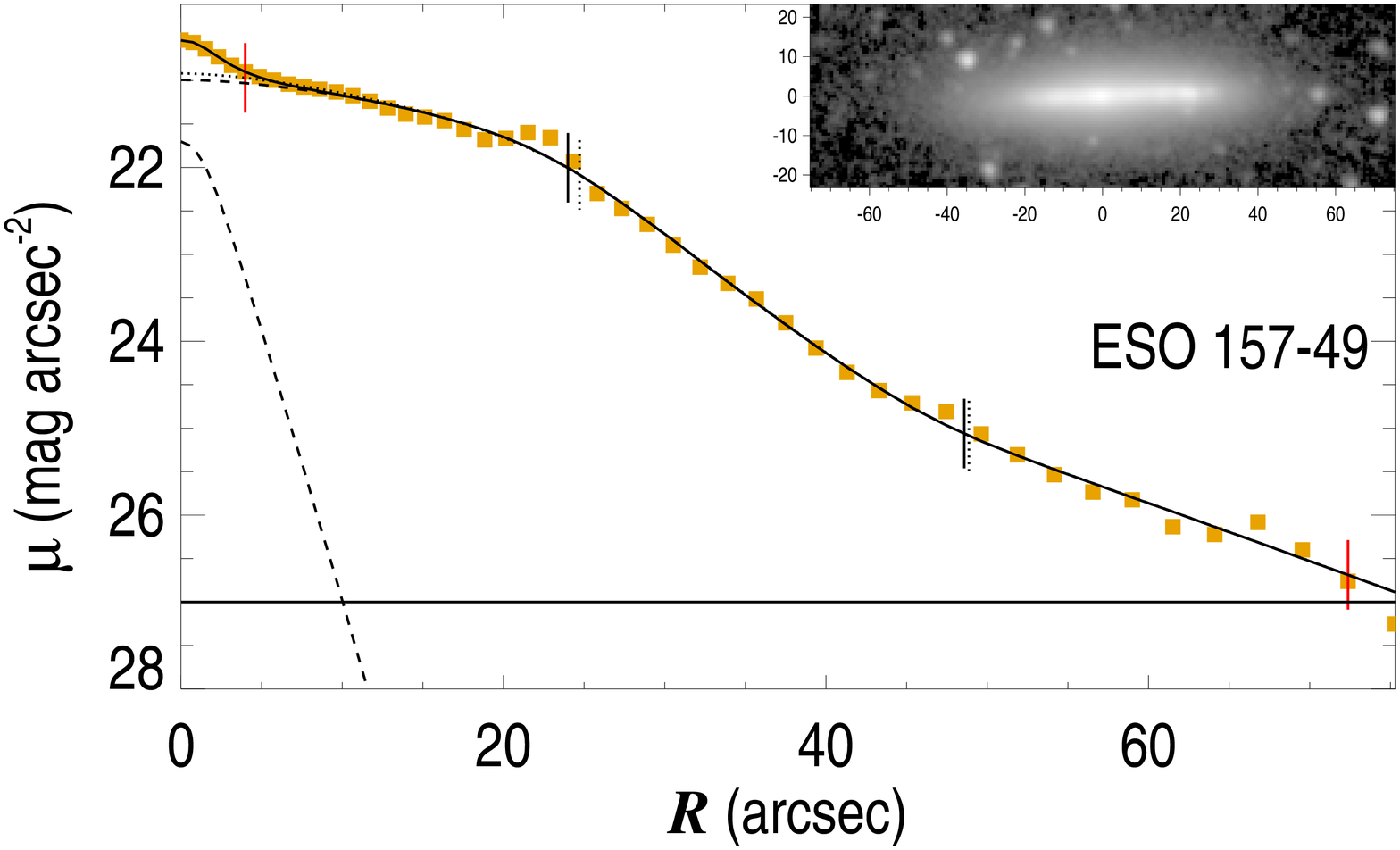}\\
\end{figure}

\begin{figure}
  \includegraphics[width=0.45\textwidth]{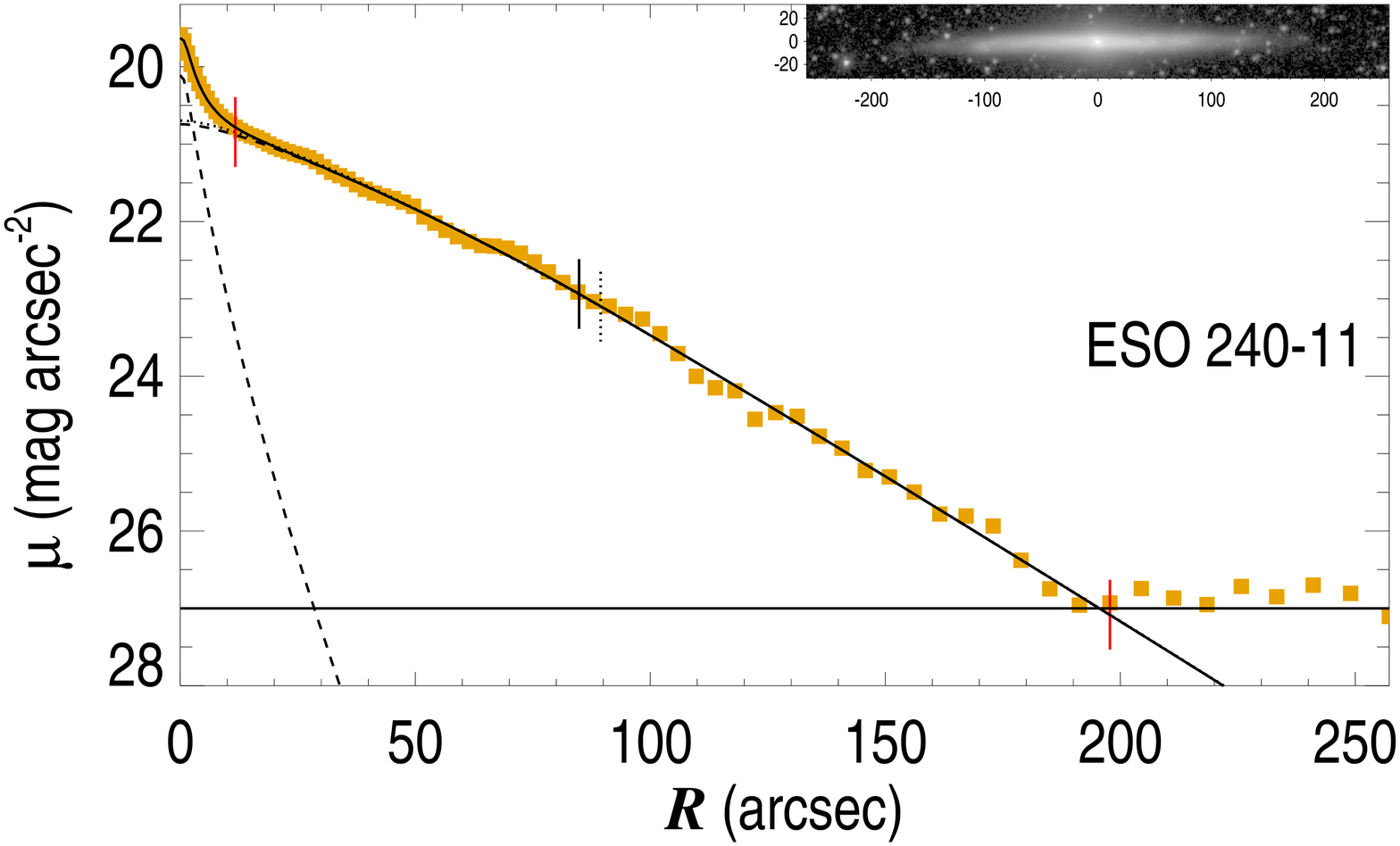}\\
\end{figure}

\begin{figure}
  \includegraphics[width=0.45\textwidth]{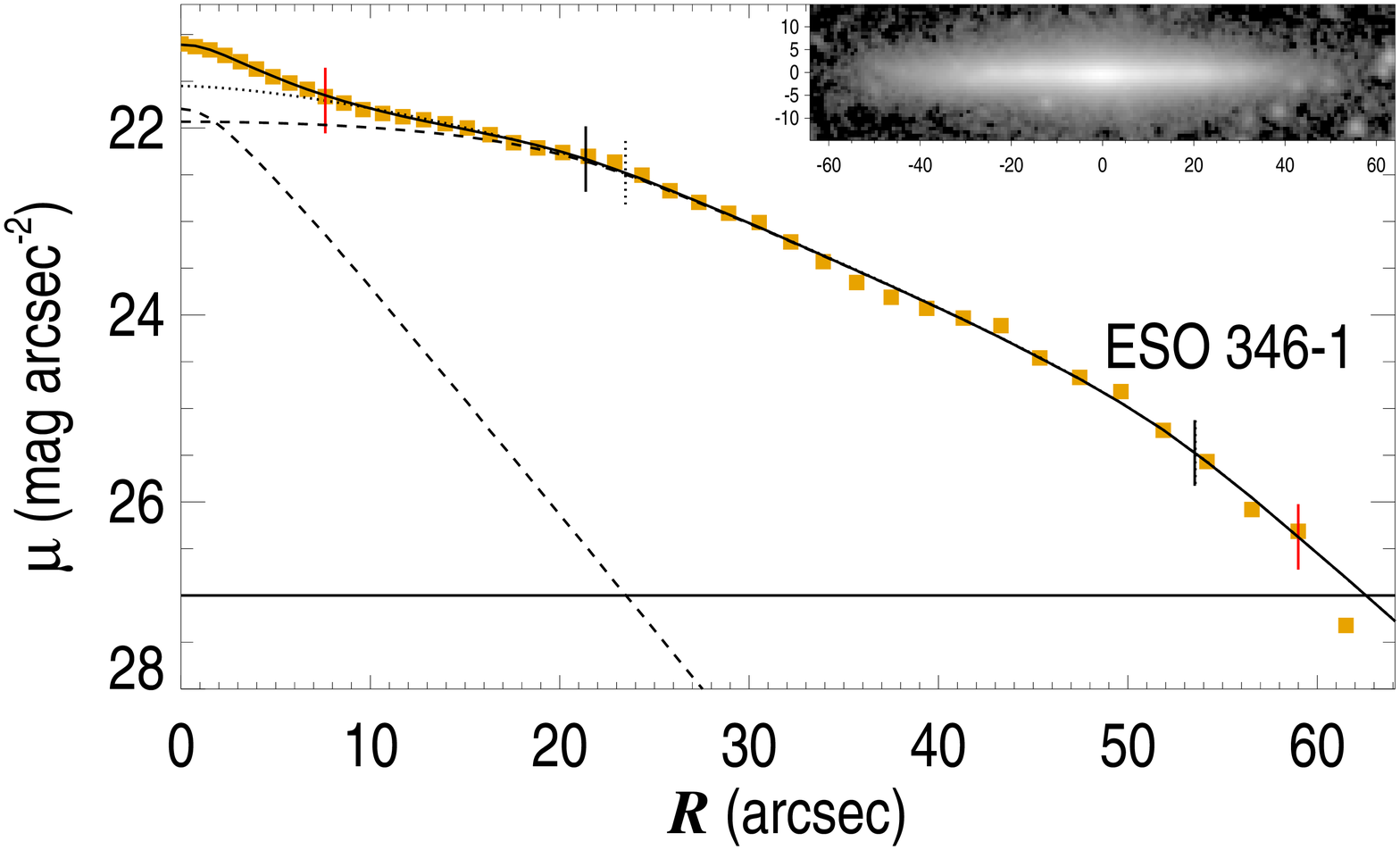}\\
\end{figure}

\begin{figure}
  \includegraphics[width=0.45\textwidth]{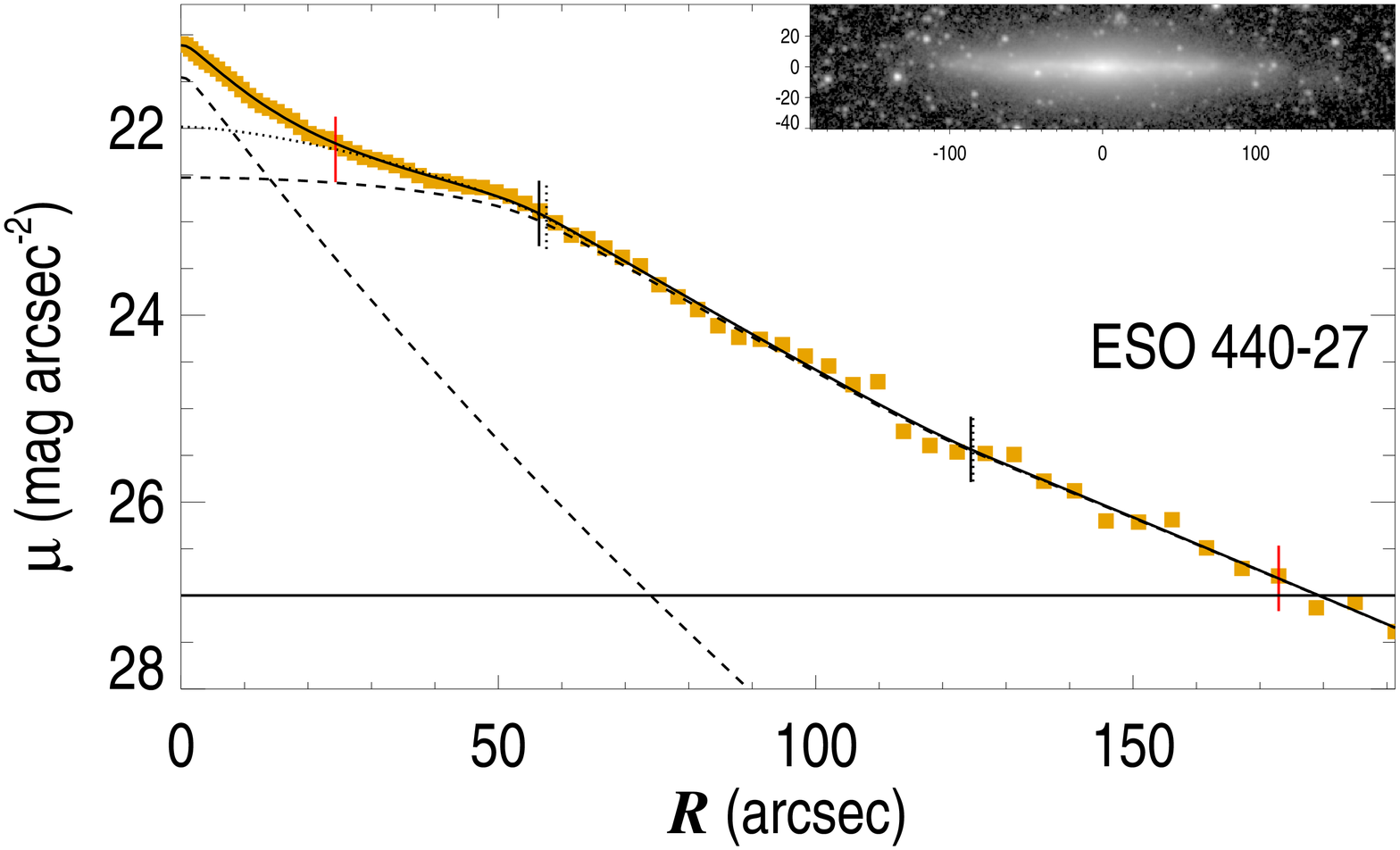}\\
\end{figure}

\begin{figure}
  \includegraphics[width=0.45\textwidth]{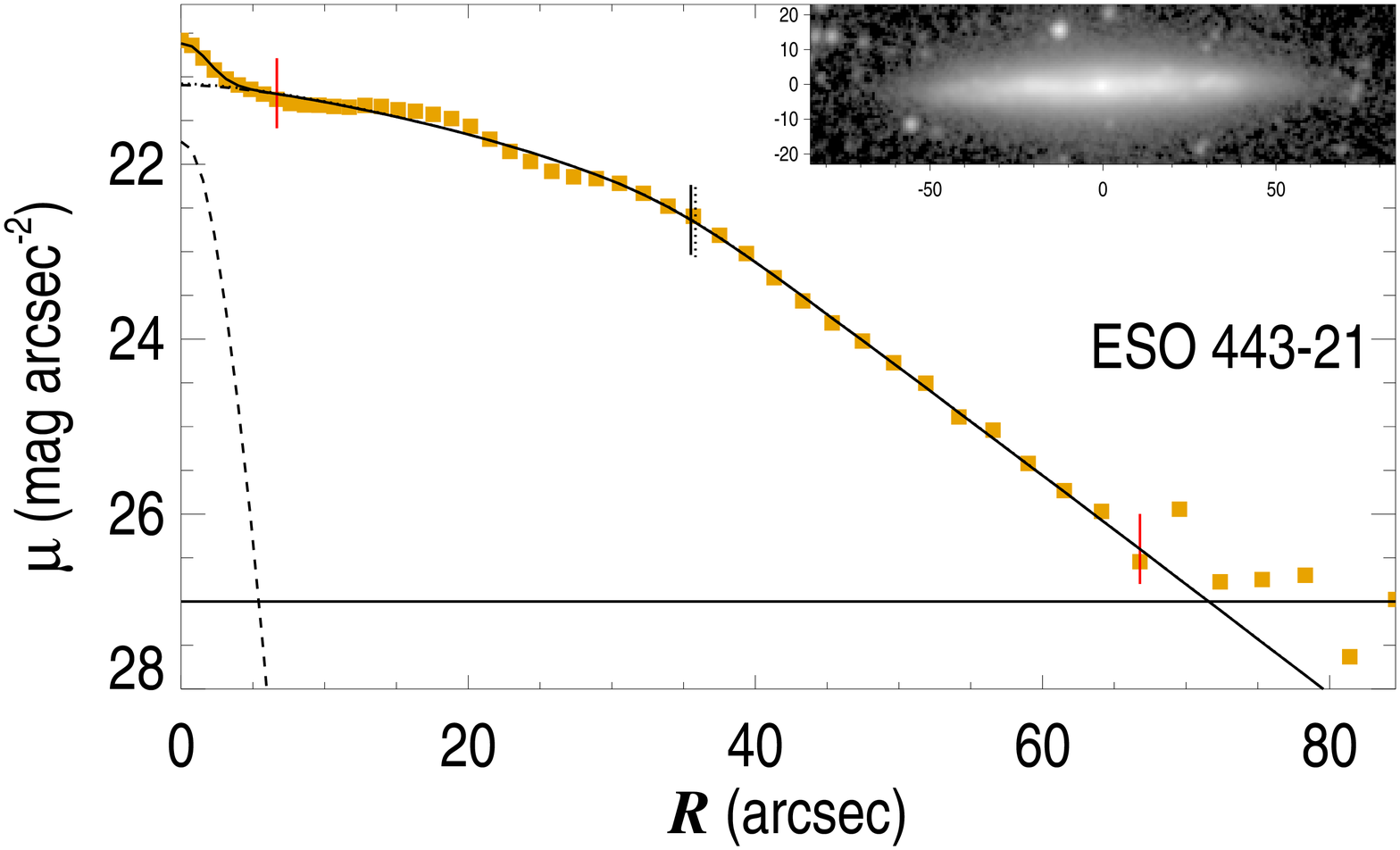}\\
\end{figure}

\begin{figure}
  \includegraphics[width=0.45\textwidth]{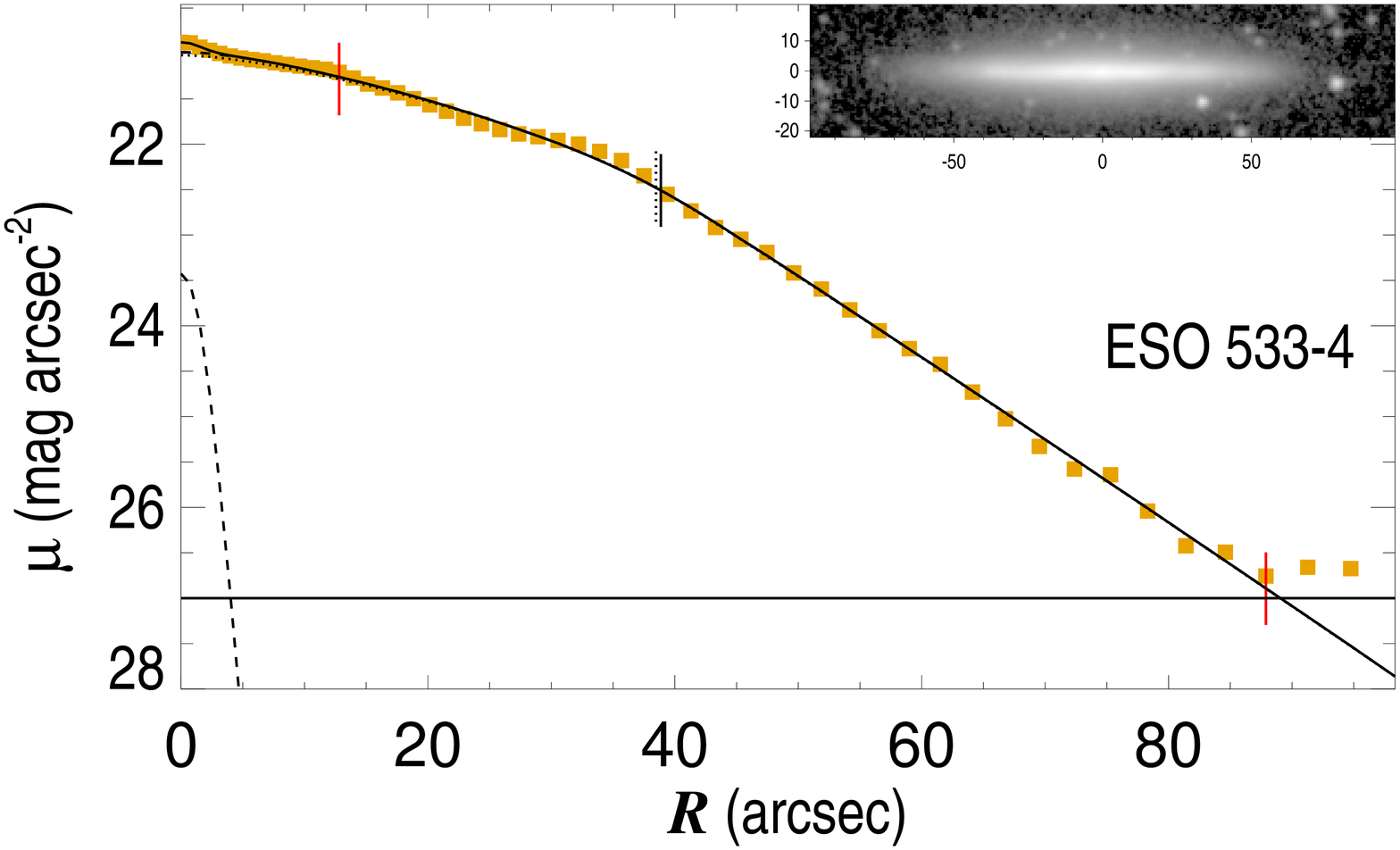}\\
\end{figure}

\begin{figure}
  \includegraphics[width=0.45\textwidth]{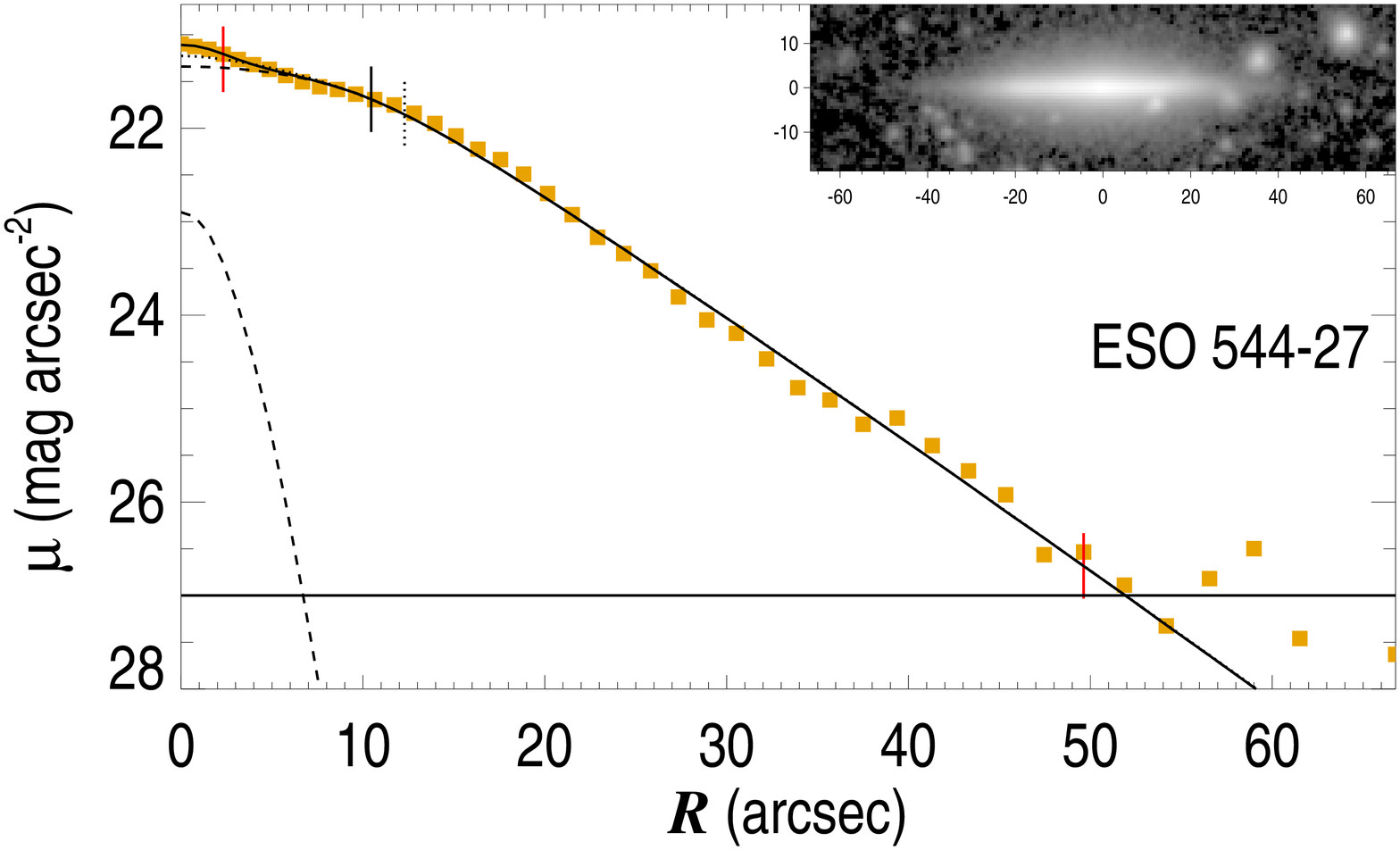}\\
\end{figure}

\begin{figure}
  \includegraphics[width=0.45\textwidth]{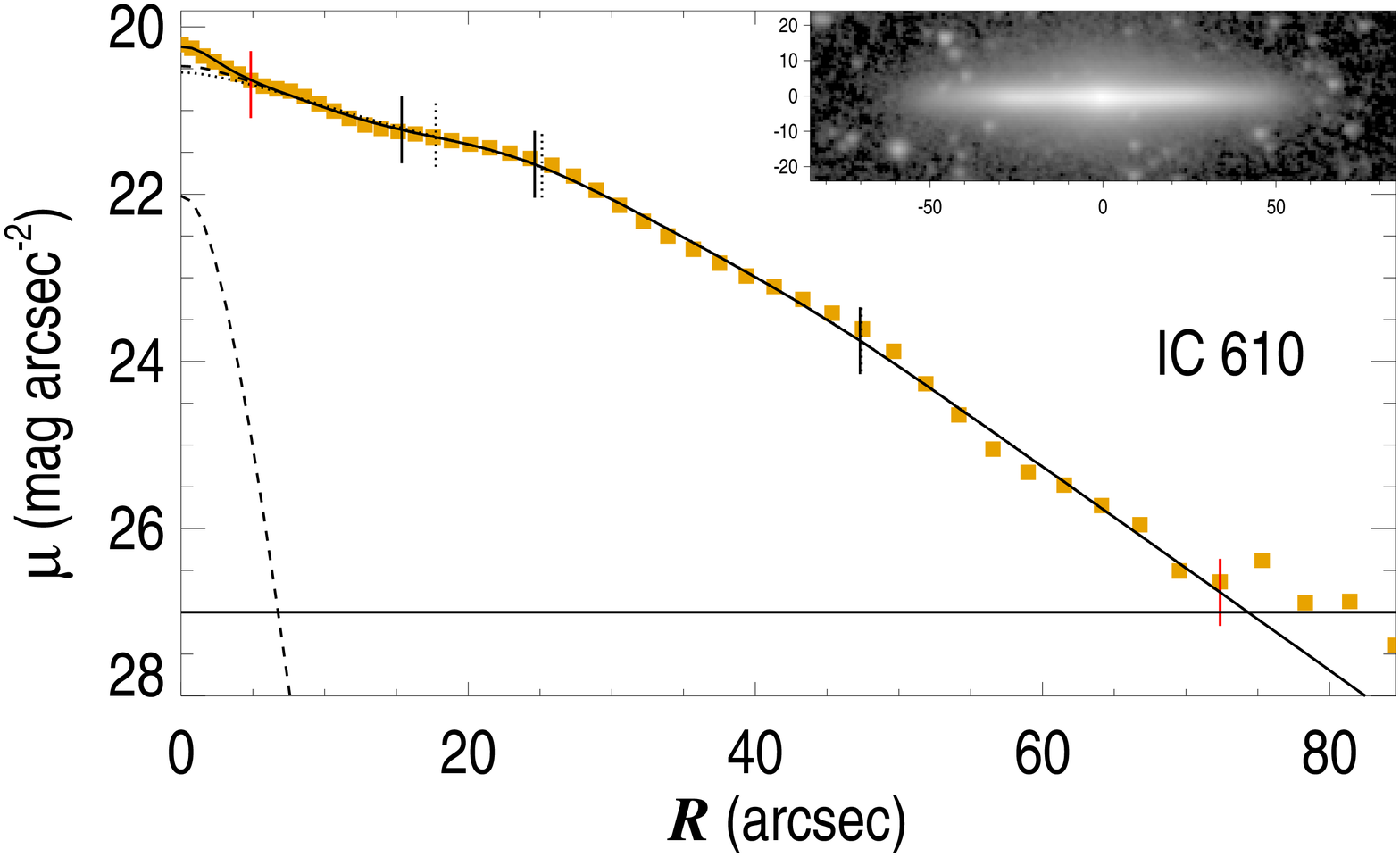}\\
\end{figure}

\clearpage

\begin{figure}
  \includegraphics[width=0.45\textwidth]{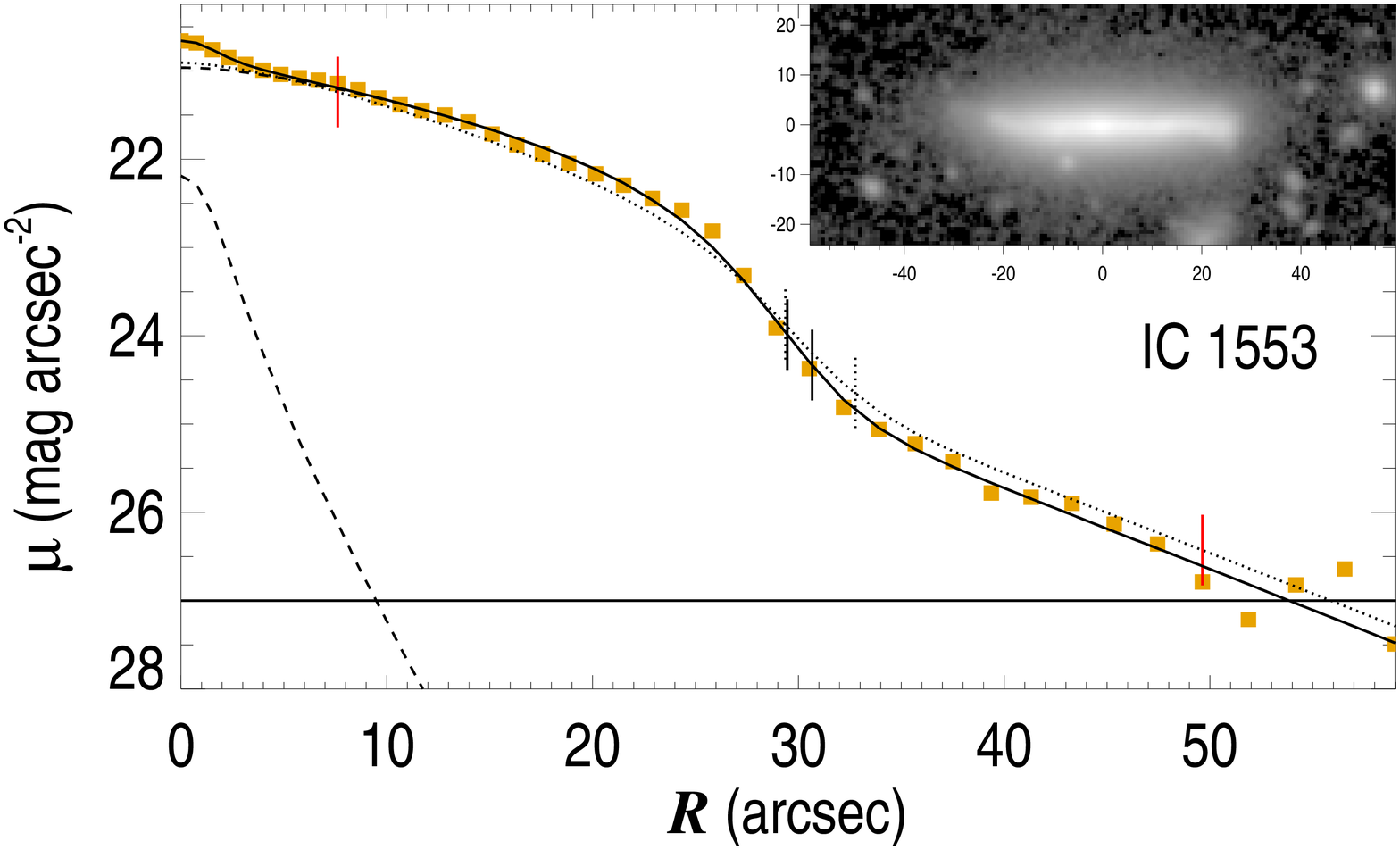}\\
\end{figure}

\begin{figure}
  \includegraphics[width=0.45\textwidth]{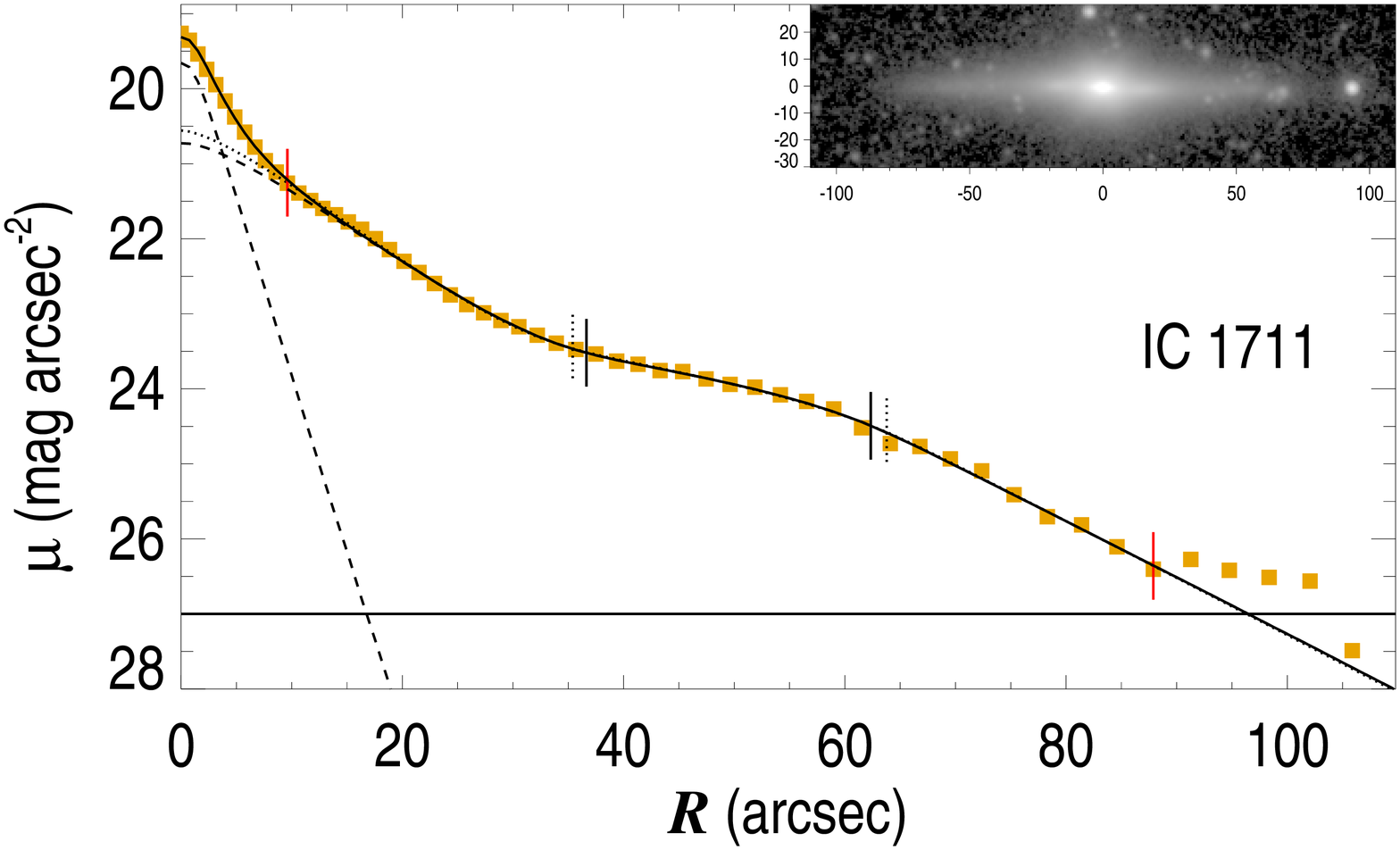}\\
\end{figure}

\begin{figure}
  \includegraphics[width=0.45\textwidth]{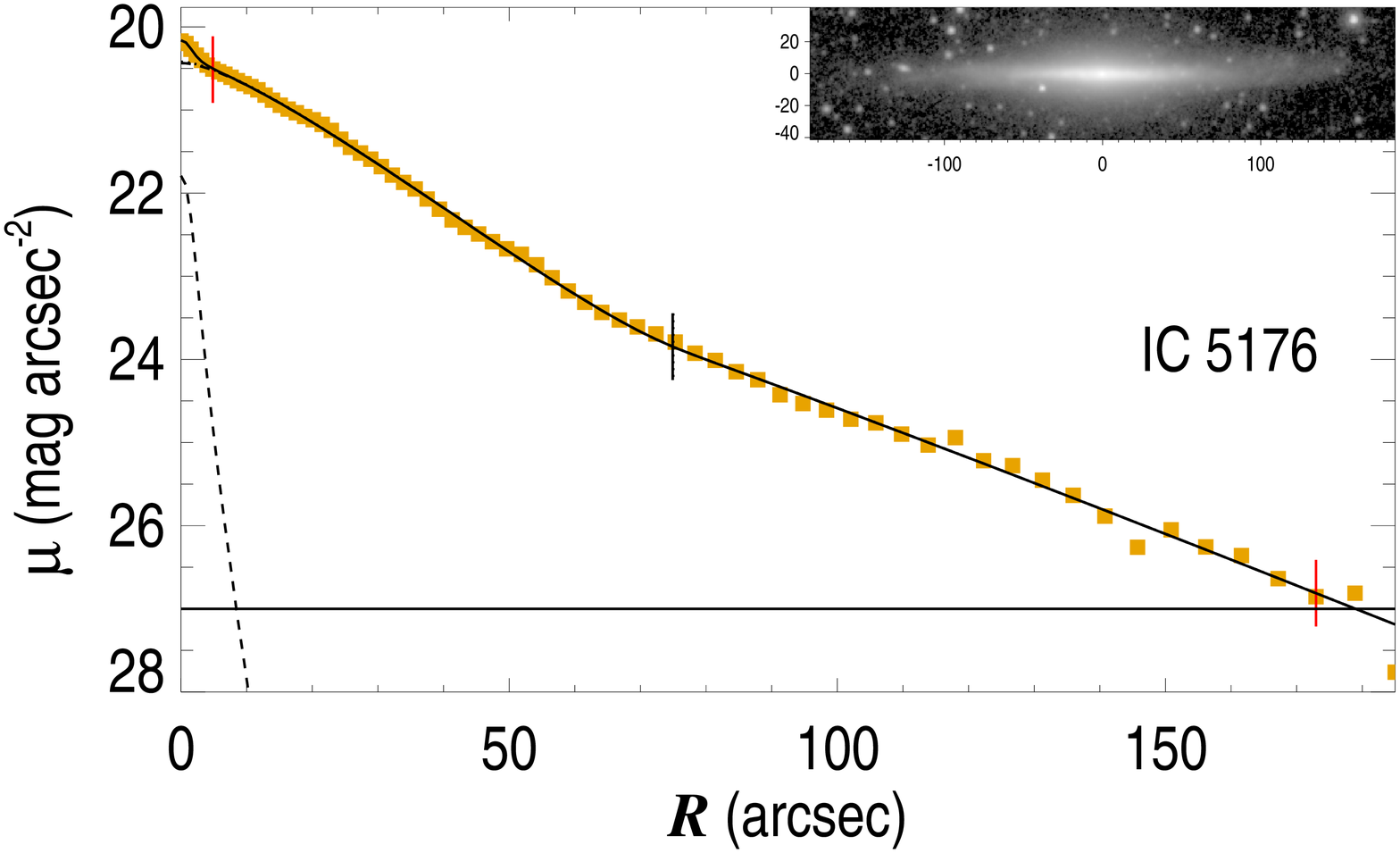}\\
\end{figure}

\begin{figure}
  \includegraphics[width=0.45\textwidth]{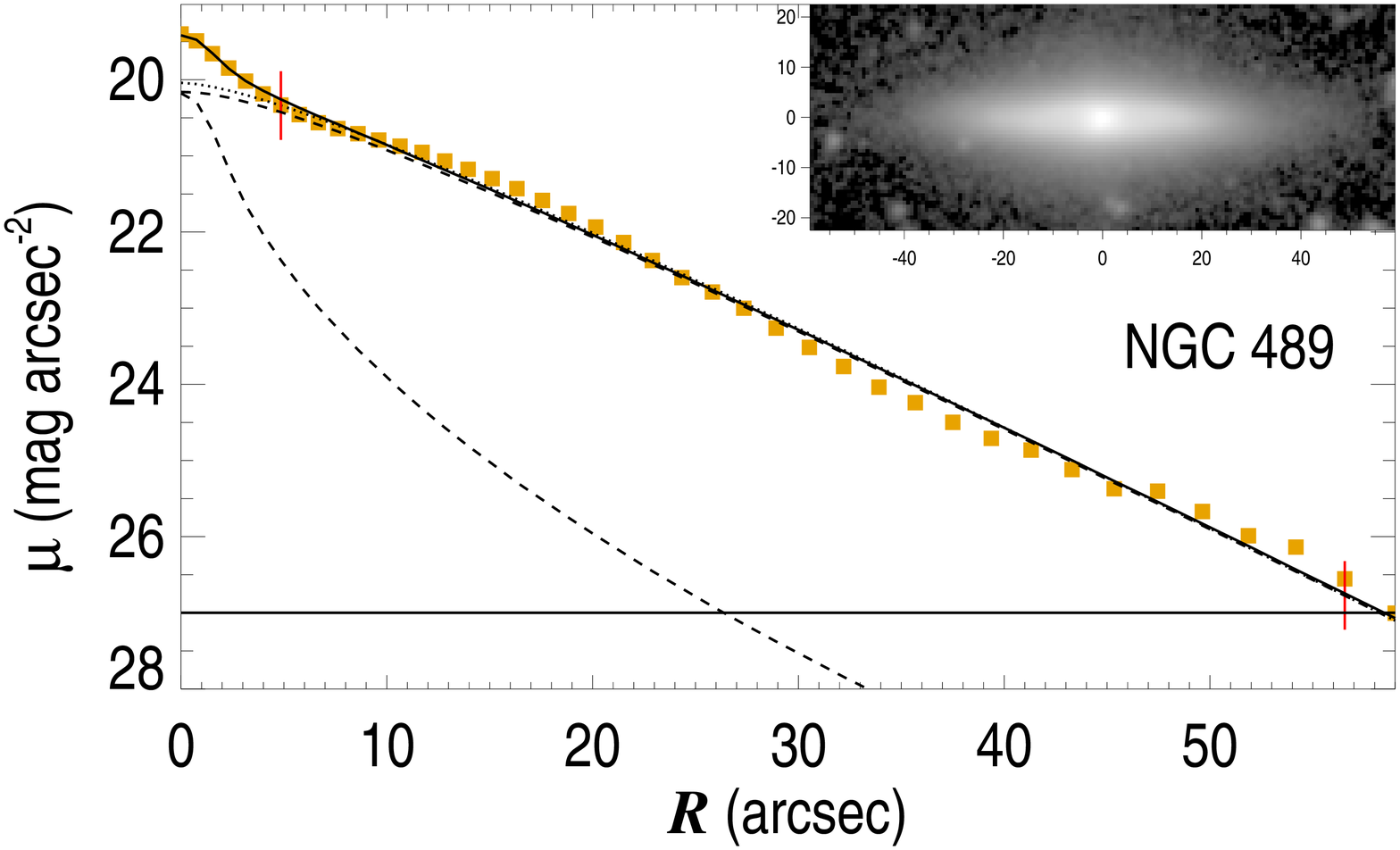}\\
\end{figure}

\begin{figure}
  \includegraphics[width=0.45\textwidth]{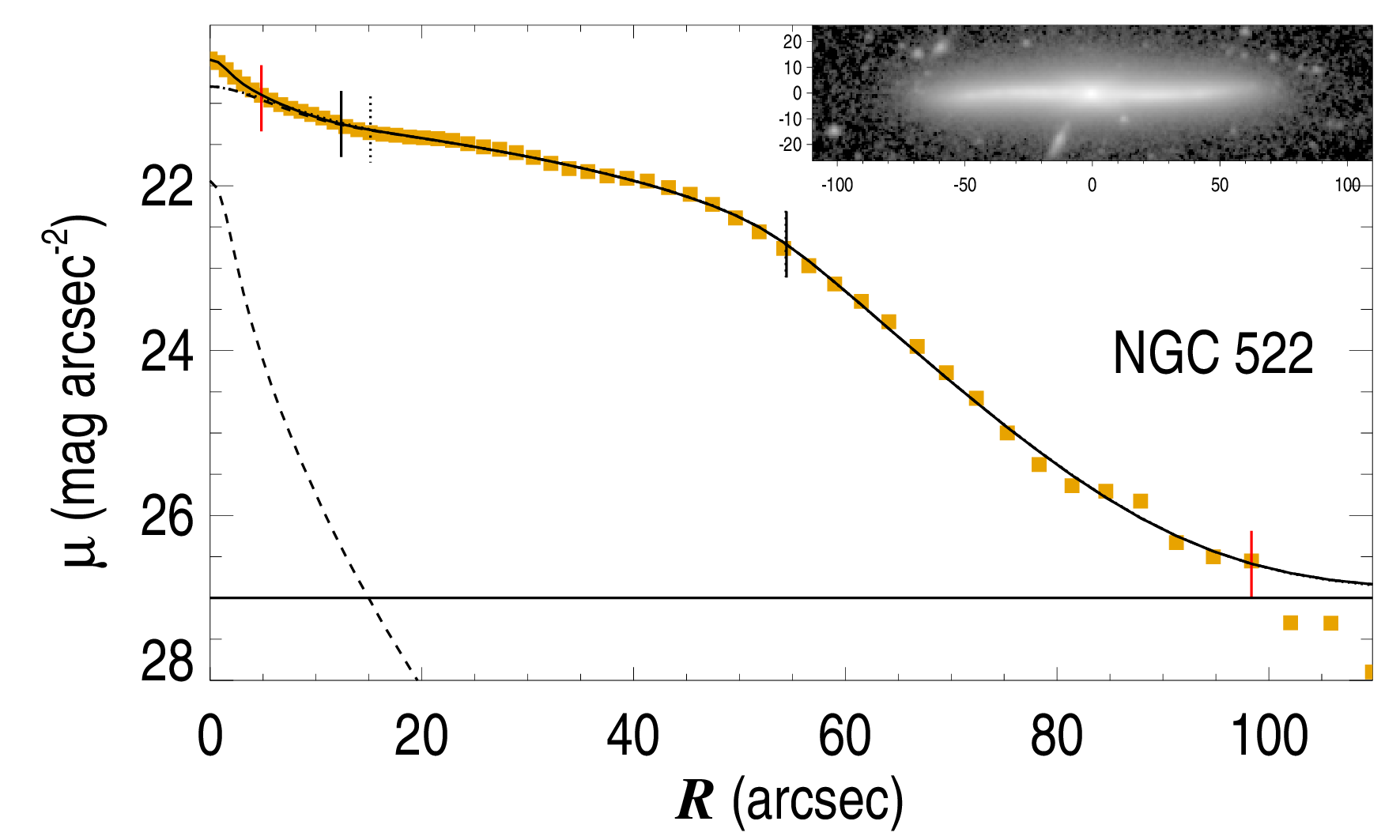}\\
\end{figure}

\begin{figure}
  \includegraphics[width=0.45\textwidth]{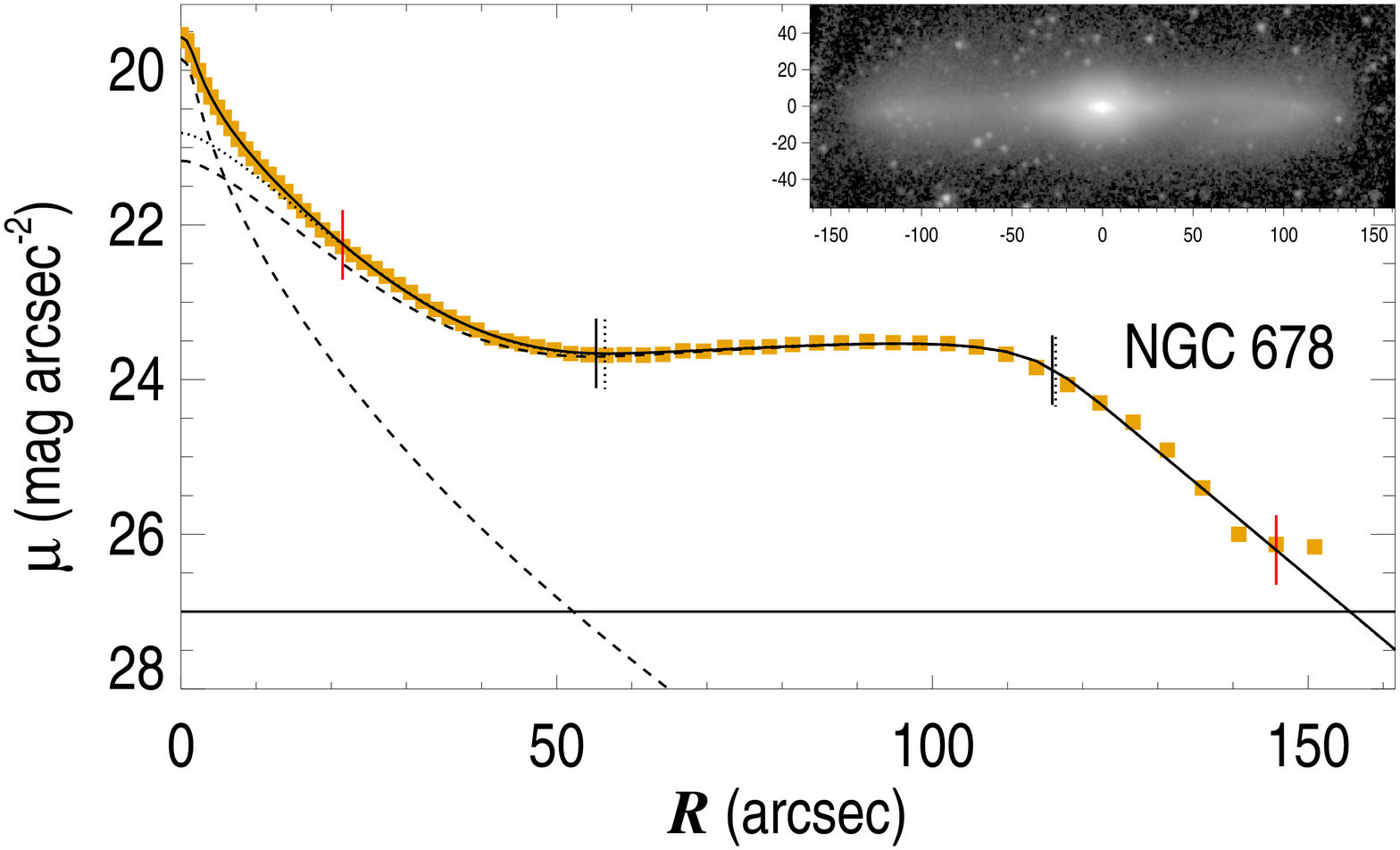}\\
\end{figure}

\begin{figure}
  \includegraphics[width=0.45\textwidth]{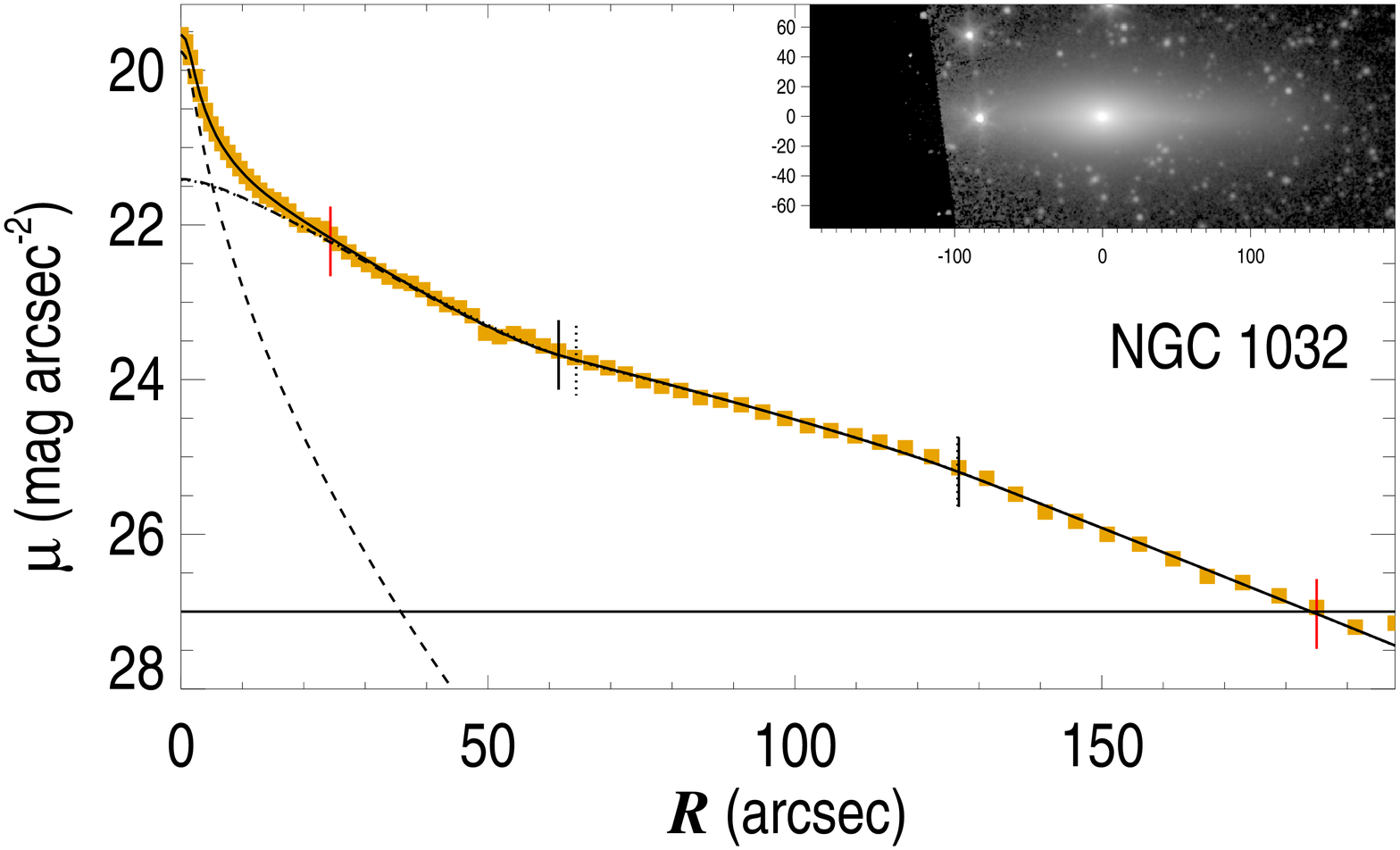}\\
\end{figure}

\begin{figure}
  \includegraphics[width=0.45\textwidth]{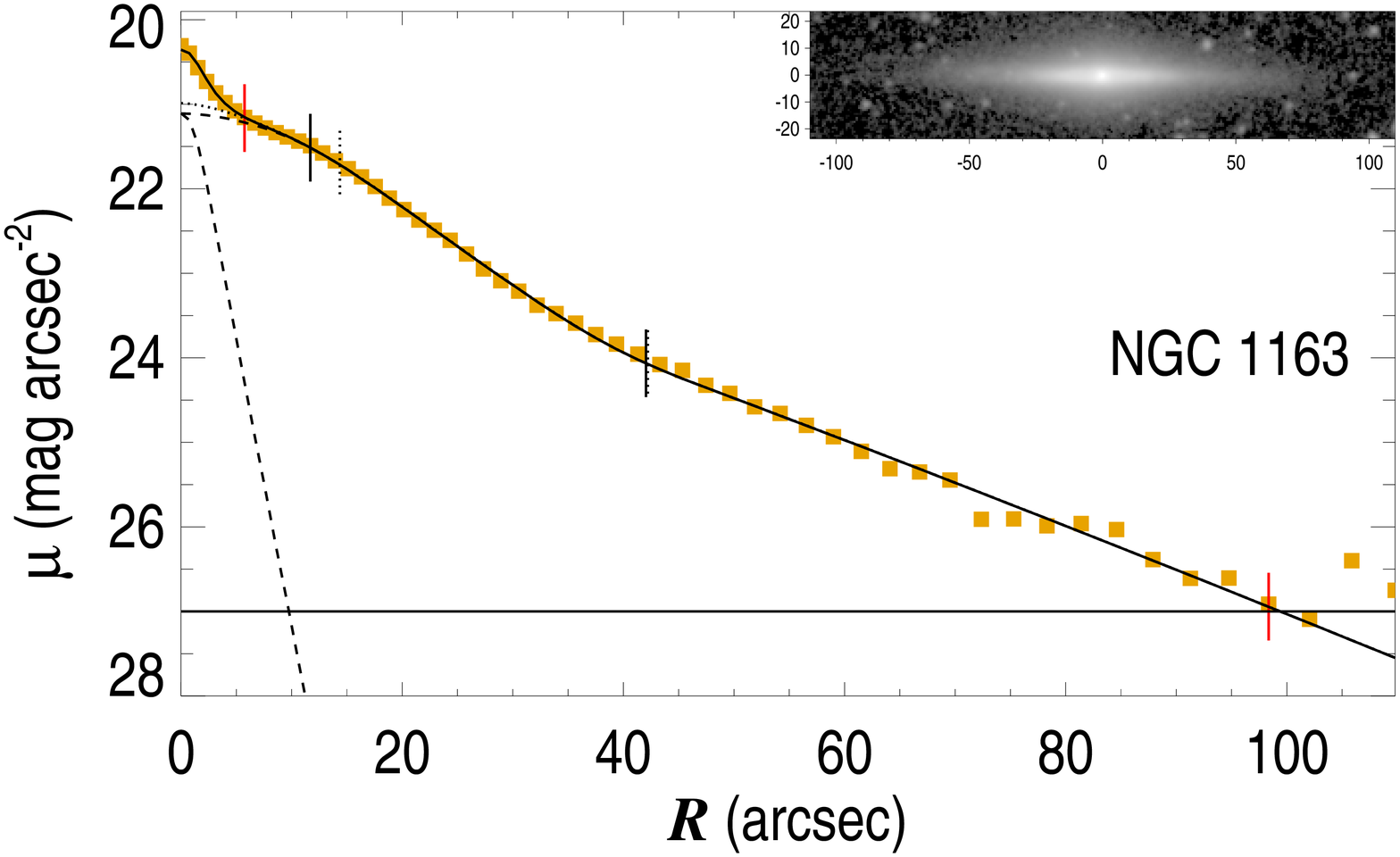}\\
\end{figure}

\clearpage

\begin{figure}
  \includegraphics[width=0.45\textwidth]{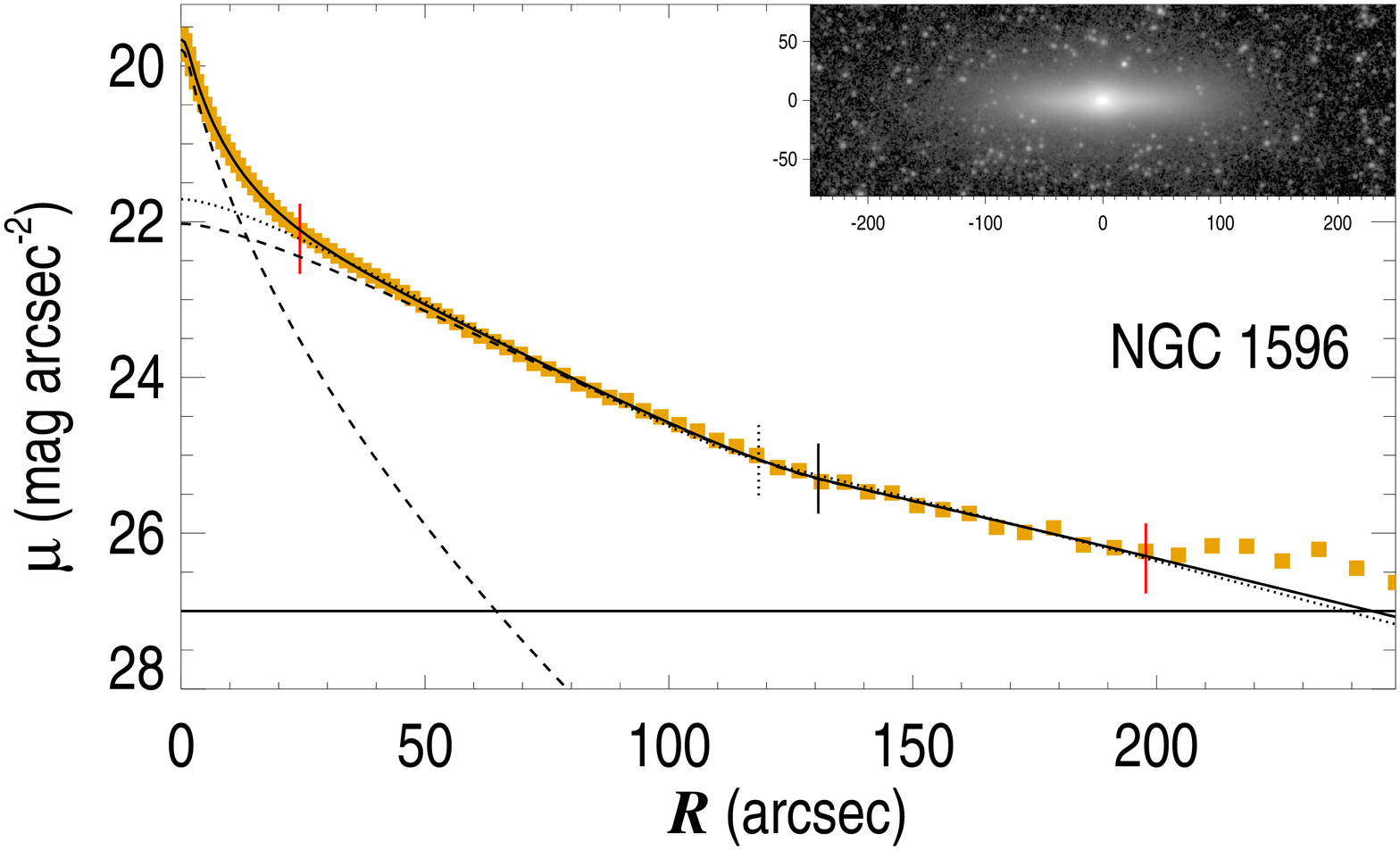}\\
\end{figure}

\begin{figure}
  \includegraphics[width=0.45\textwidth]{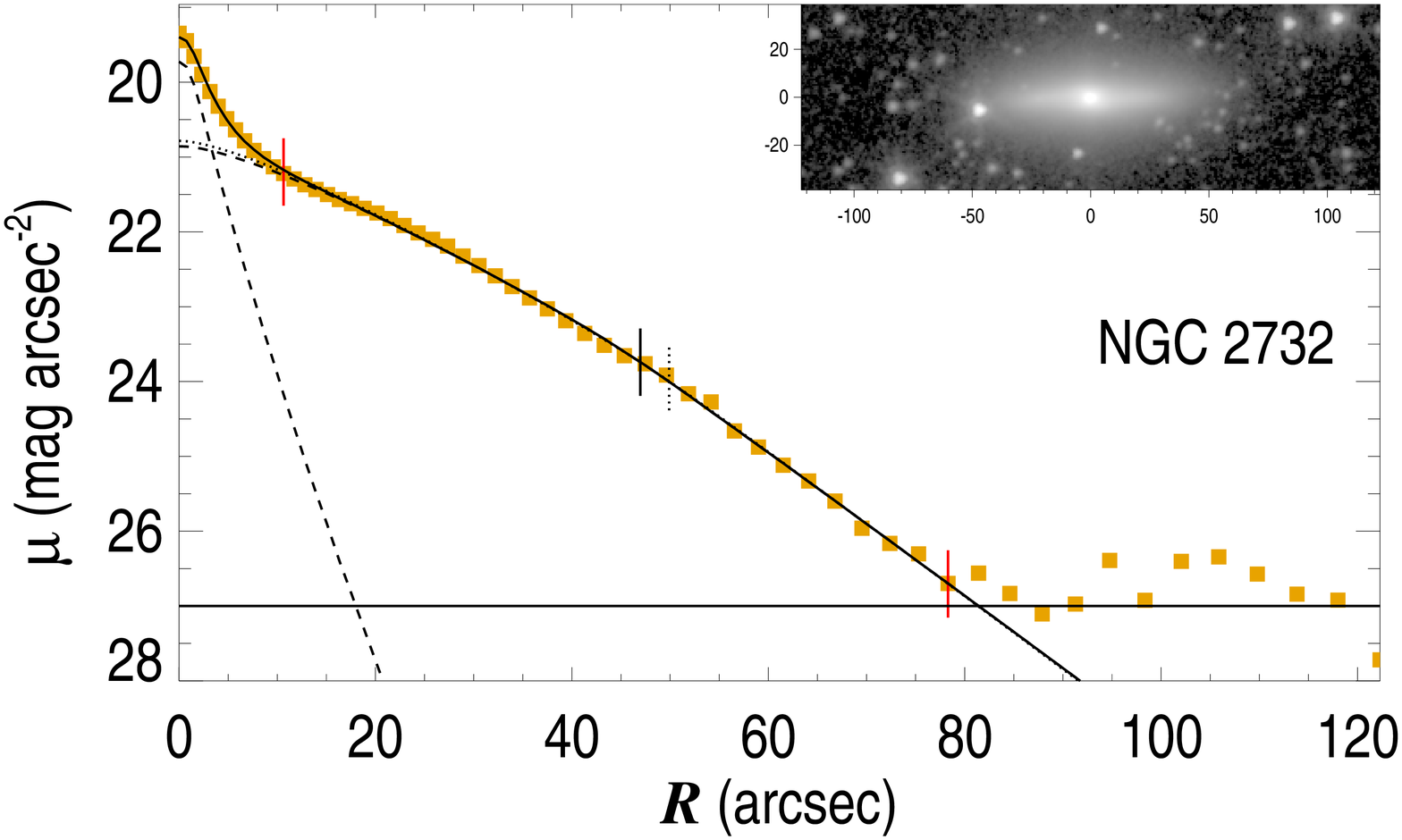}\\
\end{figure}

\begin{figure}
  \includegraphics[width=0.45\textwidth]{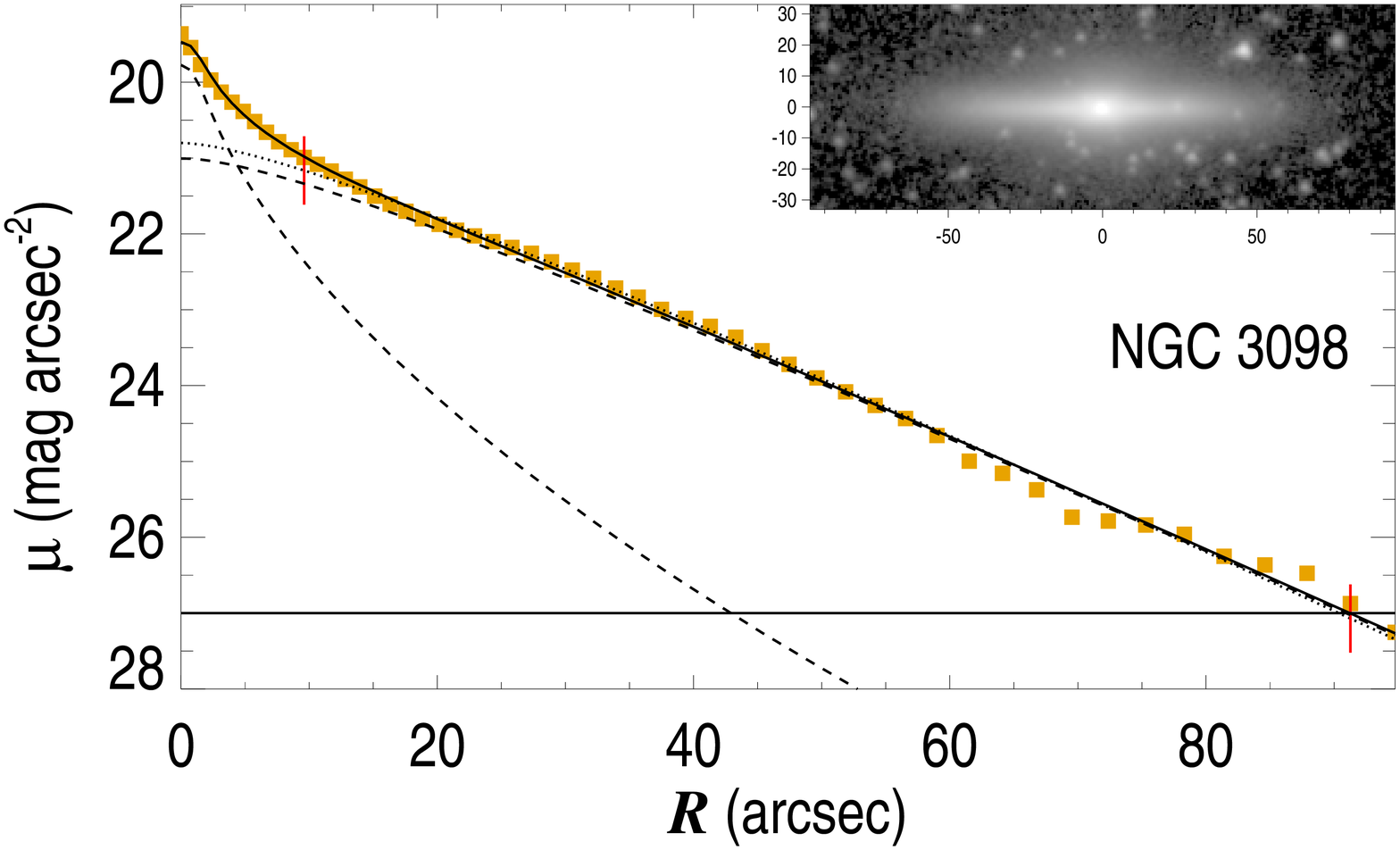}\\
\end{figure}

\begin{figure}
  \includegraphics[width=0.45\textwidth]{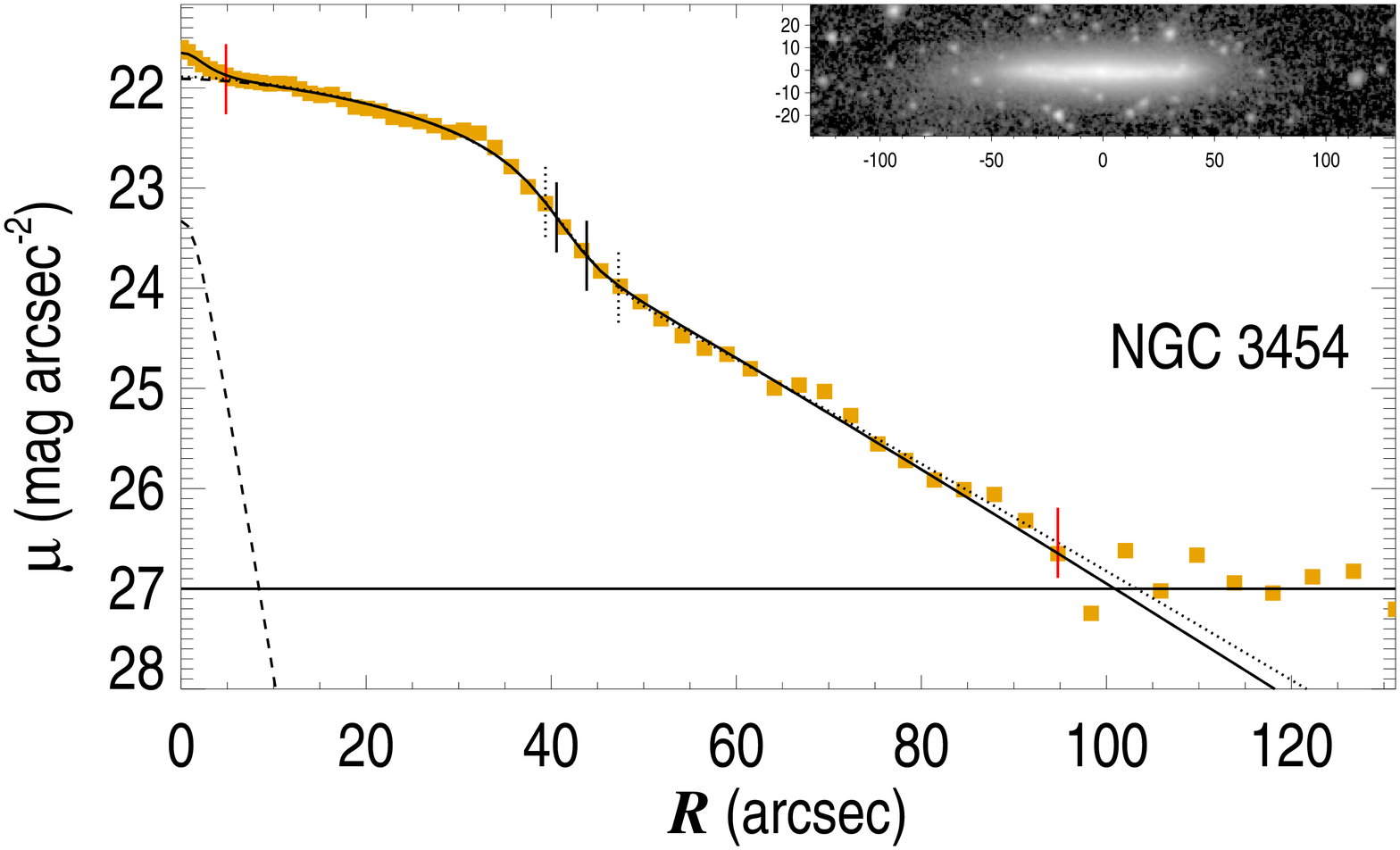}\\
\end{figure}

\begin{figure}
  \includegraphics[width=0.45\textwidth]{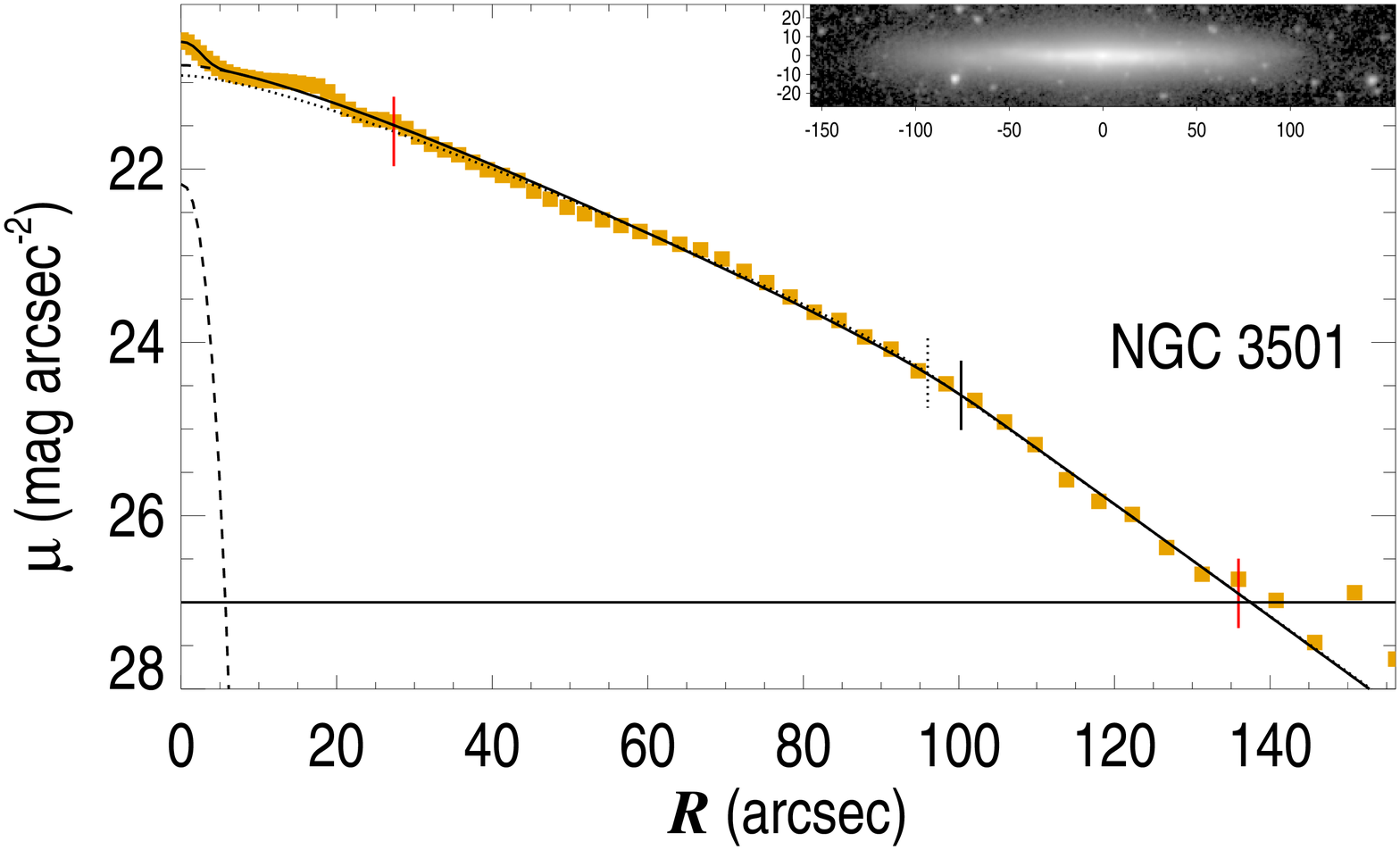}\\
\end{figure}

\begin{figure}
  \includegraphics[width=0.45\textwidth]{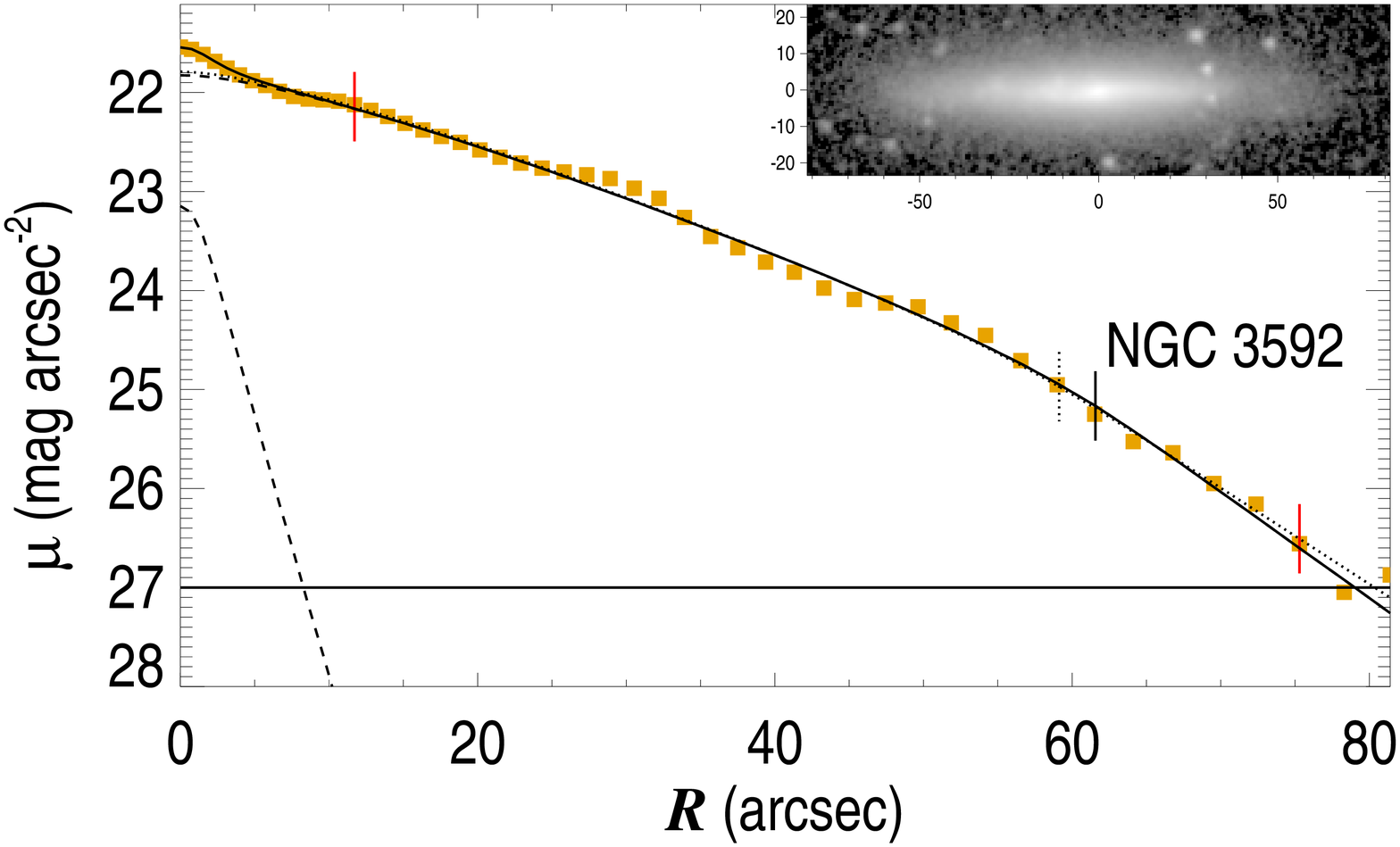}\\
\end{figure}

\begin{figure}
  \includegraphics[width=0.45\textwidth]{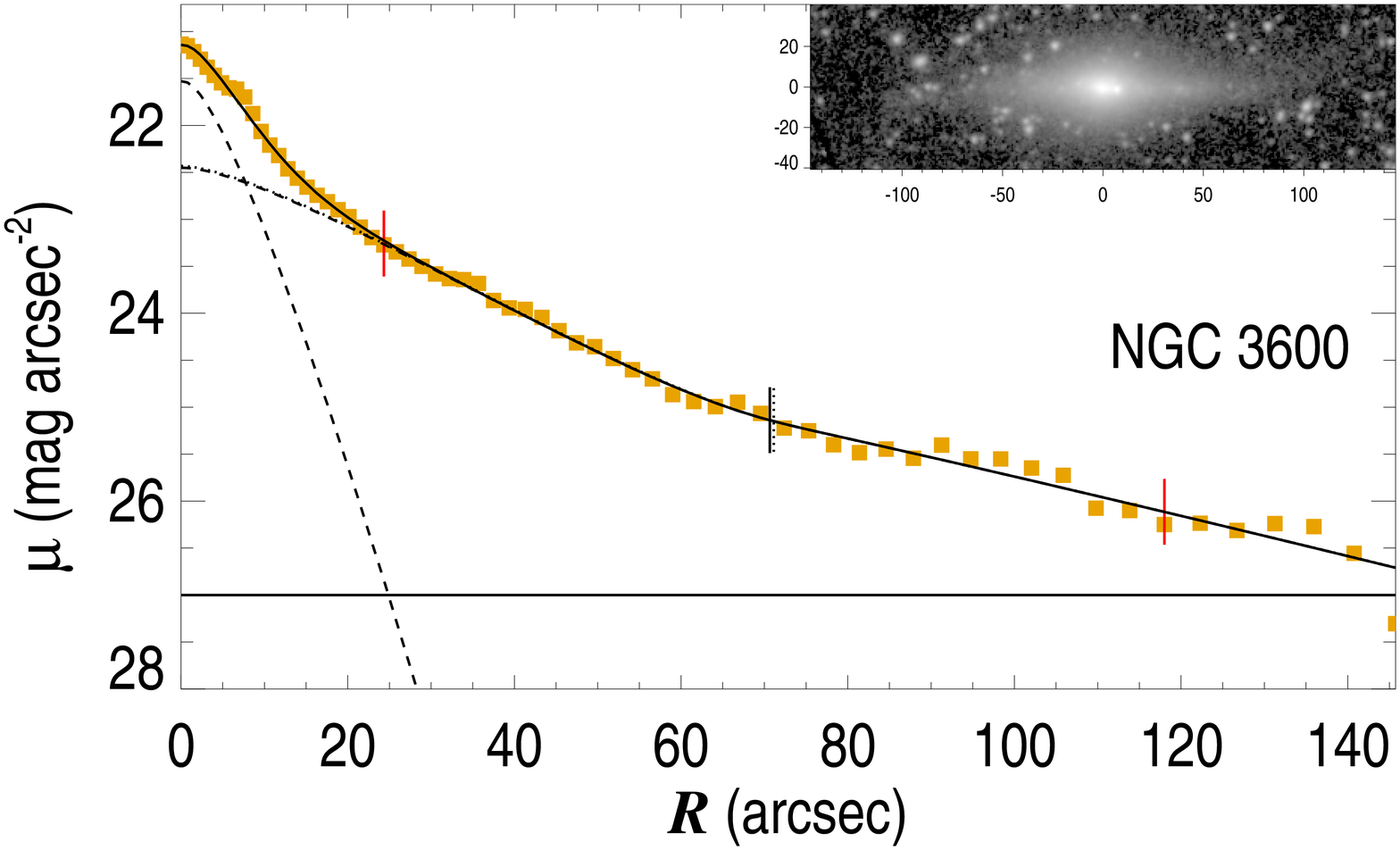}\\
\end{figure}

\begin{figure}
  \includegraphics[width=0.45\textwidth]{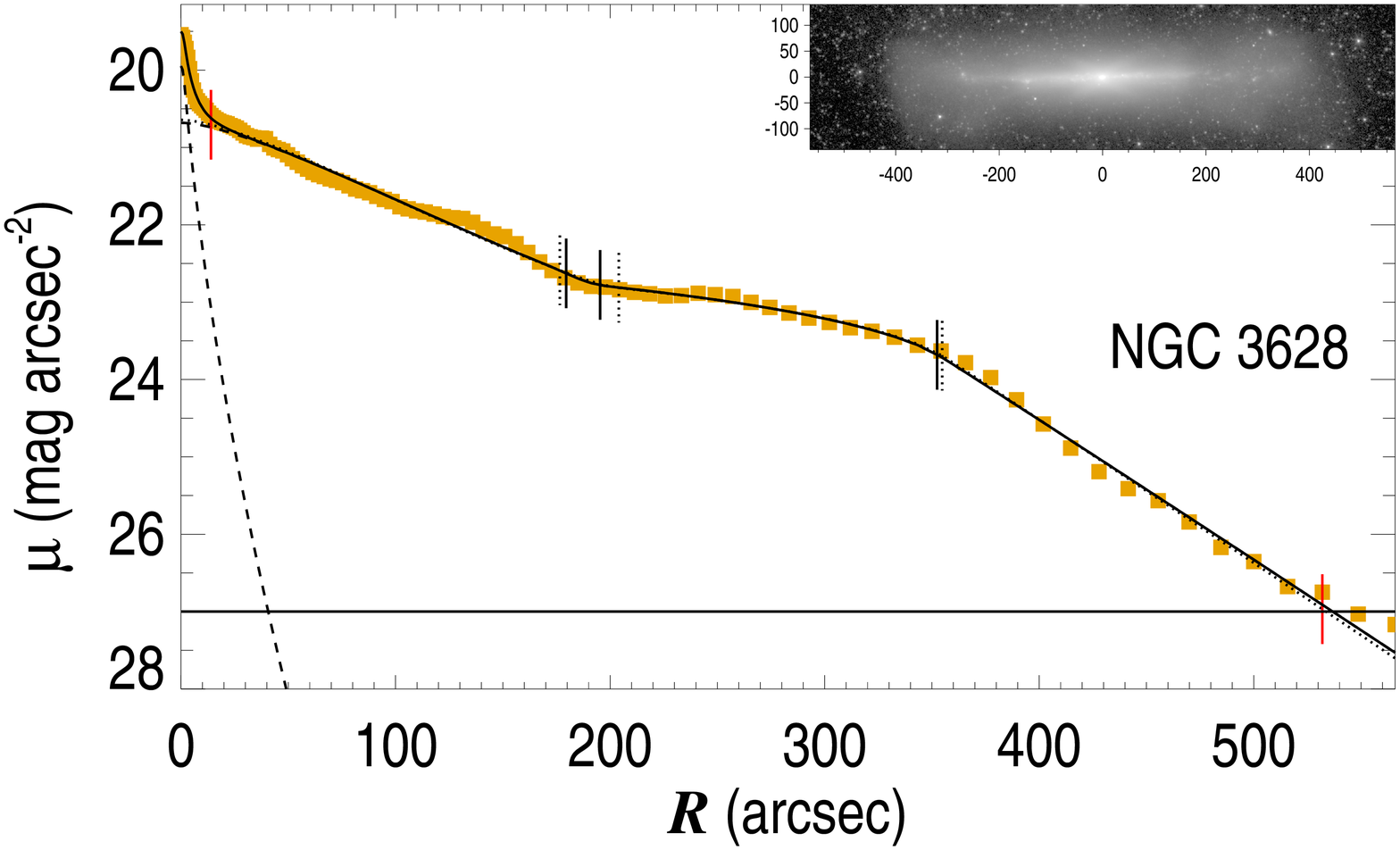}\\
\end{figure}

\clearpage

\begin{figure}
  \includegraphics[width=0.45\textwidth]{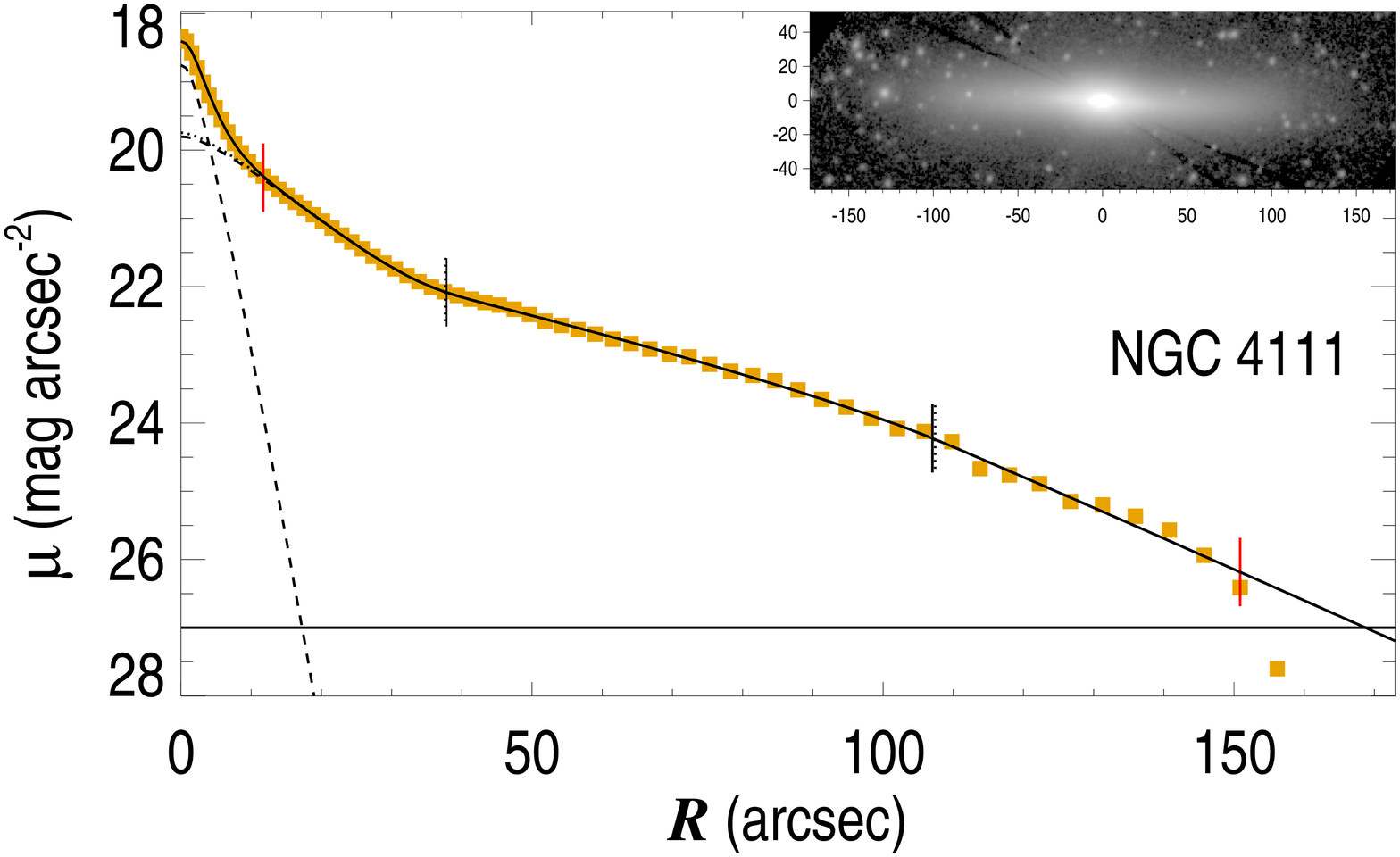}\\
\end{figure}

\begin{figure}
  \includegraphics[width=0.45\textwidth]{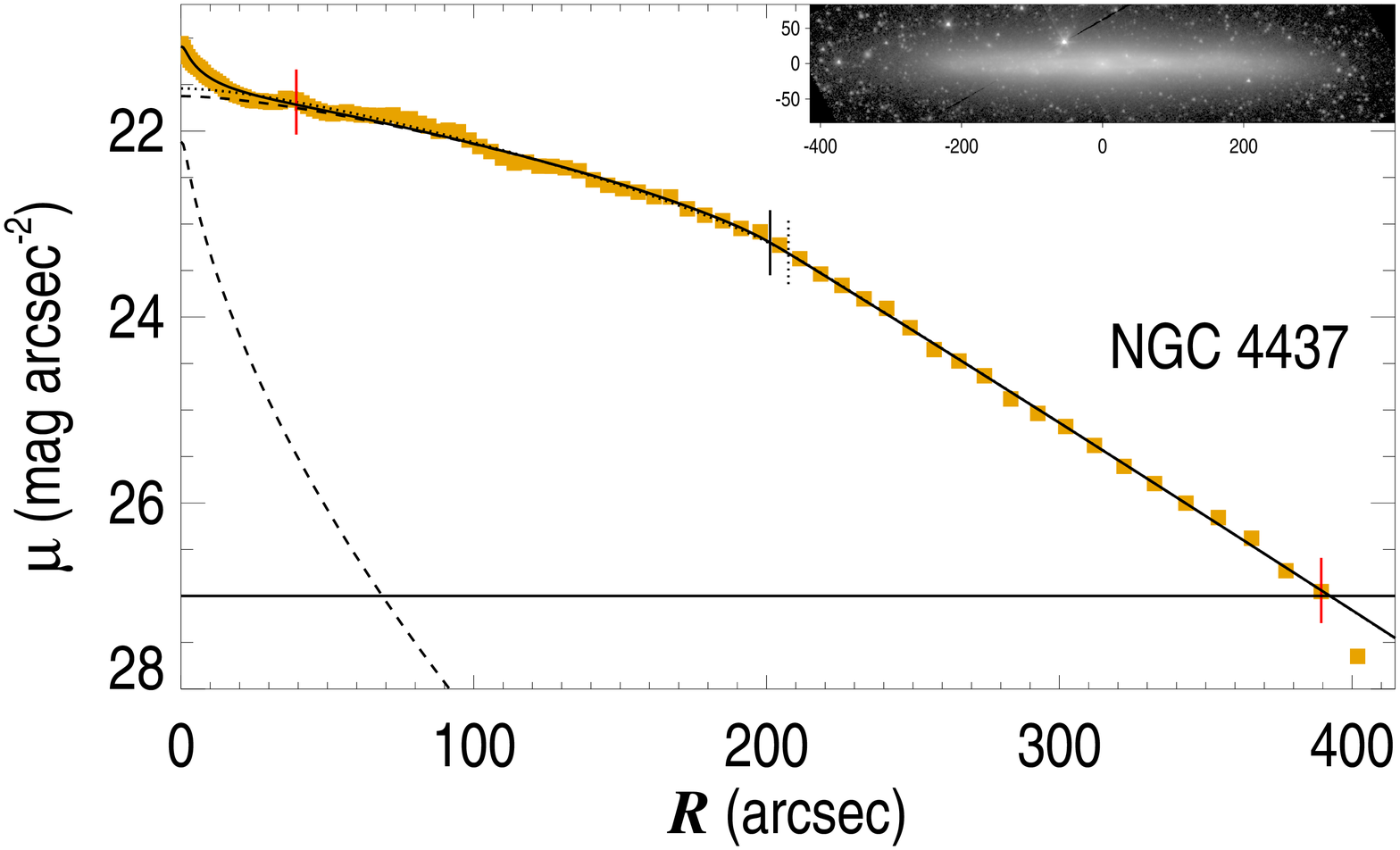}\\
\end{figure}

\begin{figure}
  \includegraphics[width=0.45\textwidth]{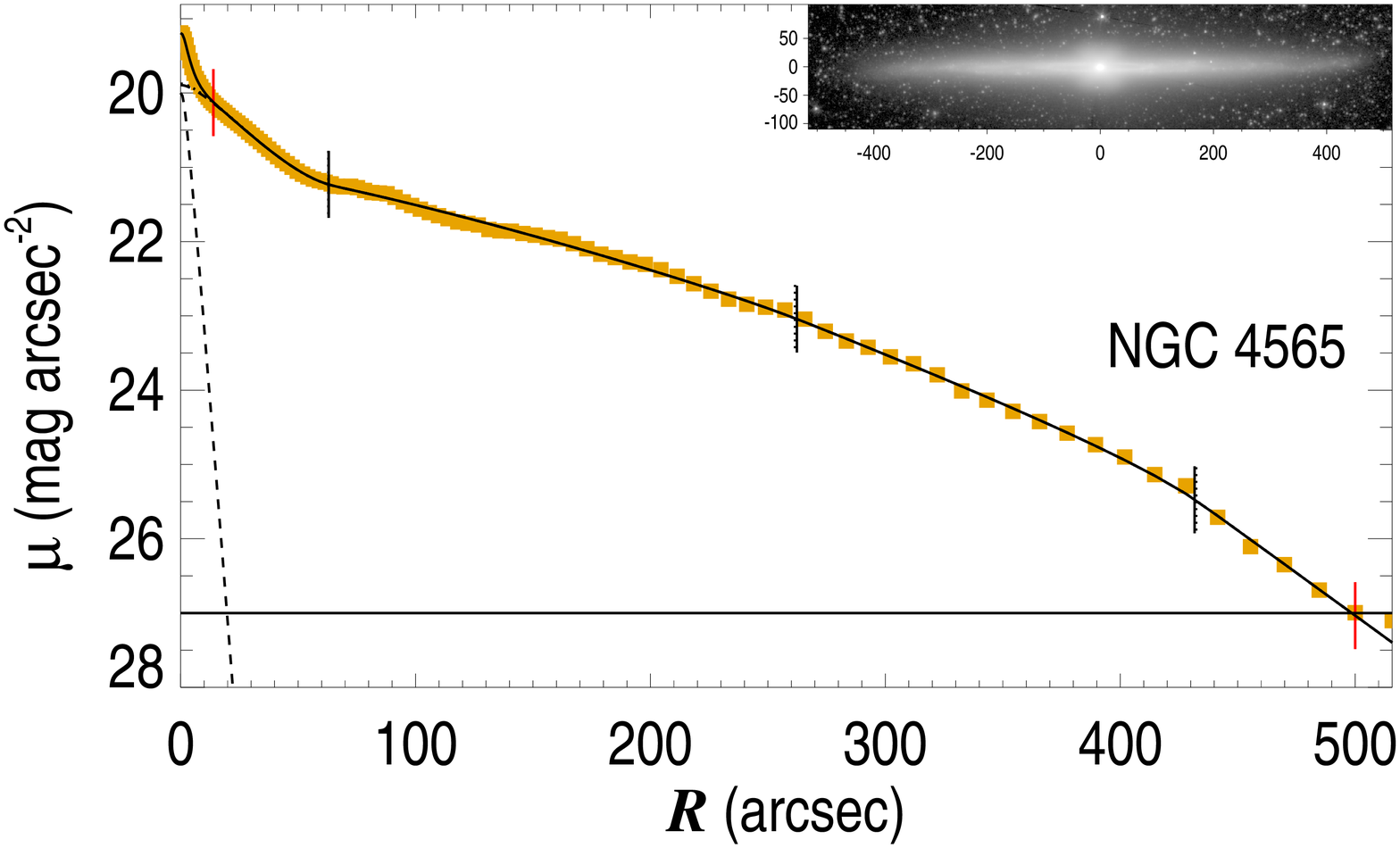}\\
\end{figure}

\begin{figure}
  \includegraphics[width=0.45\textwidth]{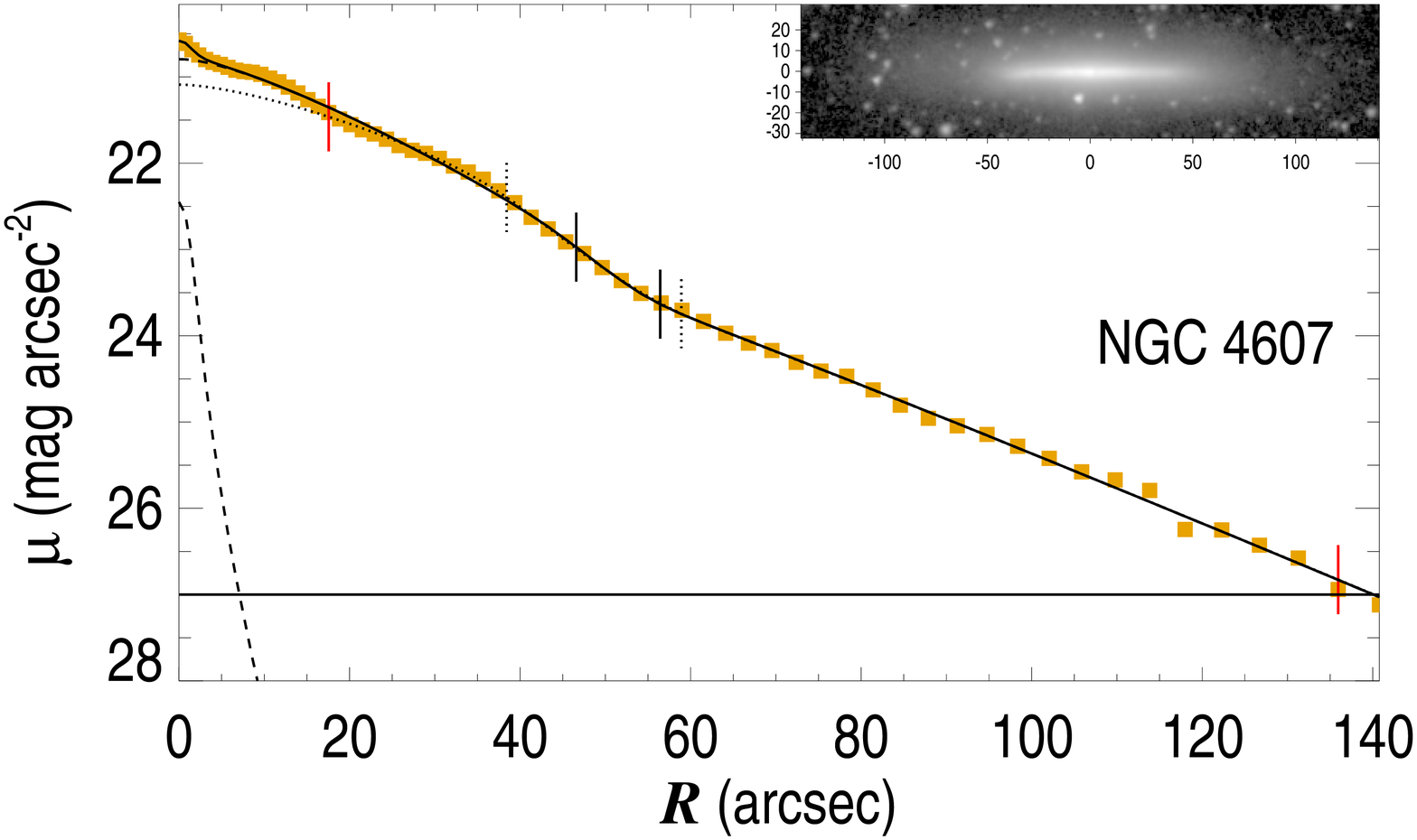}\\
\end{figure}

\begin{figure}
  \includegraphics[width=0.45\textwidth]{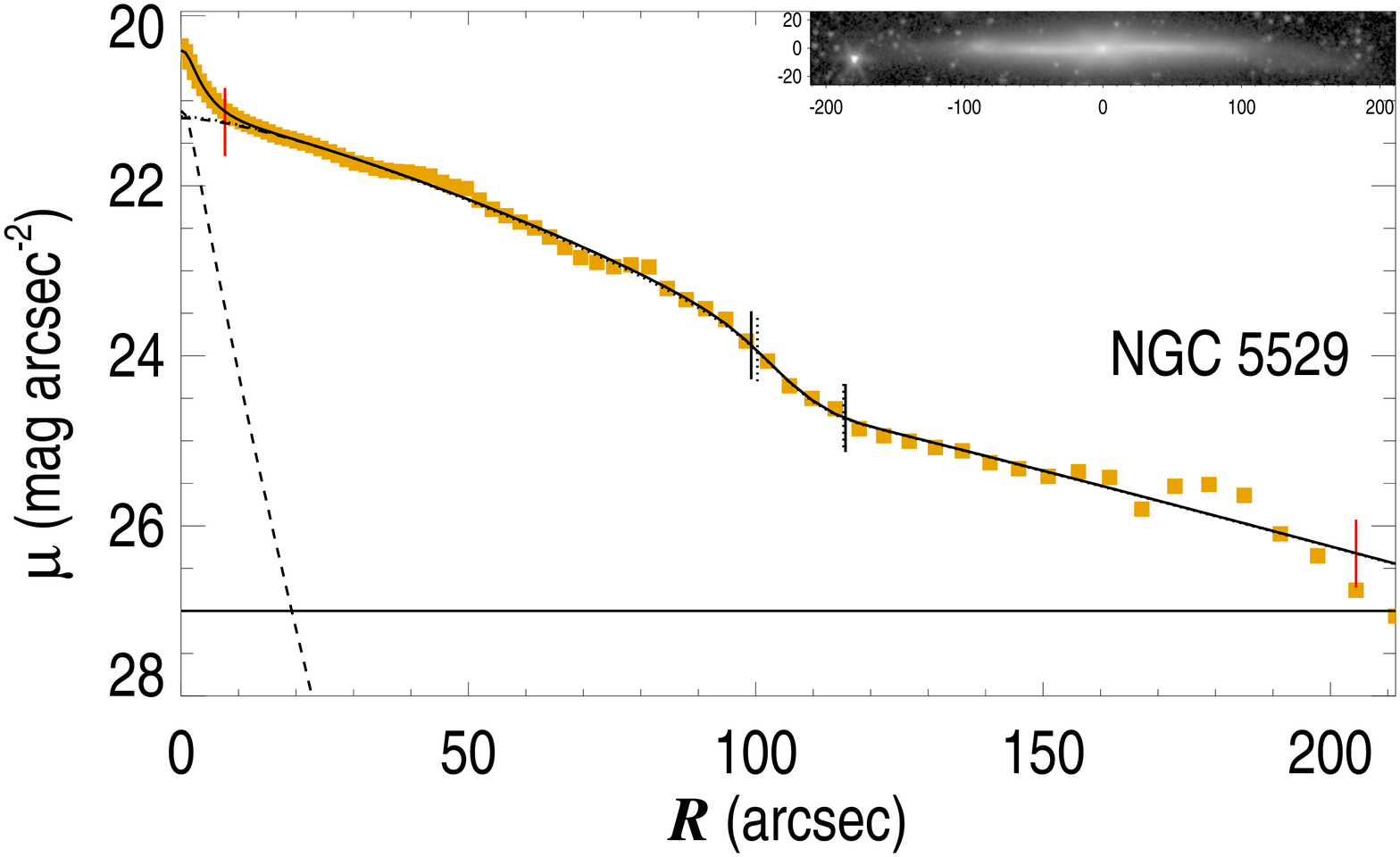}\\
\end{figure}

\begin{figure}
  \includegraphics[width=0.45\textwidth]{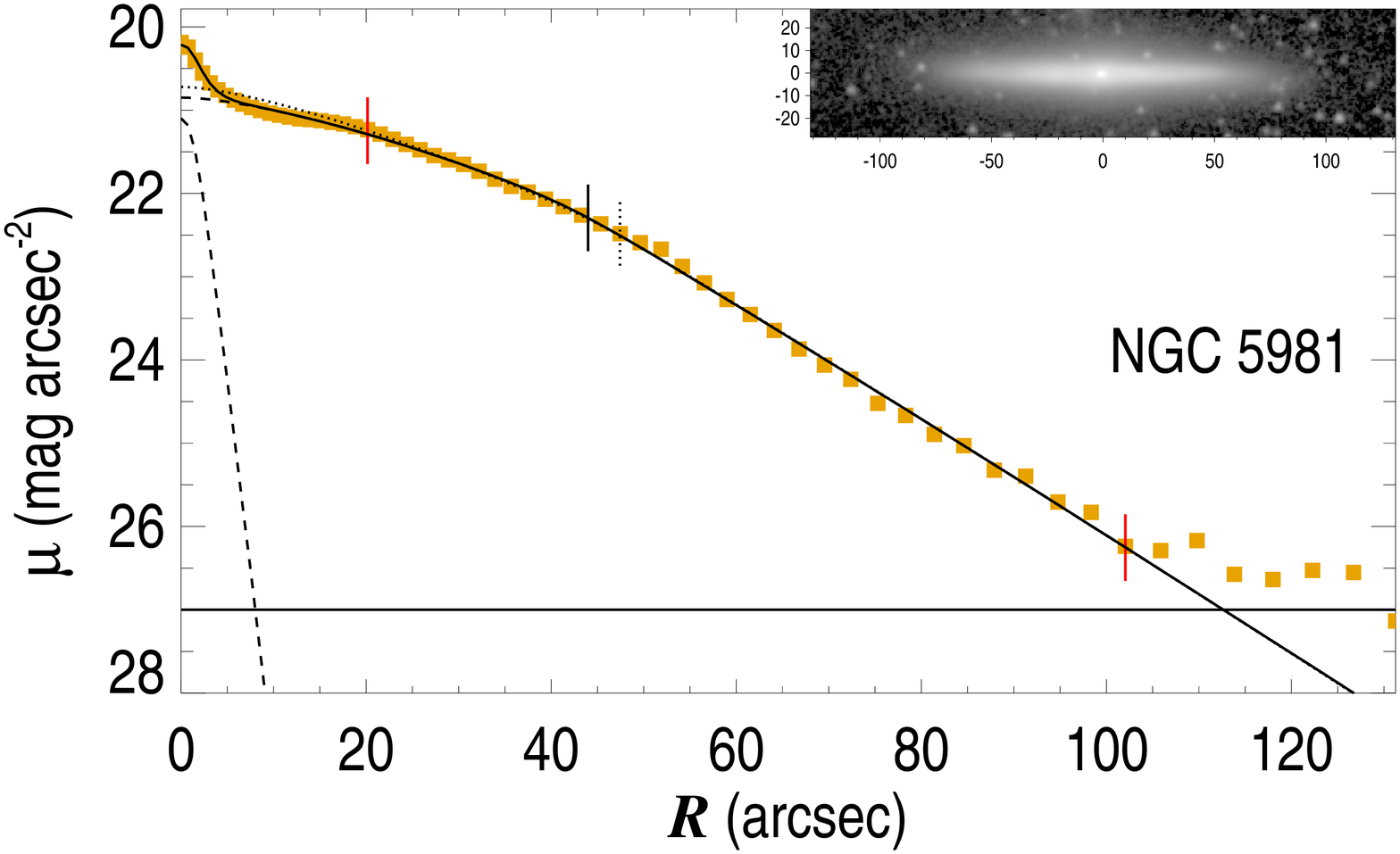}\\
\end{figure}

\begin{figure}
  \includegraphics[width=0.45\textwidth]{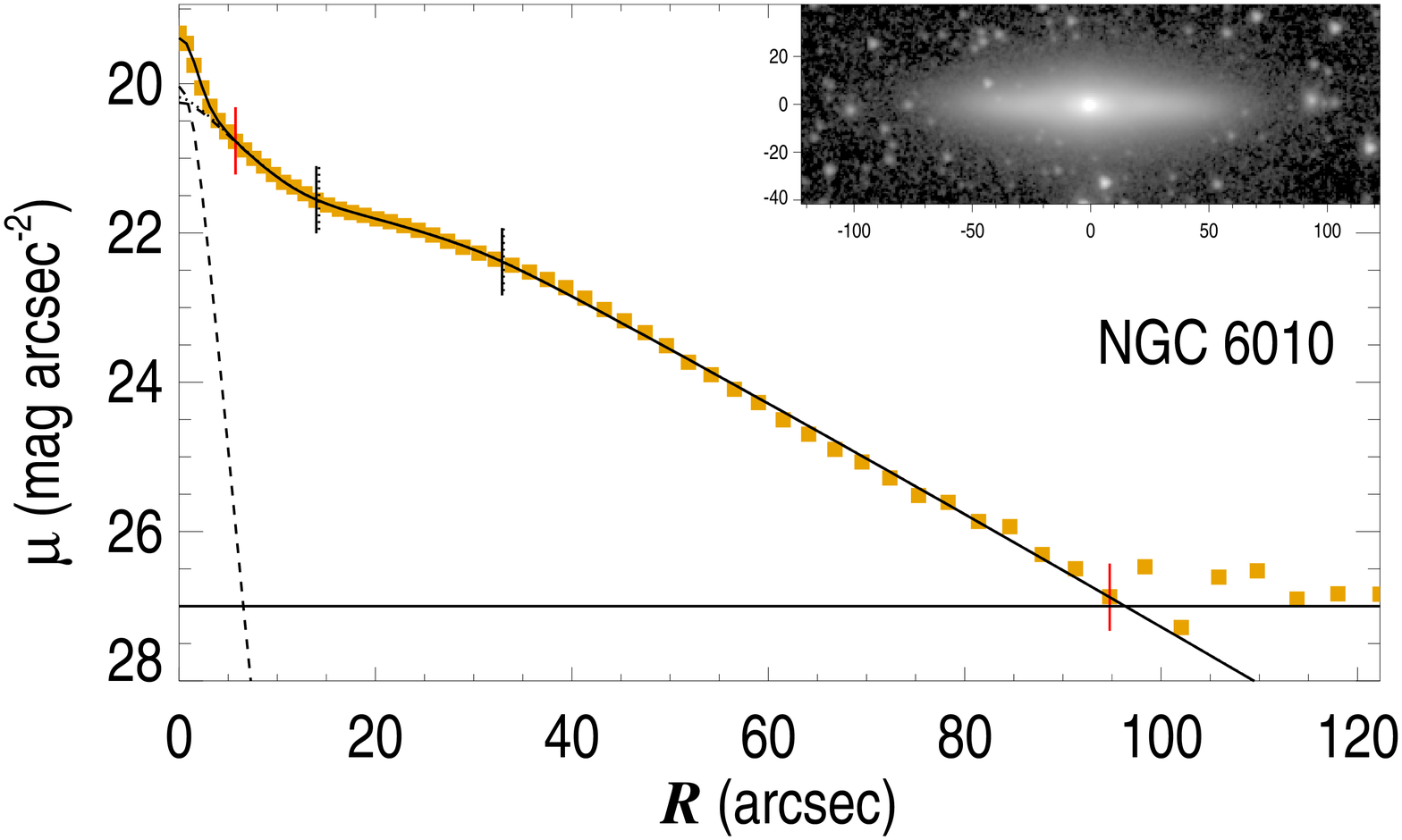}\\
\end{figure}

\begin{figure}
  \includegraphics[width=0.45\textwidth]{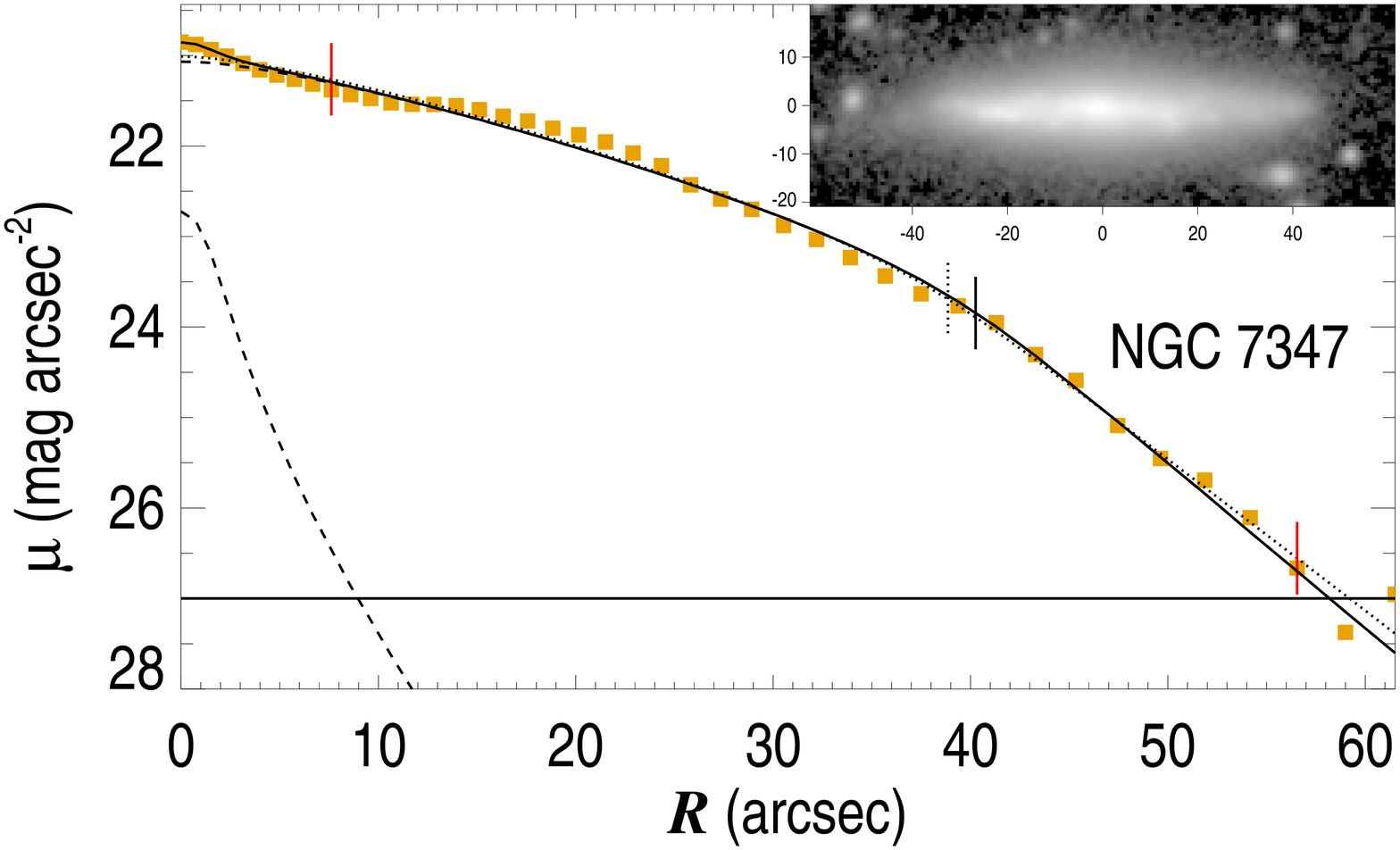}\\
\end{figure}

\clearpage

\begin{figure}
  \includegraphics[width=0.45\textwidth]{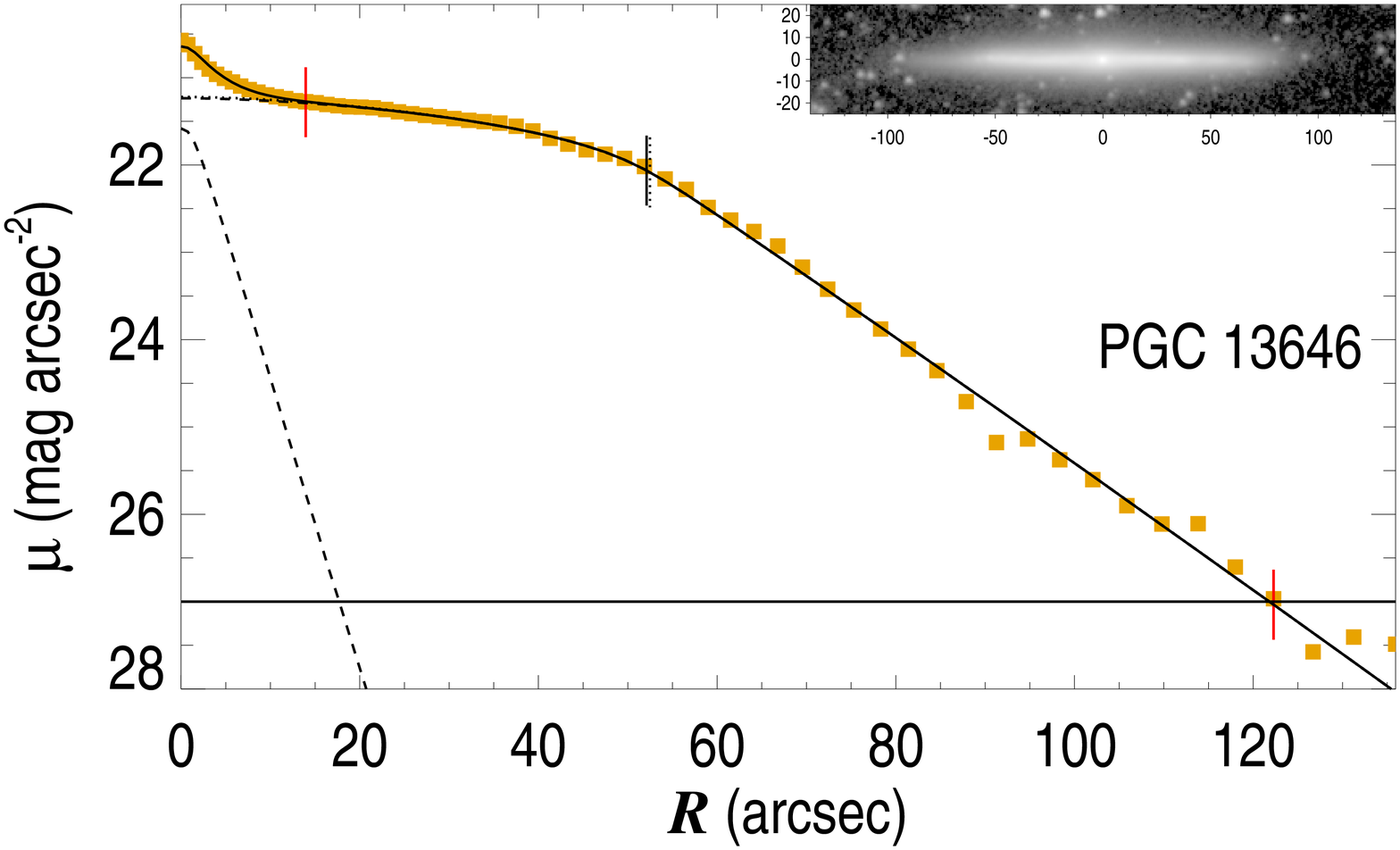}\\
\end{figure}

\begin{figure}
  \includegraphics[width=0.45\textwidth]{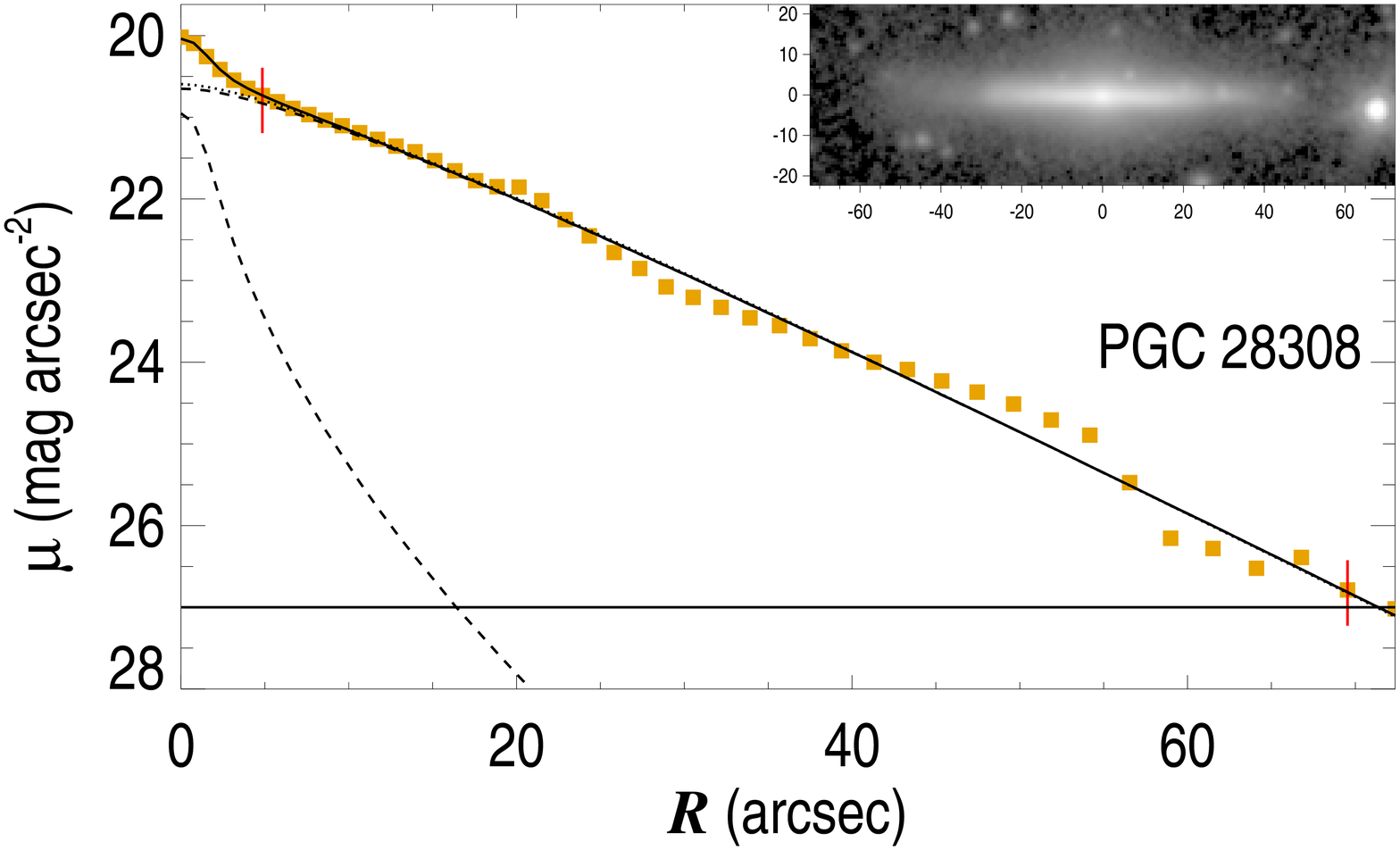}\\
\end{figure}

\begin{figure}
  \includegraphics[width=0.45\textwidth]{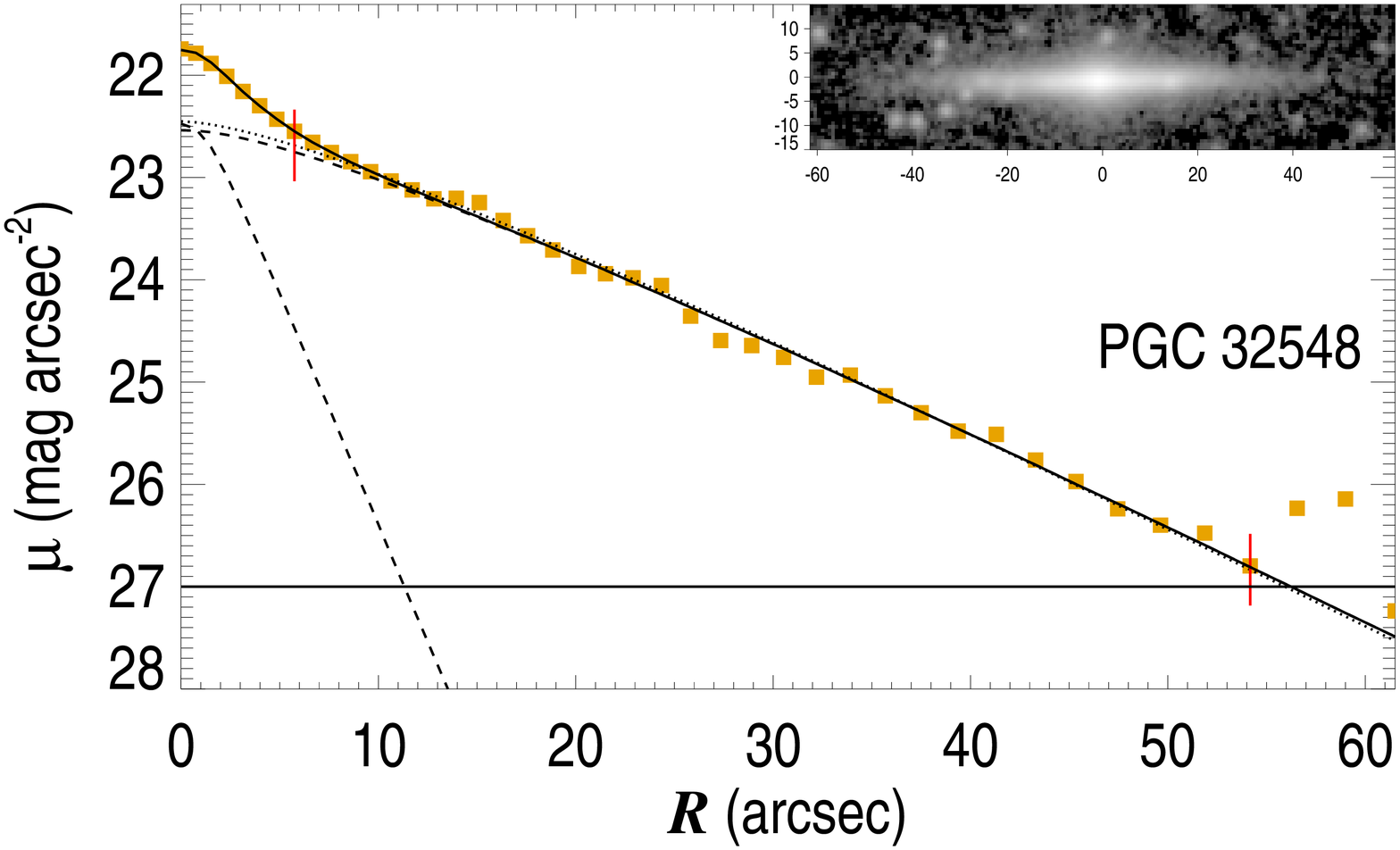}\\
\end{figure}

\begin{figure}
  \includegraphics[width=0.45\textwidth]{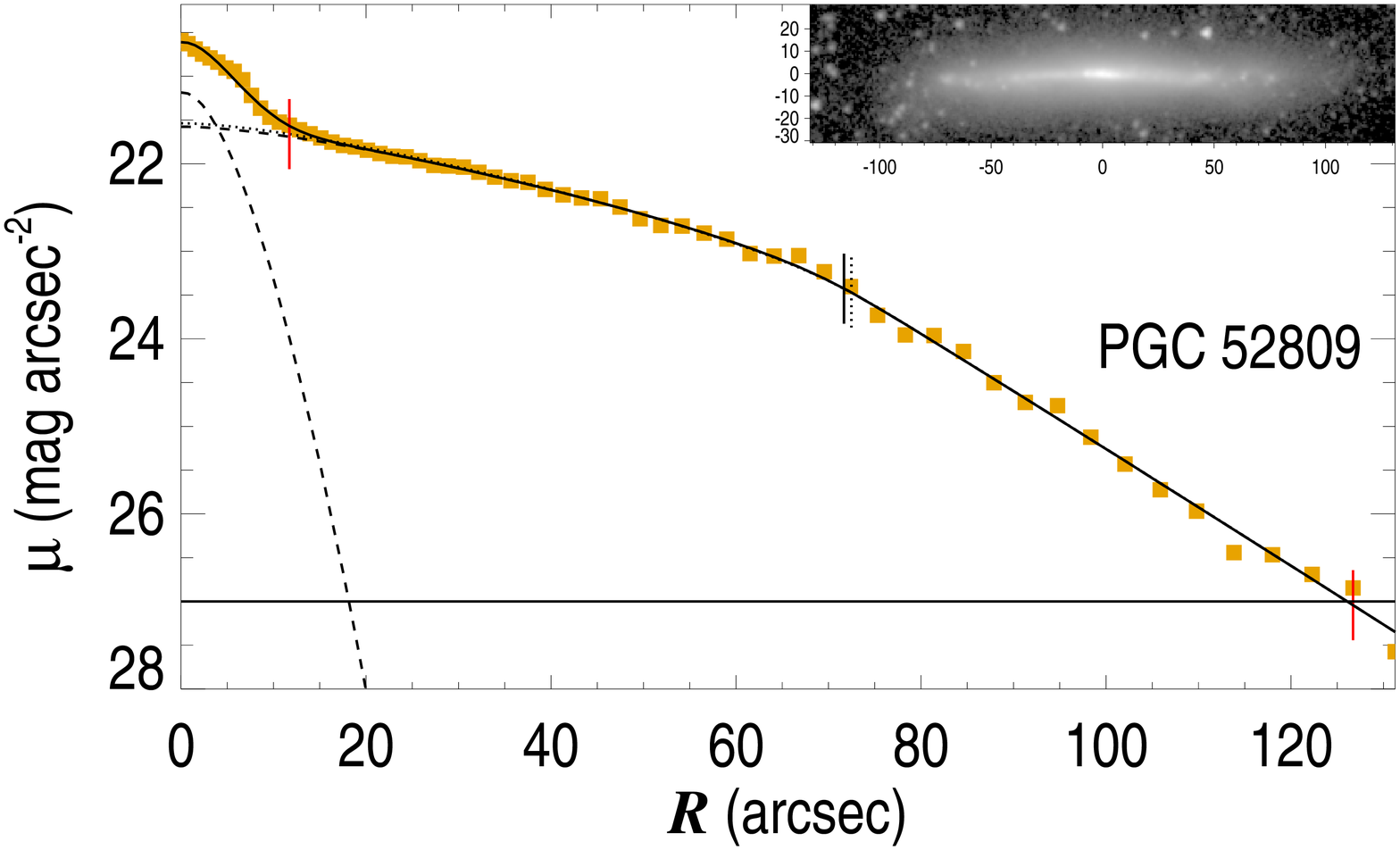}\\
\end{figure}

\begin{figure}
  \includegraphics[width=0.45\textwidth]{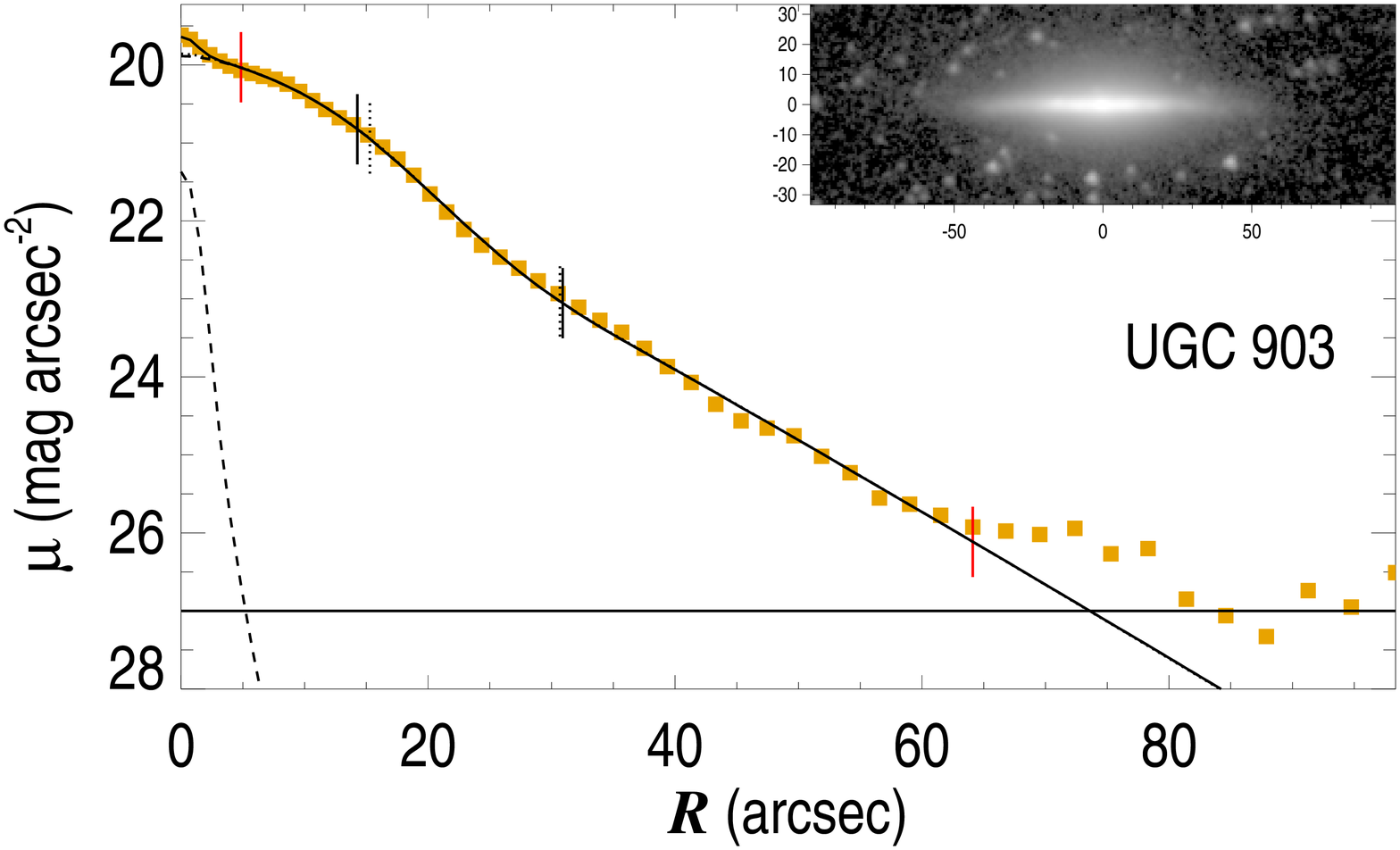}\\
\end{figure}

\begin{figure}
  \includegraphics[width=0.45\textwidth]{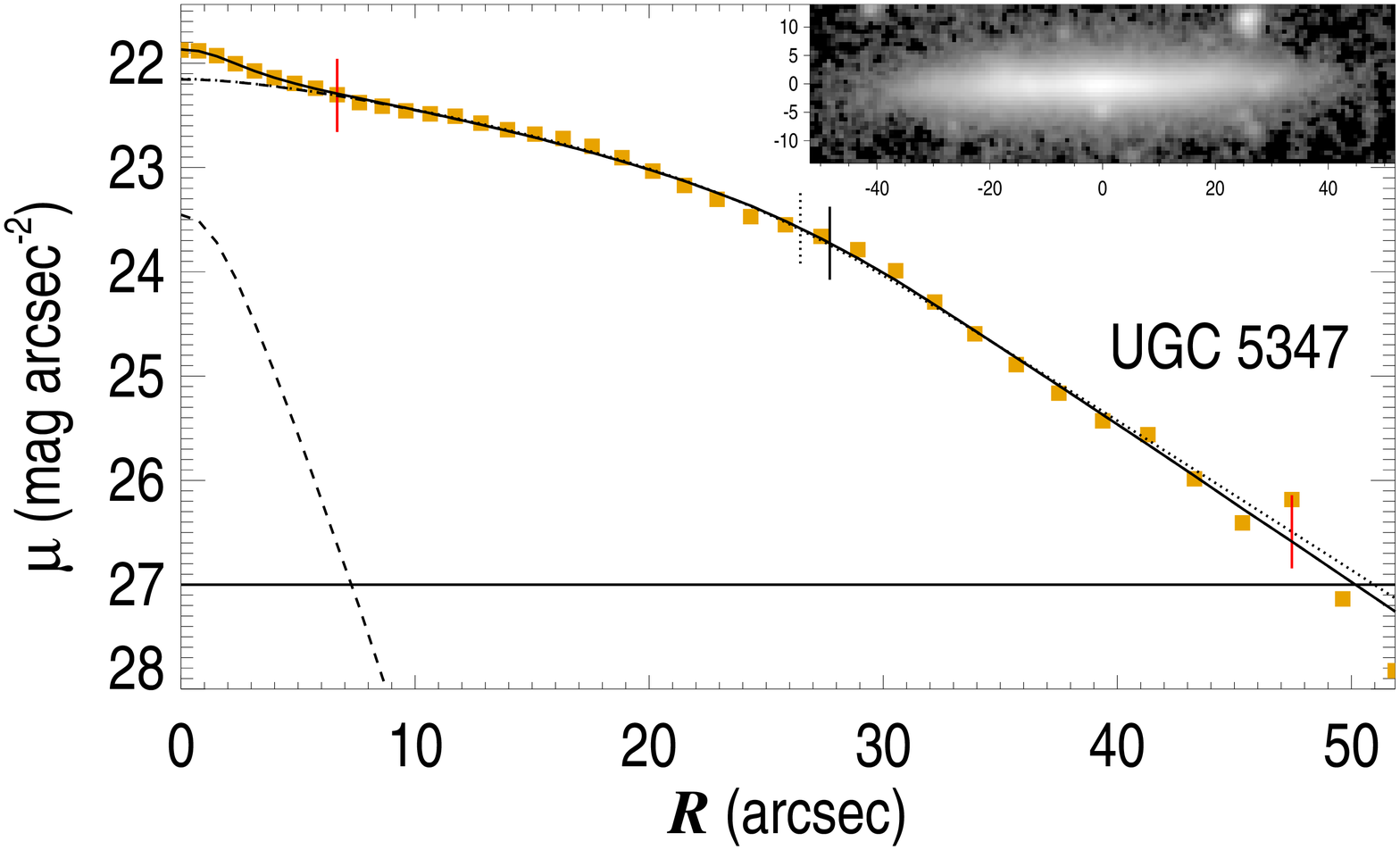}\\
\end{figure}

\begin{figure}
  \includegraphics[width=0.45\textwidth]{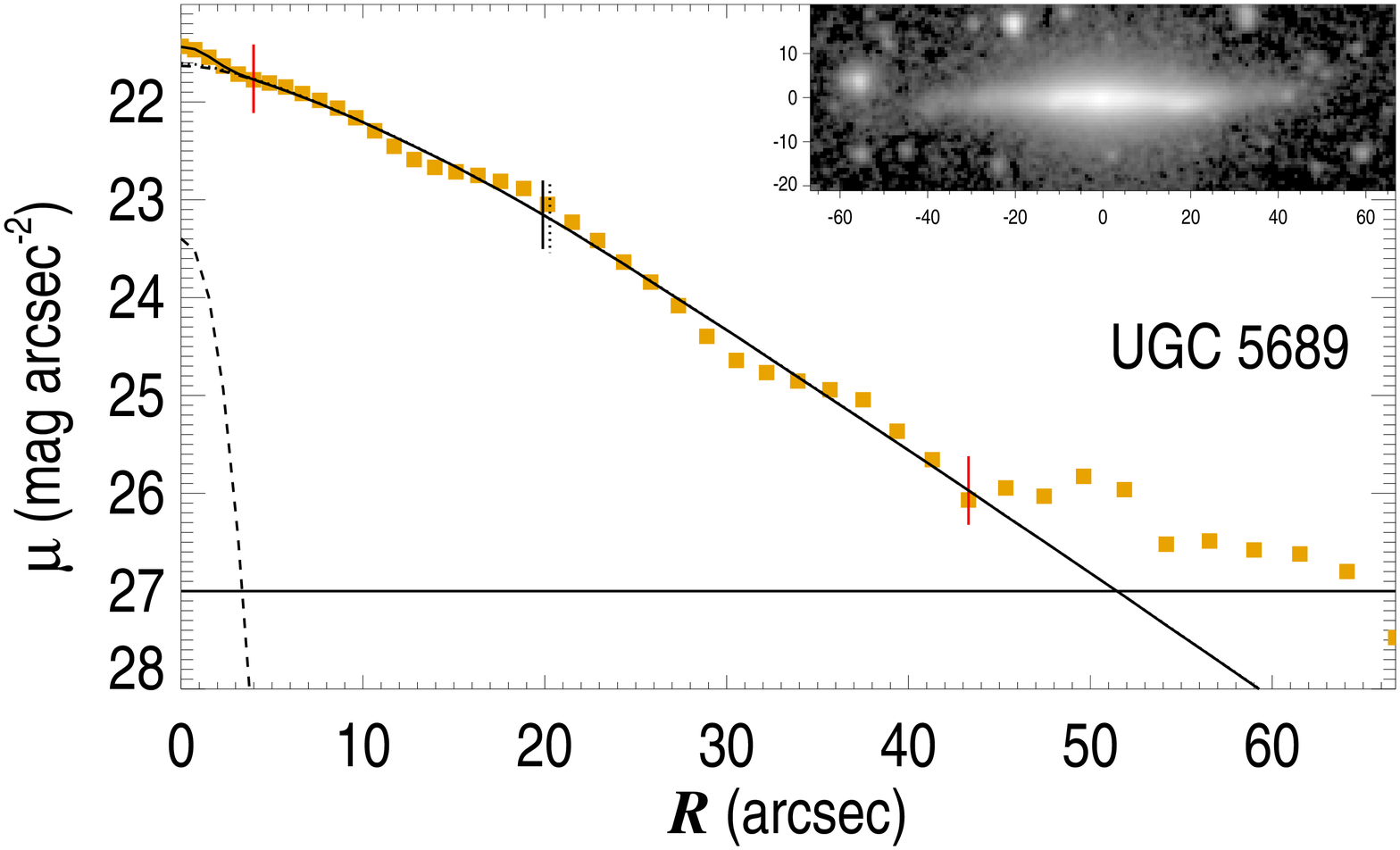}\\
\end{figure}

\begin{figure}
  \includegraphics[width=0.45\textwidth]{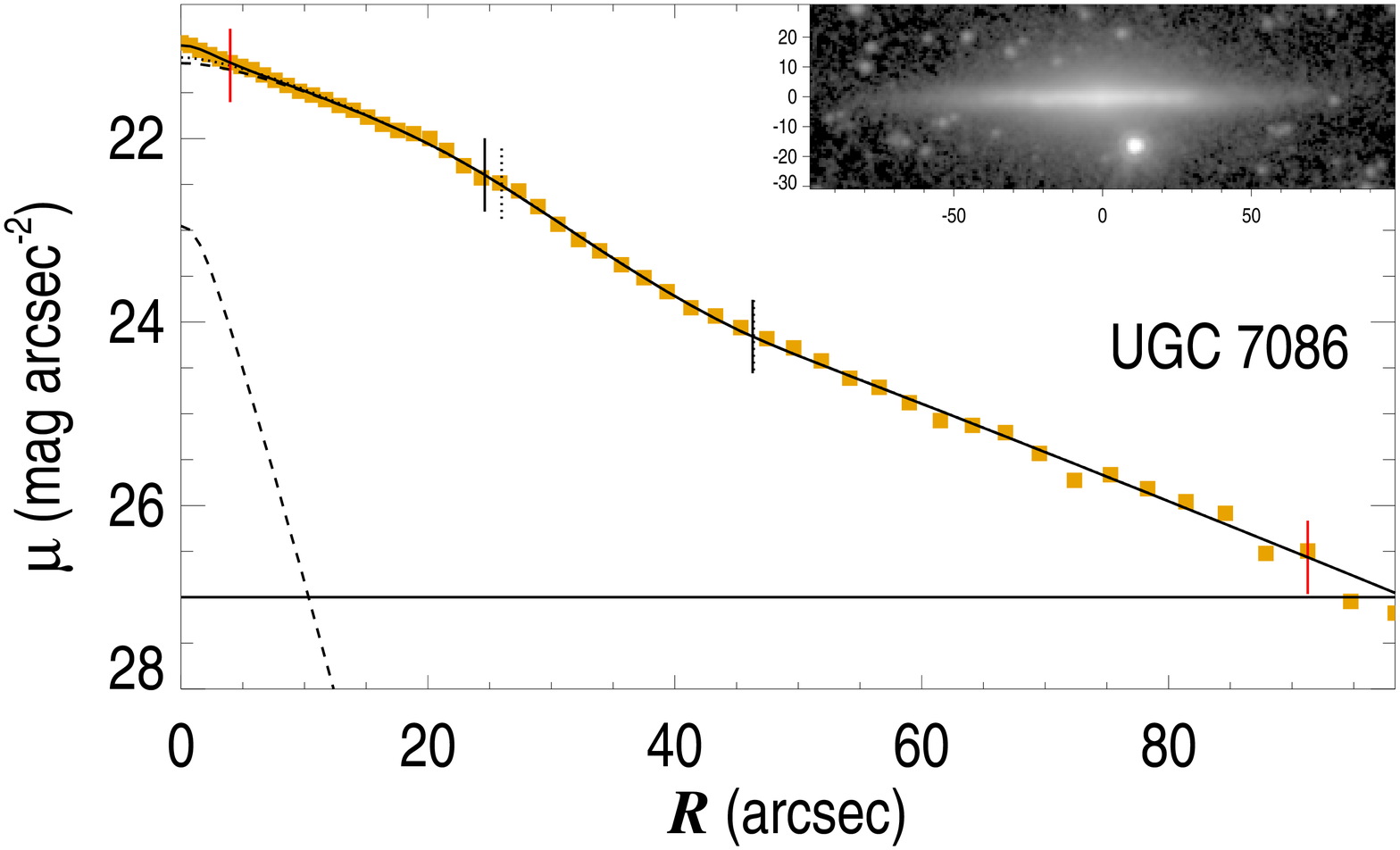}\\
\end{figure}

\clearpage

\begin{figure}
  \includegraphics[width=0.45\textwidth]{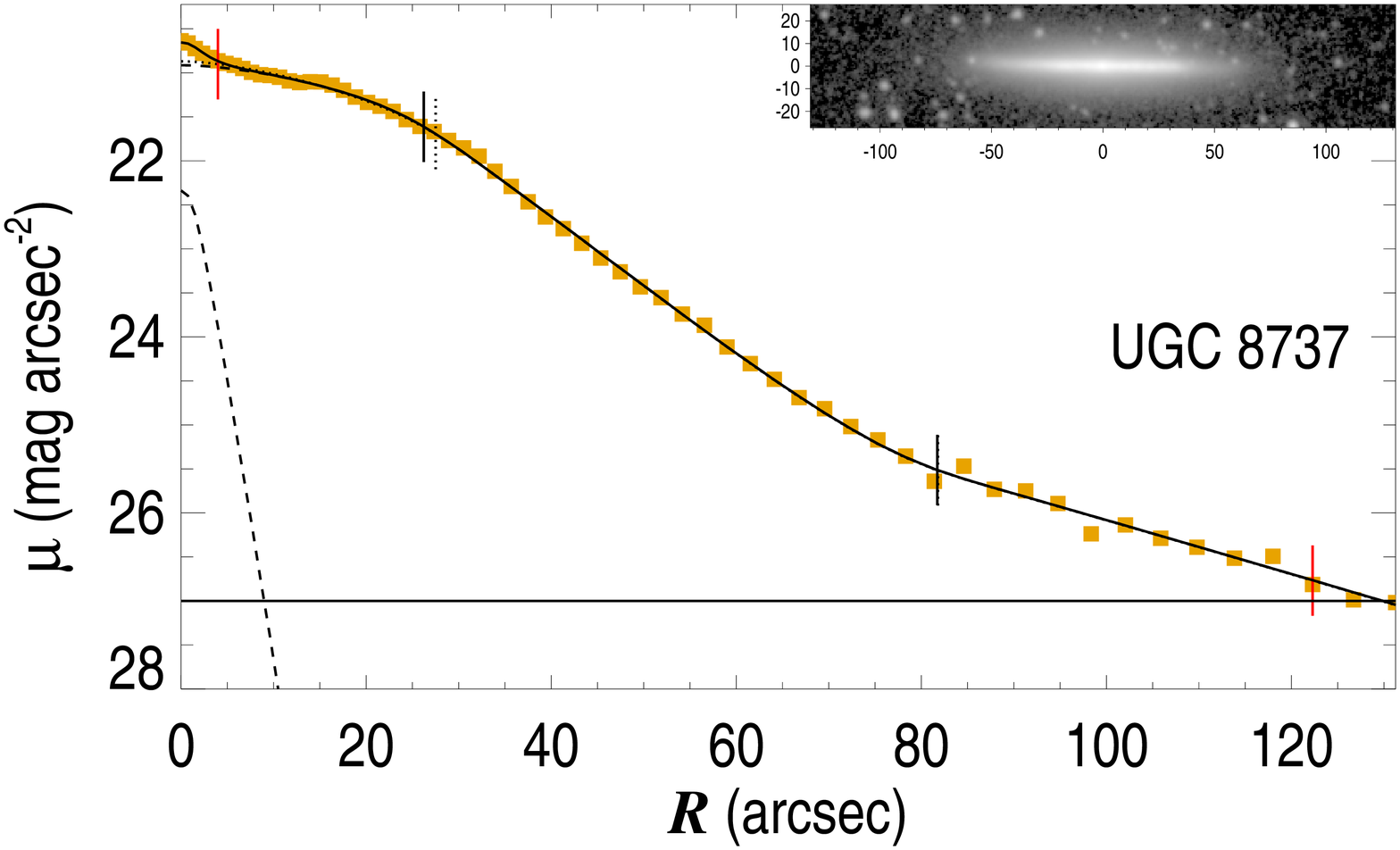}\\
\end{figure}

\begin{figure}
  \includegraphics[width=0.45\textwidth]{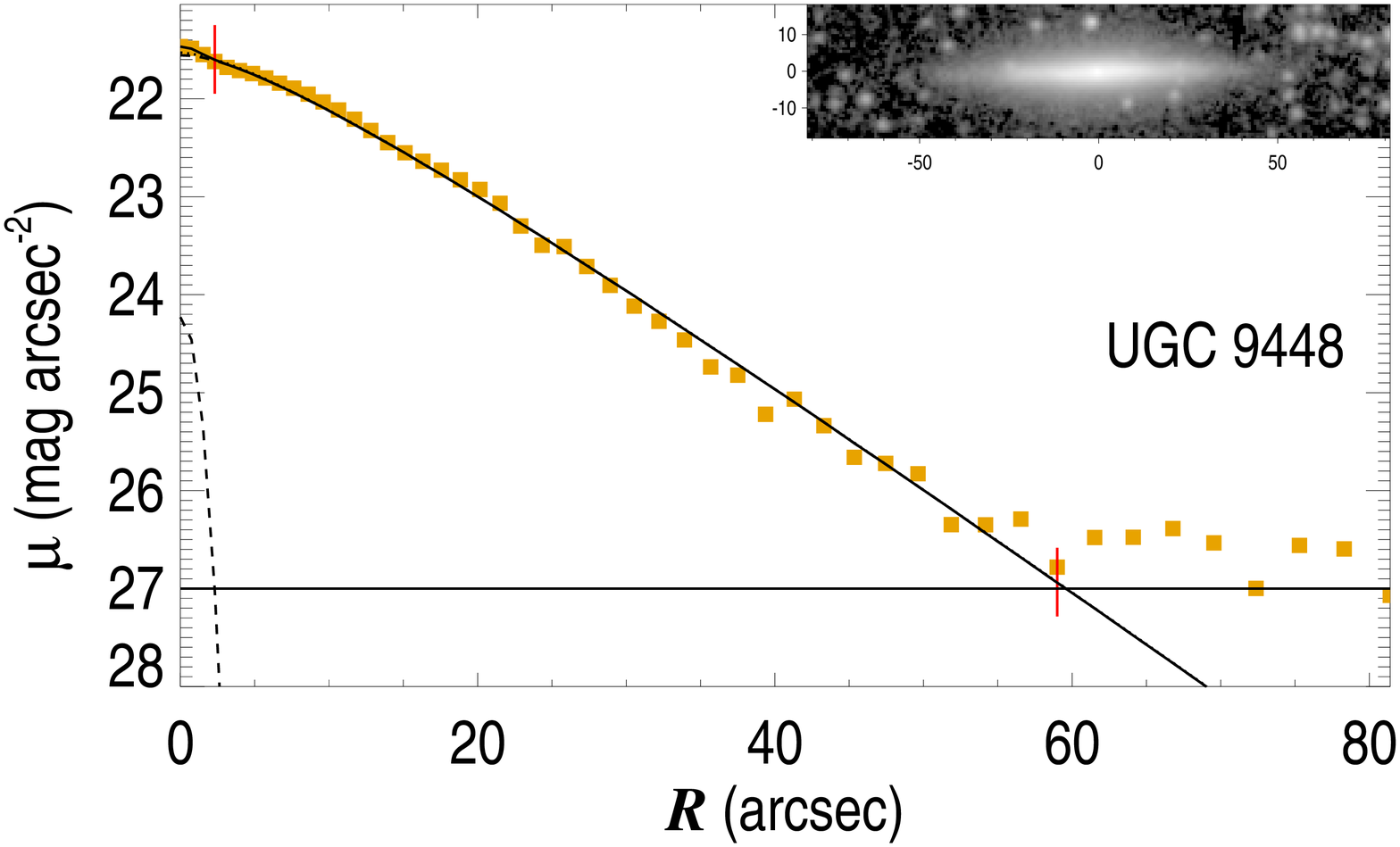}\\
\end{figure}

\begin{figure}
  \includegraphics[width=0.45\textwidth]{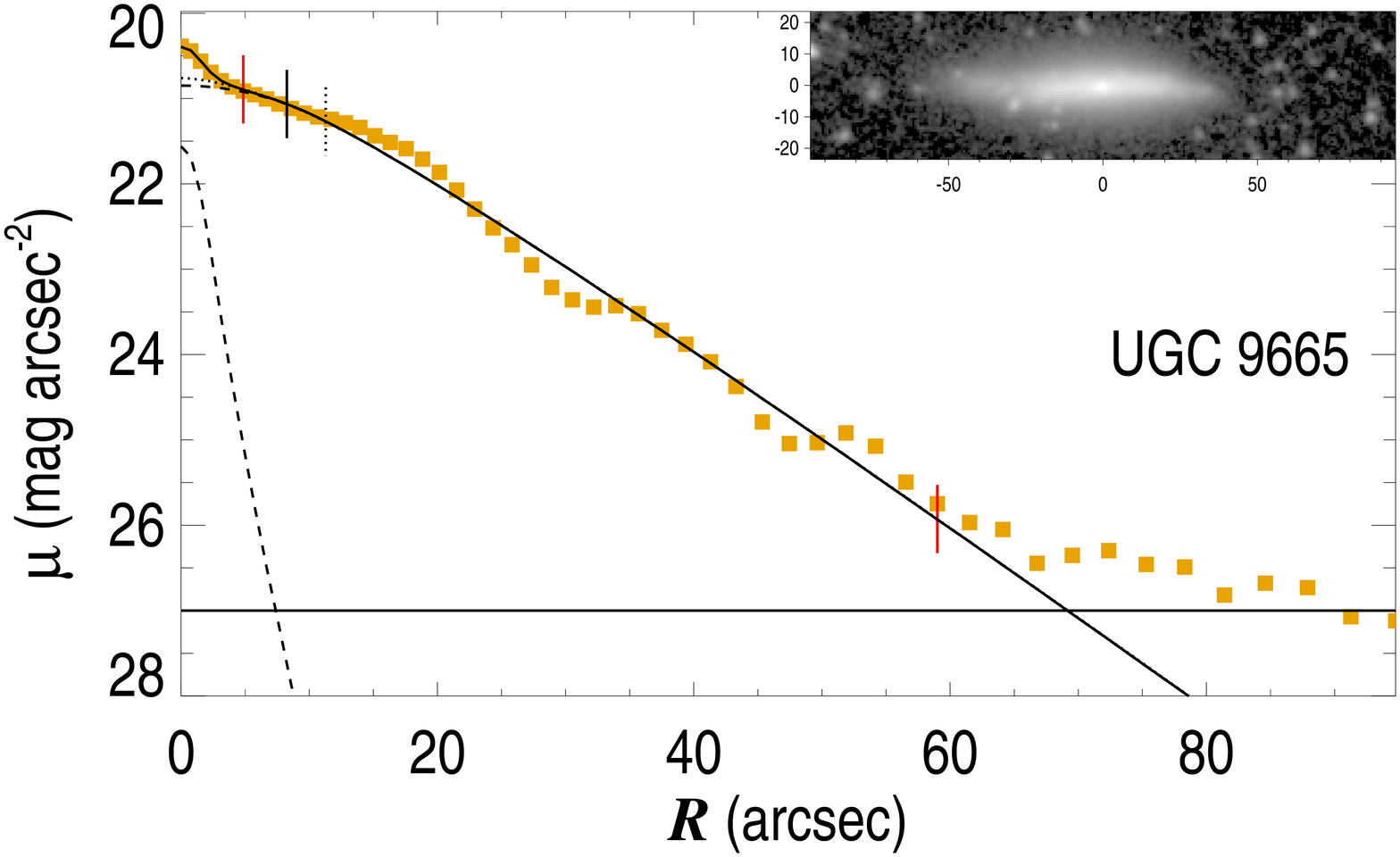}\\
\end{figure}

\begin{figure}
  \includegraphics[width=0.45\textwidth]{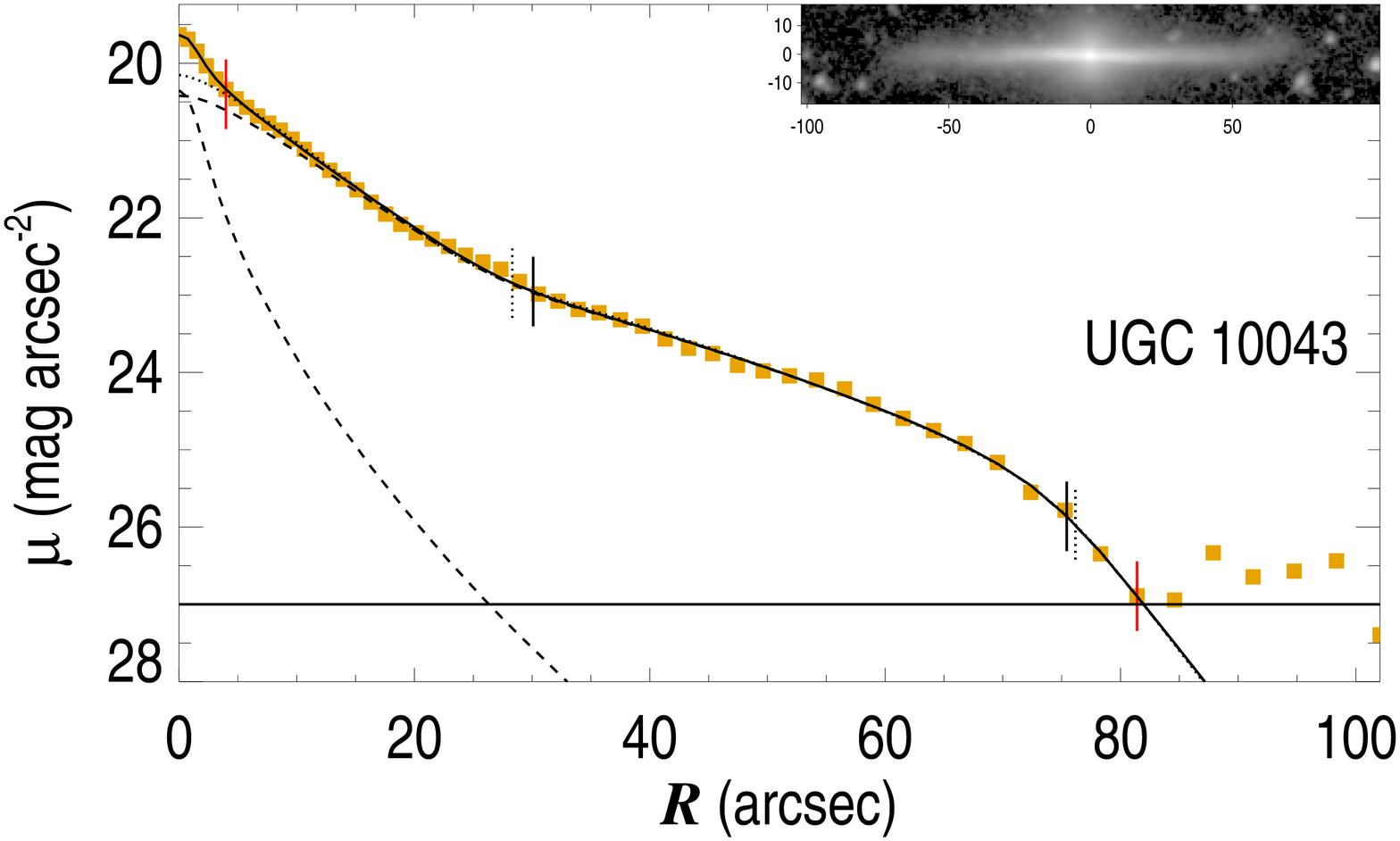}\\
\end{figure}

\begin{figure}
  \includegraphics[width=0.45\textwidth]{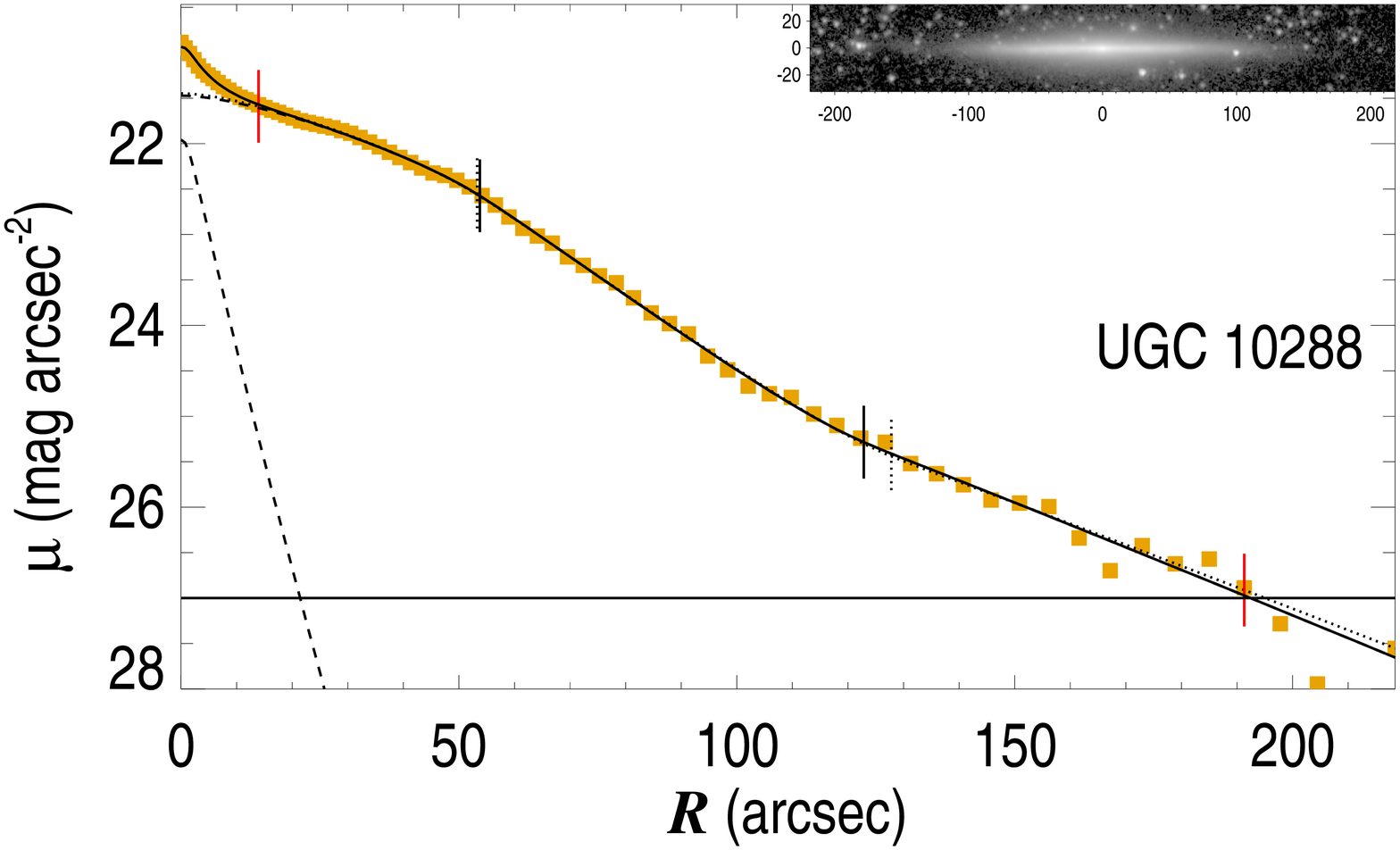}\\
\end{figure}

\begin{figure}
  \includegraphics[width=0.45\textwidth]{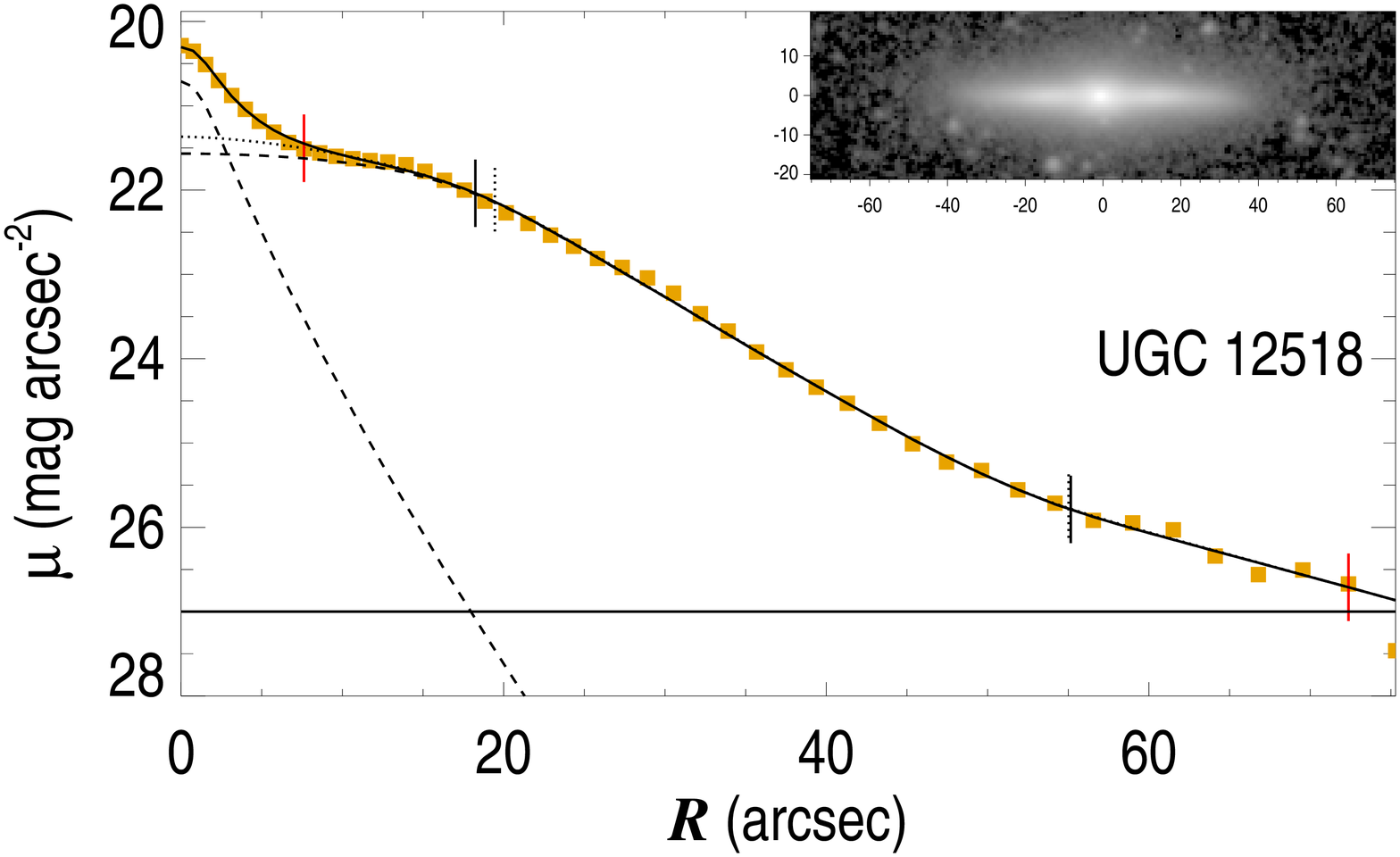}\\
\end{figure}

\begin{figure}
  \includegraphics[width=0.45\textwidth]{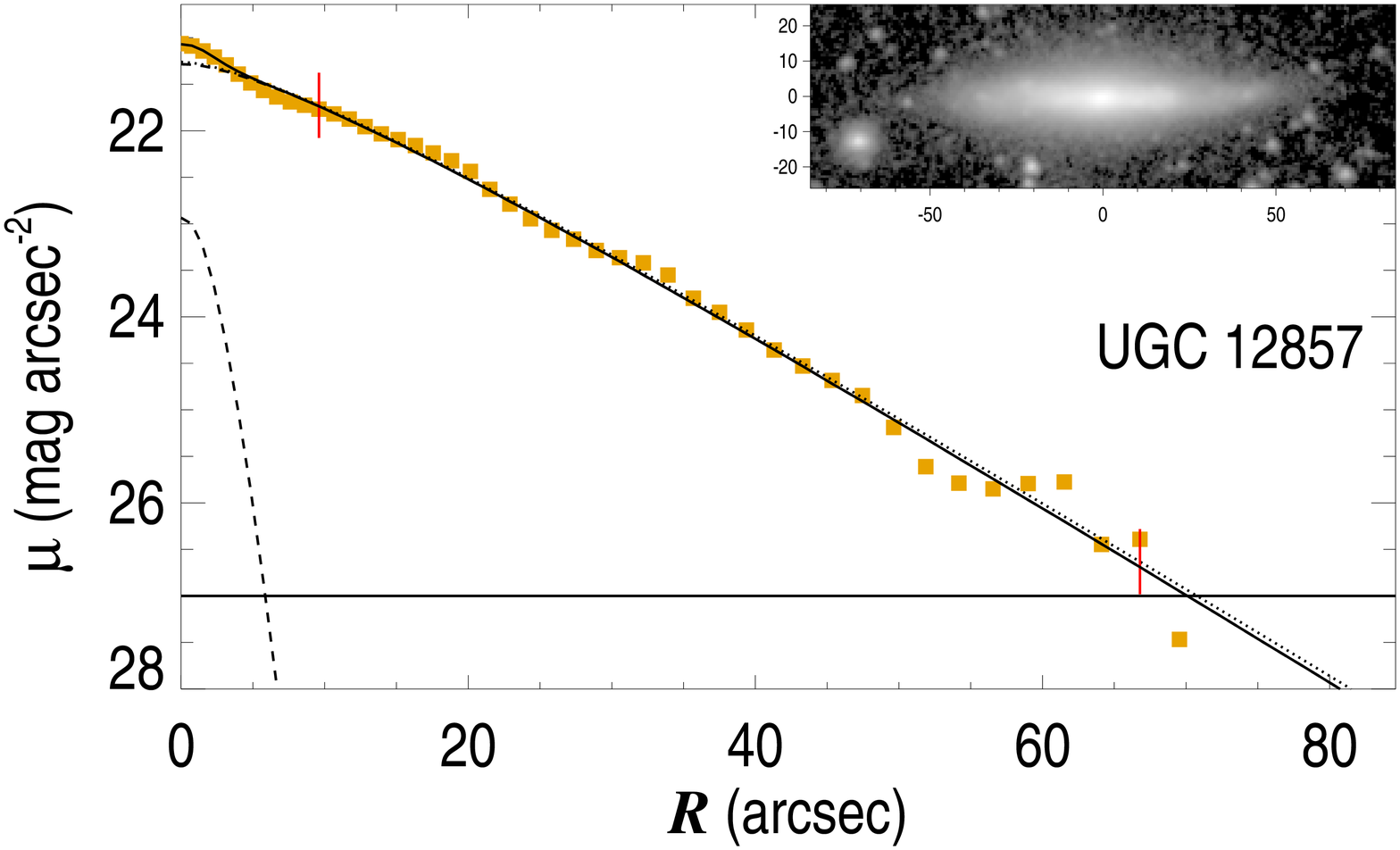}\\
\end{figure}

\end{document}